%% file: main.tex
\documentclass[10pt,journal,compsoc]{IEEEtran}
\pdfoutput=1

\input{localstyle}

\HideNotes

\begin{document}

\title{The Mason Test: A Defense Against Sybil Attacks in Wireless Networks
Without Trusted Authorities}

\author{Yue~Liu, David~R.~Bild, Robert~P.~Dick, Z.~Morley~Mao,
  and~Dan~S.~Wallach
  \IEEEcompsocitemizethanks{
    \IEEEcompsocthanksitem Y. Liu, D. R. Bild, R. P. Dick, and
    Z. M. Mao are with the Department of Electrical and Computer
    Engineering, University of Michigan, Ann Arbor,
    MI 48109.\protect\\
    E-mail: \{liuyue,drbild,dickrp,zmao\}@umich.edu
    \IEEEcompsocthanksitem D. S. Wallach is with the Department of Computer Science, Rice University, Houston, TX 77005.\protect\\
    E-mail: dwallach@cs.rice.edu}%
  \thanks{This work was supported by the National Science Foundation
    under award TC-0964545.}%
}

\IEEEcompsoctitleabstractindextext{
\input{abs}
}

\maketitle
\IEEEdisplaynotcompsoctitleabstractindextext

\input{intro}

\input{background}
\input{separation}
\input{algorithm}
\input{predictability}

\input{motion}
\input{mason}
\input{evaluation}

\input{discussion}

\input{conclusion}

\bibliographystyle{IEEEtran}
\bibliography{robbib,rdgroup} 

\end{document}

%% file: localstyle.tex
\usepackage[utf8]{inputenc}
\usepackage[T1]{fontenc}
\usepackage{float} 
\usepackage{microtype}
\usepackage{amsthm,amsmath,amssymb}
\usepackage{booktabs}
\usepackage{siunitx}
\usepackage{multirow}
\usepackage{tikz}

\usepackage[bookmarks=false,hidelinks]{hyperref}

\usepackage{RobStd}
\usepackage{algorithm}
\usepackage{algpseudocode}

\newcommand*\circled[2][char]{\tikz[baseline=(#1.base)]{
            \node[shape=circle,draw,inner sep=0.5pt] (#1) {#2};}}

\makeatletter
\let\IEEEmakecaption\@makecaption
\makeatother
\usepackage[textfont=sf,font=normal]{subcaption}
\makeatletter
\let\@makecaption\IEEEmakecaption
\makeatother

\usepackage{fixltx2e}

\graphicspath{{fig/}}

\newcommand{\figinput}[1]{%
\providecommand{\fns}{\footnotesize}%
\providecommand{\sss}{\scriptsize}%
\providecommand{\ts}{\tiny}%
{\fontsize{10}{12}\selectfont{\textsf{\input{fig/#1}}}}%
\let\fns\undefined%
\let\ss\undefined
\let\ts\undefined
}

\theoremstyle{plain}
\newtheorem{theorem}{Theorem}
\newtheorem{lemma}{Lemma}

\theoremstyle{definition}
\newtheorem{restrict}{Restriction}
\newtheorem{condition}{Condition}
\newtheorem*{assumption}{Assumption}
\newtheorem*{definition}{Definition}

\newcommand{\vect}[2][]{\ensuremath{\vec{\mathbf{#2}}_\mathbf{#1}}}

\newcommand{\bigO}{\ensuremath{O}}

\DeclareMathOperator{\signalprintdistance}{d}

\DeclareSIUnit\dBm{dBm} 

%% file: abs.tex
\begin{abstract}
  Wireless networks are vulnerable to Sybil attacks, in which a
  malicious node poses as many identities in order to gain
  disproportionate influence.  Many defenses based on spatial
  variability of wireless channels exist, but depend either on
  detailed, multi-tap channel estimation---something not exposed on
  commodity 802.11 devices---or valid RSSI observations from multiple
  trusted sources, e.g., corporate access points---something not
  directly available in ad hoc and delay-tolerant networks with
  potentially malicious neighbors. We extend these techniques to be
  practical for wireless ad hoc networks of commodity 802.11 devices.
  Specifically, we propose two efficient methods for separating the
  valid RSSI observations of behaving nodes from those falsified by
  malicious participants.  Further, we note that prior signalprint
  methods are easily defeated by mobile attackers and develop an
  appropriate challenge-response defense.  Finally, we present the
  Mason test, the first implementation of these techniques for ad hoc
  and delay-tolerant networks of commodity 802.11 devices. We
  illustrate its performance in several real-world scenarios.
\end{abstract}

%% file: intro.tex
\ifCLASSOPTIONcompsoc
  \noindent\raisebox{2\baselineskip}[0pt][0pt]%
  {\parbox{\columnwidth}{\section{Introduction}\label{sec:intro}%
  \global\everypar=\everypar}}%
  \vspace{-1\baselineskip}\vspace{-\parskip}\par
\else
  \section{Introduction}\label{sec:intro}\par
\fi

\Note{Sybil attacks are a problem for wireless networks}

\IEEEPARstart{T}{he} open nature of wireless ad hoc networks
(including delay-tolerant networks~\cite{hui11nov}) enables
applications ranging from collaborative environmental
sensing~\cite{xiang12apr} to emergency
communication~\cite{gardner-stephen11aug}, but introduces numerous
security concerns since participants are not vetted.  Solutions
generally rely on a majority of the participants following a
particular protocol, an assumption that often holds because physical
nodes are expensive. However, this assumption is easily broken by a
Sybil attack. A single physical entity can pretend to be multiple
participants, gaining unfair influence at low
cost~\cite{douceur02mar}. Newsome \etal survey Sybil attacks against
various protocols~\cite{newsome04apr}, illustrating the need for a
practical defense.

\Note{Very brief survey of prior work and why not useful to us}

Proposed defenses (see Levine \etal for a survey~\cite{levine06oct})
fall into two categories. \emph{Trusted certification}
methods~\cite{zhou05nov,ramkumar05mar} use a central authority to vet
potential participants and thus are not useful in open ad hoc (and
delay-tolerant) networks.  \emph{Resource testing}
methods~\cite{borisov06sep,li12oct,yu06sep,yu08may} verify the
resources (e.g., computing capability, storage capacity, real-world
social relationships, etc.) of each physical entity.  Most are easily
defeated in ad hoc networks of resource-limited mobile devices by
attackers with access to greater resources, e.g., workstations or data
centers.

\Note{Prior signalprint work requires trust or hardware modifications}

One useful class of defenses is based on the natural spatial variation
in the wireless propagation channel, an implicit resource.  Channel
responses are uncorrelated over distances greater than half the
transmission wavelength~\cite{rappaport02} (\SI{6.25}{\centi\meter}
for \SI{2.4}{\giga\hertz} 802.11), so two transmissions with the same
channel response are likely from the same location and
device~\cite{haeberlen04sep,xiao09sep}. However, the existing Sybil
defenses based on this observation are not directly usable in open ad
hoc networks of commodity devices.

\begin{figure*}
  \centering 
  \begin{subfigure}[t]{0.3\textwidth}
    \centering
    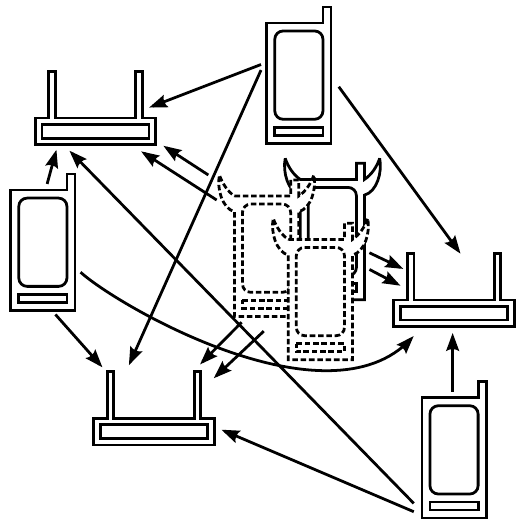
    \caption{RSSI observations from trusted APs identity the Sybils,
      $S_\textrm{1}$ and $S_\textrm{2}$, from attacker $M$.}
    \label{fig:scenario-trusted}
  \end{subfigure}
  \hfill
  \begin{subfigure}[t]{0.3\textwidth}
    \centering
    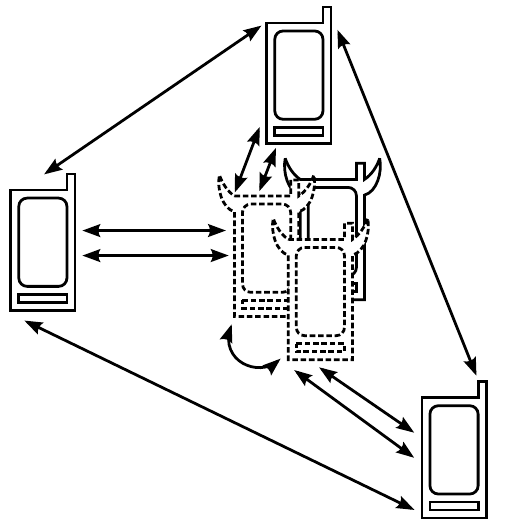
    \caption{In ad hoc networks, the participants themselves act as
      observers, but can maliciously report falsified values.}
    \label{fig:scenario-untrusted}
  \end{subfigure}
  \hfill 
  \begin{subfigure}[t]{0.3\textwidth}
    \centering
    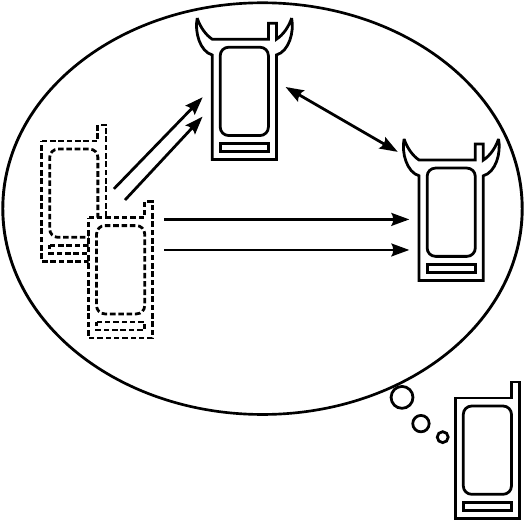
    \caption{If $I$ believes the falsified observations from
      $S_\textrm{1}$ and $S_\textrm{2}$, it will incorrectly accept
      them and reject $A$ and $B$ as Sybil.}
    \label{fig:scenario-false}
  \end{subfigure}
  \caption{Prior work~\cite{xiao09sep,faria06sep} assumes trusted RSSI
    observations, not generally available in ad hoc and delay-tolerant
    networks. We present a technique for a participant to separate
    true and false observations, enabling use in ad hoc
    networks. (Arrows point from transmitter to observer.)}
  \label{fig:scenarios}
\end{figure*}

Xiao \etal observe that in OFDM-based 802.11 the coherence bandwidth
is much smaller than the system bandwidth and thus the channel
response estimates at well-spaced frequency taps are uncorrelated,
forming a vector unique to the transmitter location and robust to
changes in transmitter power~\cite{xiao09sep}. Unfortunately commodity
802.11 devices do not expose these estimates to the driver and
operating system, restricting this technique to specialized hardware
and access points.

Commodity devices do expose an aggregate, scalar value, the received
signal strength. RSSI is not robust to changes in transmitter power,
so a vector of observations from multiple receivers---a
\emph{signalprint}---is used instead.  Several authors have proposed
such
methods~\cite{xiao09sep,faria06sep,demirbas06jun,li06sep-b,li07dec,chen10jun-b,suen05nov}
assuming trusted, true observations. In open ad hoc networks,
observations are untrusted, coming from potentially-lying neighbors,
as illustrated in \autoref{fig:scenarios}. Trust-less methods have
been proposed, but have various limitations (e.g., devices must be
non-mobile~\cite{shaohe08dec}, colluding attackers can defeat the
scheme~\cite{abbas09dec}, or are limited to outdoor environments with
predictable propagation ranges~\cite{bouassida09jul}).  Instead, a
general method to separate true and false observations is needed.

\Note{Brief overview of our separation contribution}

We observe that with high probability attackers cannot produce false
observations that make conforming identities look Sybil, due to the
unpredictability of the wireless channels. We exploit this weakness to
bound the number of misclassified identities. In cases when conforming
nodes outnumber physical attacking nodes (a major motivating factor
for the Sybil attack), we develop a notion of consistency that enables
fully-accurate classification.

\Note{Prior work doesn't deal with moving attackers}

Most past work assumes nodes are stationary, as moving attacks can
easily defeat signalprint-based detection.  As noted, but not pursued,
by Xiao \etal, successive transmissions from the same node should have
the same signalprint and attackers likely cannot quickly (i.e., in
milliseconds) switch between precise positions~\cite{xiao09sep}. We
develop a challenge--response protocol from this idea and study its
performance on real deployments.

\Note{Contributions}
We make the following primary contributions.
\begin{itemize}
\item We design two methods of $\bigO(n^3)$ complexity to separate
  true and false RSSI observations, enabling signalprint-based Sybil
  detection in ad hoc networks of mutually distrusting nodes.  The
  first method gives partial separation, bounding the number of
  misclassified identities. The second provides full separation, but
  works only when conforming nodes outnumber the physical attacking
  nodes.
\item We prove conditions under which a participant can fully separate
  true and false observations.
\item We develop a challenge-response protocol to detect attackers
  attempting to use motion to defeat the signalprint-based Sybil
  defense.
\item We describe the Mason test, a practical protocol for Sybil
  defense based on these ideas.  We implemented the Mason test as a
  Linux kernel module for 802.11 ad hoc
  networks\footnote{\url{http://github.com/EmbeddedAtUM/mason/}} and
  characterize its performance in real-world scenarios.
\end{itemize}

\Note{ The remainder of the paper is organized as follows.
  \autoref{sec:background} describes our general problem formulation
  and briefly reviews necessary background information.
  \autoref{sec:separation} proves conditions under which neighbors
  reporting false observations can be detected and
  \autoref{sec:algorithm} describes efficient algorithms for the
  separation.  \autoref{sec:motion} describes our challenge--response
  detection scheme for moving attackers. \autoref{sec:mason} presents
  the Mason test, a complete protocol based on these ideas for Sybil
  detection in open ad hoc networks and \autoref{sec:evaluation}
  illustrates its effectiveness in several real-world deployments. }

%% file: fig/scenario-trusted.pdf_tex
\begingroup%
  \makeatletter%
  \providecommand\color[2][]{%
    \errmessage{(Inkscape) Color is used for the text in Inkscape, but the package 'color.sty' is not loaded}%
    \renewcommand\color[2][]{}%
  }%
  \providecommand\transparent[1]{%
    \errmessage{(Inkscape) Transparency is used (non-zero) for the text in Inkscape, but the package 'transparent.sty' is not loaded}%
    \renewcommand\transparent[1]{}%
  }%
  \providecommand\rotatebox[2]{#2}%
  \ifx\svgwidth\undefined%
    \setlength{\unitlength}{151.2bp}%
    \ifx\svgscale\undefined%
      \relax%
    \else%
      \setlength{\unitlength}{\unitlength * \real{\svgscale}}%
    \fi%
  \else%
    \setlength{\unitlength}{\svgwidth}%
  \fi%
  \global\let\svgwidth\undefined%
  \global\let\svgscale\undefined%
  \makeatother%
  \begin{picture}(1,1)%
    \put(0,0){\includegraphics[width=\unitlength]{scenario-trusted.pdf}}%
    \put(0.86473923,0.12414906){\color[rgb]{0,0,0}\makebox(0,0)[b]{\smash{I}}}%
    \put(0.08167747,0.5196685){\color[rgb]{0,0,0}\makebox(0,0)[b]{\smash{A}}}%
    \put(0.56878254,0.8379775){\color[rgb]{0,0,0}\makebox(0,0)[b]{\smash{B}}}%
    \put(0.50749638,0.46748473){\color[rgb]{0,0,0}\makebox(0,0)[b]{\smash{S\textsubscript{1}}}}%
    \put(0.63330801,0.58717015){\color[rgb]{0,0,0}\makebox(0,0)[b]{\smash{M}}}%
    \put(0.61000973,0.4357035){\color[rgb]{0,0,0}\makebox(0,0)[b]{\smash{S\textsubscript{2}}}}%
  \end{picture}%
\endgroup%

%% file: fig/scenario-untrusted.pdf_tex
\begingroup%
  \makeatletter%
  \providecommand\color[2][]{%
    \errmessage{(Inkscape) Color is used for the text in Inkscape, but the package 'color.sty' is not loaded}%
    \renewcommand\color[2][]{}%
  }%
  \providecommand\transparent[1]{%
    \errmessage{(Inkscape) Transparency is used (non-zero) for the text in Inkscape, but the package 'transparent.sty' is not loaded}%
    \renewcommand\transparent[1]{}%
  }%
  \providecommand\rotatebox[2]{#2}%
  \ifx\svgwidth\undefined%
    \setlength{\unitlength}{151.2bp}%
    \ifx\svgscale\undefined%
      \relax%
    \else%
      \setlength{\unitlength}{\unitlength * \real{\svgscale}}%
    \fi%
  \else%
    \setlength{\unitlength}{\svgwidth}%
  \fi%
  \global\let\svgwidth\undefined%
  \global\let\svgscale\undefined%
  \makeatother%
  \begin{picture}(1,1)%
    \put(0,0){\includegraphics[width=\unitlength]{scenario-untrusted.pdf}}%
    \put(0.86473923,0.12414906){\color[rgb]{0,0,0}\makebox(0,0)[b]{\smash{I}}}%
    \put(0.08167747,0.5196685){\color[rgb]{0,0,0}\makebox(0,0)[b]{\smash{A}}}%
    \put(0.56878254,0.8379775){\color[rgb]{0,0,0}\makebox(0,0)[b]{\smash{B}}}%
    \put(0.50749638,0.46748473){\color[rgb]{0,0,0}\makebox(0,0)[b]{\smash{S\textsubscript{1}}}}%
    \put(0.63330801,0.58717015){\color[rgb]{0,0,0}\makebox(0,0)[b]{\smash{M}}}%
    \put(0.61000973,0.4357035){\color[rgb]{0,0,0}\makebox(0,0)[b]{\smash{S\textsubscript{2}}}}%
  \end{picture}%
\endgroup%

%% file: fig/scenario-false.pdf_tex
\begingroup%
  \makeatletter%
  \providecommand\color[2][]{%
    \errmessage{(Inkscape) Color is used for the text in Inkscape, but the package 'color.sty' is not loaded}%
    \renewcommand\color[2][]{}%
  }%
  \providecommand\transparent[1]{%
    \errmessage{(Inkscape) Transparency is used (non-zero) for the text in Inkscape, but the package 'transparent.sty' is not loaded}%
    \renewcommand\transparent[1]{}%
  }%
  \providecommand\rotatebox[2]{#2}%
  \ifx\svgwidth\undefined%
    \setlength{\unitlength}{151.2bp}%
    \ifx\svgscale\undefined%
      \relax%
    \else%
      \setlength{\unitlength}{\unitlength * \real{\svgscale}}%
    \fi%
  \else%
    \setlength{\unitlength}{\svgwidth}%
  \fi%
  \global\let\svgwidth\undefined%
  \global\let\svgscale\undefined%
  \makeatother%
  \begin{picture}(1,1)%
    \put(0,0){\includegraphics[width=\unitlength]{scenario-false.pdf}}%
    \put(0.14057849,0.61554924){\color[rgb]{0,0,0}\makebox(0,0)[b]{\smash{A}}}%
    \put(0.92826141,0.12362332){\color[rgb]{0,0,0}\makebox(0,0)[b]{\smash{I}}}%
    \put(0.2293287,0.46973884){\color[rgb]{0,0,0}\makebox(0,0)[b]{\smash{B}}}%
    \put(0.85843356,0.58683268){\color[rgb]{0,0,0}\makebox(0,0)[b]{\smash{S\textsubscript{2}}}}%
    \put(0.46434388,0.81698851){\color[rgb]{0,0,0}\makebox(0,0)[b]{\smash{S\textsubscript{1}}}}%
  \end{picture}%
\endgroup%

%% file: background.tex
\section{Problem Formulation and \\Background}
\label{sec:background}

In this section, we define our problem, summarize the solution
framework, describe our attack model, and briefly review the
signalprint method.

\subsection{Problem Formulation}
\label{sec:problem}

\begin{figure*}
  \centering 
  \begin{subfigure}[t]{0.3\textwidth}
    \centering
    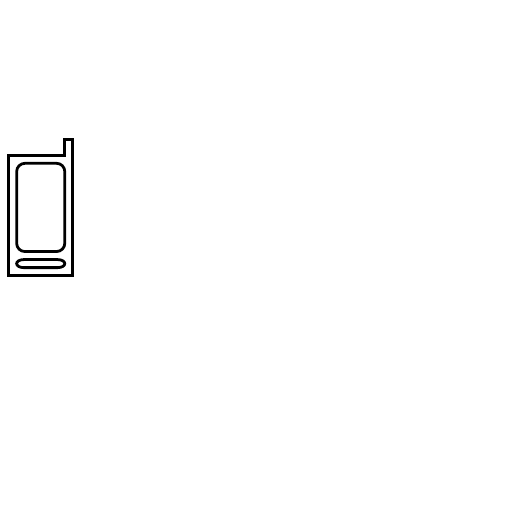
    \caption{Nodes record their observed RSSIs of probes broadcast by
      neighbors. A and B have sent; C, D, and E are next.}
    \label{fig:formulation-hello}
  \end{subfigure}
  \hfill
  \begin{subfigure}[t]{0.3\textwidth}
    \centering
    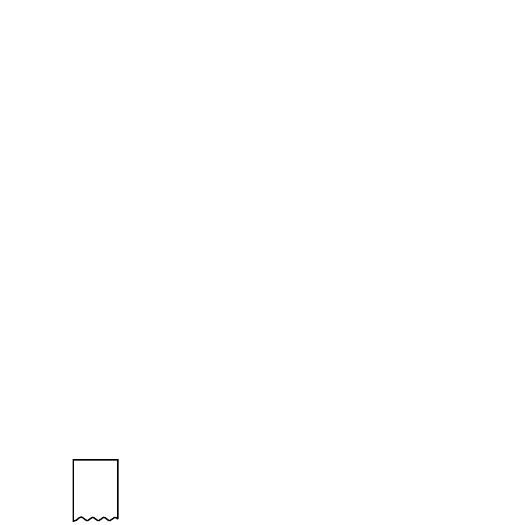
    \caption{RSSI observations are shared among all
      participants. Malicious nodes may lie about their observations.}
    \label{fig:formulation-exchange}
  \end{subfigure}
  \hfill
  \begin{subfigure}[t]{0.3\textwidth}
    \centering
    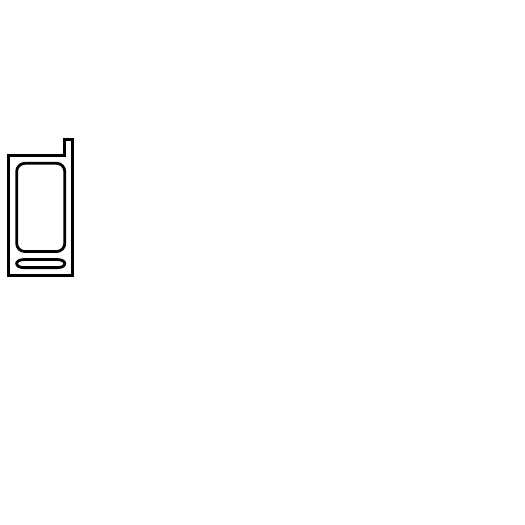
    \caption{Each participant selects a subset of the observations to
      form signalprints for Sybil detection.}
    \label{fig:formulation-comparison}
  \end{subfigure}
  \caption{The solution framework for signalprint-based Sybil
    detection in ad hoc networks. This paper fleshes out this concept
    into a safe and secure protocol, the Mason test.}
  \label{fig:formulation}
\end{figure*}

Our high-level goal is to allow a wireless network participant to
occasionally determine which of its one-hop neighbors are
non-Sybil. These identities may safely participate in other
protocols. In mobile networks, the process must be repeated
occasionally (e.g., once per hour) as the topology changes. Safety is
more important than system performance, so nearly all Sybil identities
should be identified, but some non-Sybils may be rejected.

Prior work showed the effectiveness of signalprint techniques with
trusted RSSI observers.  We extend those methods to work without a
priori trust in any observers.  As illustrated in
\autoref{fig:formulation}, we assume an arbitrary identity (or
condition) starts the process.  Participants take turns broadcasting a
probe packet while all others record the observed RSSIs.  These
observations are then shared, although malicious nodes may
lie. Finally each participant individually selects a (hopefully
truthful) subset of observers for signalprint-based Sybil
classification.

This paper presents our method for truthful subset selection and
fleshes out this framework into a usable, safe, and secure protocol.
As with any system intended for real-world use, we had to carefully
balance system complexity and potential security weaknesses.
\autoref{sec:discussion} discusses these choices and related potential
concerns.

\subsection{Attack Model}
\label{sec:attackmodel}

We assume attackers have the following capabilities and restrictions.
\begin{enumerate}
\item Attackers may collude through arbitrary side channels.
\item Attackers may accumulate information, e.g., RSSIs, across
  multiple rounds of the Mason test.
\item Attackers have limited ability to predict RSSI observations of
  other nodes, e.g., \SI{7}{\dBm} uncertainty (see
  \autoref{sec:predictability}), precluding fine-grained
  pre-characterization.
\item Attackers can control transmit power for each packet, but not
  precisely or quickly steer the output in a desired direction, e.g.,
  beam-forming.
\item Attackers cannot quickly and precisely switch between multiple
  positions, e.g., they do not have high-speed, automated
  electromechanical control.
\end{enumerate}
These capabilities and restrictions model attacking nodes that are
commodity devices, a cheaper attack vector than distributing
specialized hardware. These devices could be obtained by compromising
those owned by normal network participants or directly deployed by the
attacker.

One common denial-of-service (DOS) attack in wireless
networks---jamming the channel---cannot be defended against by
commodity devices.  Thus, we do not defend against other
more-complicated DOS attacks. However, note that ad hoc and
delay-tolerant networks are much more resistant than infrastructured
networks to such attacks, because a single attack can affect only a
small portion of the network.

Notably, we assume attackers do not have per-antenna control of MIMO
(Multiple-Input and Multiple-Output)~\cite{gesbert03apr} devices.
Such control would defeat the signalprint method (even with trusted
observers), but is not a feasible attack.  Commodity MIMO devices
(e.g., 802.11n adapters) do not expose this control to software and
thus are not suitable attack vectors.  Distributing specialized
MIMO-capable hardware over large portions of the network would be
prohibitively expensive.

We believe that the signalprint method can be extended to MIMO systems
(see our technical report for an overview~\cite{liu13dec}), but doing
so is beyond the scope of this work.  Our focus is extending
signalprint-based methods to ad hoc networks of commodity devices by
removing the requirement for trusted observations.

\subsection{Review of Signalprints}
\label{sec:signalprints}

\begin{figure}
  \begin{minipage}{0.46\linewidth}
    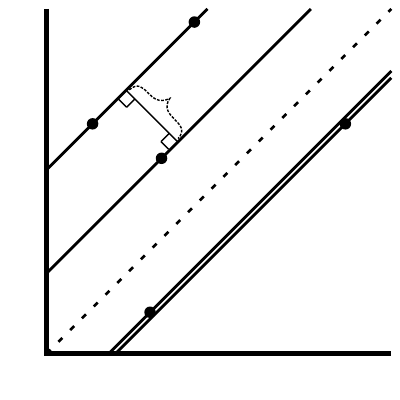
    \caption{Sybils, $A$--$B$ and $D$--$E$, occupy nearby slope-1
      lines.}
    \label{fig:signalprint-plot}
  \end{minipage}
  \hfill
  \begin{minipage}{0.46\linewidth}
    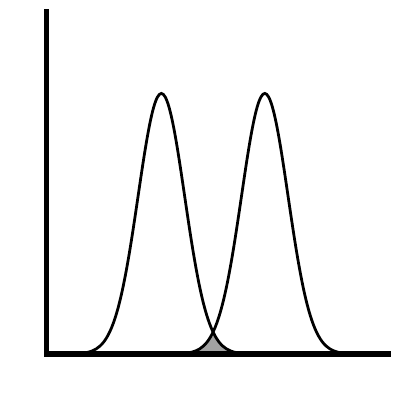
    \caption{The distance threshold trades false positives for
      negatives.}
    \label{fig:signalprint-dists}
  \end{minipage}
\end{figure}

We briefly review the signalprint method. See prior work for
details~\cite{xiao09sep,demirbas06jun}. A \emph{signalprint} is a
vector of RSSIs at multiple observers for a single transmission.
Ignoring noise, the vector of received powers (in logarithmic units,
e.g., \si{\dBm}) at multiple receivers for a given transmission can be
modeled~\cite{rappaport02} as $\vect{s} = \vect{h} + p\vect{1},$ where
$p$ is the transmit power and $\vect{h}$ is the attenuation vector, a
function of the channel amplitude response and the receiver
characteristics.  Transmissions from different locations have
uncorrelated signalprints, as the channel responses are likely
uncorrelated. Those from the same location, however, share a channel
response and will be correlated.  That is, for two transmissions $a$
and $b$ from the same location with transmit powers $p_a$ and $p_b =
p_a + c,$ the signalprints $\vect[b]{s} = \vect{h} + p_a\vect{1}$ and
$\vect[b]{s} = \vect{h} + (p_a+c)\vect{1}$ are related as $\vect[b]{s}
= \vect[a]{s} + c\vect{1}.$ In other words, all observers see the same
RSSI difference $c$ for the two transmissions.

This is illustrated geometrically in \autoref{fig:signalprint-plot}
for a two-receiver signalprint. $A$ and $B$ are Sybil, while $C$ is
not.  $D$ and $E$ are also Sybil, but due to noise the signalprints
are not perfectly correlated.  Instead, signalprints occupying lines
closer than some threshold are taken to be Sybil.
\begin{definition}
  The \emph{signalprint distance} $\signalprintdistance(\vect[a]{s},
  \vect[b]{s})$ between two signalprints $\vect[a]{s}$ and
  $\vect[b]{s}$ is the perpendicular distance between the slope-1
  lines containing them.  Letting
  \[\vect{w} \triangleq \vect[a]{s} - \vect[b]{s}\] be the distance vector between the
  signalprints and \[\vect[\perp]{v} \triangleq \vect{w} -
  \frac{\vect{w}\cdot\vect{1}}{\|\vect{1}\|^2}\vect{1}\] be the vector
  rejection of $\vect{w}$ from $\vect{1}$, then \[\signalprintdistance(\vect[a]{s},
  \vect[b]{s}) = \|\vect[\perp]{v}\|.\]
\end{definition}
As shown in \autoref{fig:signalprint-dists}, the distance
distributions for Sybil and non-Sybil identities overlap, so the
threshold choice trades false positives for negatives.  A good
threshold can detect at least 99.9\% of Sybils while accepting at
least 95\% of non-Sybils~\cite{xiao09sep,demirbas06jun}.

%% file: fig/formulation-hello.pdf_tex
\begingroup%
  \makeatletter%
  \providecommand\color[2][]{%
    \errmessage{(Inkscape) Color is used for the text in Inkscape, but the package 'color.sty' is not loaded}%
    \renewcommand\color[2][]{}%
  }%
  \providecommand\transparent[1]{%
    \errmessage{(Inkscape) Transparency is used (non-zero) for the text in Inkscape, but the package 'transparent.sty' is not loaded}%
    \renewcommand\transparent[1]{}%
  }%
  \providecommand\rotatebox[2]{#2}%
  \ifx\svgwidth\undefined%
    \setlength{\unitlength}{151.2bp}%
    \ifx\svgscale\undefined%
      \relax%
    \else%
      \setlength{\unitlength}{\unitlength * \real{\svgscale}}%
    \fi%
  \else%
    \setlength{\unitlength}{\svgwidth}%
  \fi%
  \global\let\svgwidth\undefined%
  \global\let\svgscale\undefined%
  \makeatother%
  \begin{picture}(1,1)%
    \put(0,0){\includegraphics[width=\unitlength,page=1]{formulation-hello.pdf}}%
    \put(0.07759887,0.58595979){\color[rgb]{0,0,0}\makebox(0,0)[b]{\smash{A}}}%
    \put(0,0){\includegraphics[width=\unitlength,page=2]{formulation-hello.pdf}}%
    \put(-0.29665948,0.618456){\color[rgb]{0,0,0}\makebox(0,0)[lb]{\smash{}}}%
    \put(0.60824803,0.84116715){\color[rgb]{0,0,0}\makebox(0,0)[b]{\smash{B}}}%
    \put(0.59374306,0.45929117){\color[rgb]{0,0,0}\makebox(0,0)[b]{\smash{C}}}%
    \put(0.30678756,0.16203542){\color[rgb]{0,0,0}\makebox(0,0)[b]{\smash{D}}}%
    \put(0.92178103,0.12981015){\color[rgb]{0,0,0}\makebox(0,0)[b]{\smash{E}}}%
    \put(0,0){\includegraphics[width=\unitlength,page=3]{formulation-hello.pdf}}%
    \put(0.71577655,0.46787387){\color[rgb]{0,0,0}\makebox(0,0)[b]{\smash{{\ts A: 12}}}}%
    \put(0,0){\includegraphics[width=\unitlength,page=4]{formulation-hello.pdf}}%
    \put(0.42877201,0.18108033){\color[rgb]{0,0,0}\makebox(0,0)[b]{\smash{{\ts A: 25}}}}%
    \put(0,0){\includegraphics[width=\unitlength,page=5]{formulation-hello.pdf}}%
    \put(0.79737031,0.10015353){\color[rgb]{0,0,0}\makebox(0,0)[b]{\smash{{\ts A: 31}}}}%
    \put(0,0){\includegraphics[width=\unitlength,page=6]{formulation-hello.pdf}}%
    \put(0.73618385,0.91836532){\color[rgb]{0,0,0}\makebox(0,0)[b]{\smash{{\ts A: 20}}}}%
    \put(0,0){\includegraphics[width=\unitlength,page=7]{formulation-hello.pdf}}%
    \put(0.07711576,0.41488373){\color[rgb]{0,0,0}\makebox(0,0)[b]{\smash{{\ts B: 19}}}}%
    \put(0.42877201,0.14349818){\color[rgb]{0,0,0}\makebox(0,0)[b]{\smash{{\ts B: 18}}}}%
    \put(0.71577655,0.42857143){\color[rgb]{0,0,0}\makebox(0,0)[b]{\smash{{\ts B: 17}}}}%
    \put(0.79737031,0.0611188){\color[rgb]{0,0,0}\makebox(0,0)[b]{\smash{{\ts B: 27}}}}%
  \end{picture}%
\endgroup%

%% file: fig/formulation-exchange.pdf_tex
\begingroup%
  \makeatletter%
  \providecommand\color[2][]{%
    \errmessage{(Inkscape) Color is used for the text in Inkscape, but the package 'color.sty' is not loaded}%
    \renewcommand\color[2][]{}%
  }%
  \providecommand\transparent[1]{%
    \errmessage{(Inkscape) Transparency is used (non-zero) for the text in Inkscape, but the package 'transparent.sty' is not loaded}%
    \renewcommand\transparent[1]{}%
  }%
  \providecommand\rotatebox[2]{#2}%
  \ifx\svgwidth\undefined%
    \setlength{\unitlength}{151.2bp}%
    \ifx\svgscale\undefined%
      \relax%
    \else%
      \setlength{\unitlength}{\unitlength * \real{\svgscale}}%
    \fi%
  \else%
    \setlength{\unitlength}{\svgwidth}%
  \fi%
  \global\let\svgwidth\undefined%
  \global\let\svgscale\undefined%
  \makeatother%
  \begin{picture}(1,1)%
    \put(0,0){\includegraphics[width=\unitlength,page=1]{formulation-exchange.pdf}}%
    \put(0.18204425,0.08618116){\color[rgb]{0,0,0}\makebox(0,0)[b]{\smash{{\ts A: 12}}}}%
    \put(0.18204429,0.04859901){\color[rgb]{0,0,0}\makebox(0,0)[b]{\smash{{\ts B: 17}}}}%
    \put(0,0){\includegraphics[width=\unitlength,page=2]{formulation-exchange.pdf}}%
    \put(0.14500722,0.83750387){\color[rgb]{0,0,0}\makebox(0,0)[b]{\smash{{\ts A: 12}}}}%
    \put(0.14500726,0.79992172){\color[rgb]{0,0,0}\makebox(0,0)[b]{\smash{{\ts B: 17}}}}%
    \put(0,0){\includegraphics[width=\unitlength,page=3]{formulation-exchange.pdf}}%
    \put(0.82225589,0.89570493){\color[rgb]{0,0,0}\makebox(0,0)[b]{\smash{{\ts A: 12}}}}%
    \put(0.82225593,0.85812278){\color[rgb]{0,0,0}\makebox(0,0)[b]{\smash{{\ts B: 17}}}}%
    \put(0,0){\includegraphics[width=\unitlength,page=4]{formulation-exchange.pdf}}%
    \put(0.80404915,0.36014656){\color[rgb]{0,0,0}\makebox(0,0)[b]{\smash{{\ts A: 12}}}}%
    \put(0.80404919,0.3225644){\color[rgb]{0,0,0}\makebox(0,0)[b]{\smash{{\ts B: 17}}}}%
    \put(0,0){\includegraphics[width=\unitlength,page=5]{formulation-exchange.pdf}}%
    \put(0.80109187,0.08089014){\color[rgb]{0,0,0}\makebox(0,0)[b]{\smash{{\ts A: 12}}}}%
    \put(0.80109191,0.04330798){\color[rgb]{0,0,0}\makebox(0,0)[b]{\smash{{\ts B: 17}}}}%
    \put(0,0){\includegraphics[width=\unitlength,page=6]{formulation-exchange.pdf}}%
    \put(0.15984933,0.09519566){\color[rgb]{0,0,0}\makebox(0,0)[b]{\smash{{\ts A: 25}}}}%
    \put(0.15984937,0.0576135){\color[rgb]{0,0,0}\makebox(0,0)[b]{\smash{{\ts B: 18}}}}%
    \put(0,0){\includegraphics[width=\unitlength,page=7]{formulation-exchange.pdf}}%
    \put(0.1228123,0.84651837){\color[rgb]{0,0,0}\makebox(0,0)[b]{\smash{{\ts A: 25}}}}%
    \put(0.12281234,0.80893622){\color[rgb]{0,0,0}\makebox(0,0)[b]{\smash{{\ts B: 18}}}}%
    \put(0,0){\includegraphics[width=\unitlength,page=8]{formulation-exchange.pdf}}%
    \put(0.80006097,0.90471943){\color[rgb]{0,0,0}\makebox(0,0)[b]{\smash{{\ts A: 25}}}}%
    \put(0.80006101,0.86713727){\color[rgb]{0,0,0}\makebox(0,0)[b]{\smash{{\ts B: 18}}}}%
    \put(0,0){\includegraphics[width=\unitlength,page=9]{formulation-exchange.pdf}}%
    \put(0.78185423,0.36916105){\color[rgb]{0,0,0}\makebox(0,0)[b]{\smash{{\ts A: 25}}}}%
    \put(0.78185427,0.3315789){\color[rgb]{0,0,0}\makebox(0,0)[b]{\smash{{\ts B: 18}}}}%
    \put(0,0){\includegraphics[width=\unitlength,page=10]{formulation-exchange.pdf}}%
    \put(0.77889695,0.08990464){\color[rgb]{0,0,0}\makebox(0,0)[b]{\smash{{\ts A: 25}}}}%
    \put(0.77889699,0.05232248){\color[rgb]{0,0,0}\makebox(0,0)[b]{\smash{{\ts B: 18}}}}%
    \put(0,0){\includegraphics[width=\unitlength,page=11]{formulation-exchange.pdf}}%
    \put(0.13765442,0.10421015){\color[rgb]{0,0,0}\makebox(0,0)[b]{\smash{{\ts A: 31}}}}%
    \put(0.13765446,0.06662799){\color[rgb]{0,0,0}\makebox(0,0)[b]{\smash{{\ts B: 27}}}}%
    \put(0,0){\includegraphics[width=\unitlength,page=12]{formulation-exchange.pdf}}%
    \put(0.10061738,0.85553287){\color[rgb]{0,0,0}\makebox(0,0)[b]{\smash{{\ts A: 31}}}}%
    \put(0.10061742,0.81795071){\color[rgb]{0,0,0}\makebox(0,0)[b]{\smash{{\ts B: 27}}}}%
    \put(0,0){\includegraphics[width=\unitlength,page=13]{formulation-exchange.pdf}}%
    \put(0.77786606,0.91373393){\color[rgb]{0,0,0}\makebox(0,0)[b]{\smash{{\ts A: 31}}}}%
    \put(0.7778661,0.87615177){\color[rgb]{0,0,0}\makebox(0,0)[b]{\smash{{\ts B: 27}}}}%
    \put(0,0){\includegraphics[width=\unitlength,page=14]{formulation-exchange.pdf}}%
    \put(0.75965932,0.37817555){\color[rgb]{0,0,0}\makebox(0,0)[b]{\smash{{\ts A: 31}}}}%
    \put(0.75965936,0.3405934){\color[rgb]{0,0,0}\makebox(0,0)[b]{\smash{{\ts B: 27}}}}%
    \put(0,0){\includegraphics[width=\unitlength,page=15]{formulation-exchange.pdf}}%
    \put(0.75670204,0.09891913){\color[rgb]{0,0,0}\makebox(0,0)[b]{\smash{{\ts A: 31}}}}%
    \put(0.75670208,0.06133698){\color[rgb]{0,0,0}\makebox(0,0)[b]{\smash{{\ts B: 27}}}}%
    \put(0,0){\includegraphics[width=\unitlength,page=16]{formulation-exchange.pdf}}%
    \put(0.11545949,0.11322464){\color[rgb]{0,0,0}\makebox(0,0)[b]{\smash{{\ts A: 20}}}}%
    \put(0.11545953,0.07564249){\color[rgb]{0,0,0}\makebox(0,0)[b]{\smash{{\ts C: 16}}}}%
    \put(0,0){\includegraphics[width=\unitlength,page=17]{formulation-exchange.pdf}}%
    \put(0.07842245,0.86454737){\color[rgb]{0,0,0}\makebox(0,0)[b]{\smash{{\ts A: 20}}}}%
    \put(0.07842249,0.82696521){\color[rgb]{0,0,0}\makebox(0,0)[b]{\smash{{\ts C: 16}}}}%
    \put(0,0){\includegraphics[width=\unitlength,page=18]{formulation-exchange.pdf}}%
    \put(0.75567113,0.92274842){\color[rgb]{0,0,0}\makebox(0,0)[b]{\smash{{\ts A: 20}}}}%
    \put(0.75567117,0.88516627){\color[rgb]{0,0,0}\makebox(0,0)[b]{\smash{{\ts C: 16}}}}%
    \put(0,0){\includegraphics[width=\unitlength,page=19]{formulation-exchange.pdf}}%
    \put(0.73746439,0.38719005){\color[rgb]{0,0,0}\makebox(0,0)[b]{\smash{{\ts A: 20}}}}%
    \put(0.73746443,0.34960789){\color[rgb]{0,0,0}\makebox(0,0)[b]{\smash{{\ts C: 16}}}}%
    \put(0,0){\includegraphics[width=\unitlength,page=20]{formulation-exchange.pdf}}%
    \put(0.73450711,0.10793363){\color[rgb]{0,0,0}\makebox(0,0)[b]{\smash{{\ts A: 20}}}}%
    \put(0.73450715,0.07035148){\color[rgb]{0,0,0}\makebox(0,0)[b]{\smash{{\ts C: 16}}}}%
    \put(0,0){\includegraphics[width=\unitlength,page=21]{formulation-exchange.pdf}}%
    \put(0.09326456,0.12223913){\color[rgb]{0,0,0}\makebox(0,0)[b]{\smash{{\ts B: 19}}}}%
    \put(0.0932646,0.08465698){\color[rgb]{0,0,0}\makebox(0,0)[b]{\smash{{\ts C: 15}}}}%
    \put(0,0){\includegraphics[width=\unitlength,page=22]{formulation-exchange.pdf}}%
    \put(0.05622752,0.87356186){\color[rgb]{0,0,0}\makebox(0,0)[b]{\smash{{\ts B: 19}}}}%
    \put(0.05622756,0.83597971){\color[rgb]{0,0,0}\makebox(0,0)[b]{\smash{{\ts C: 15}}}}%
    \put(0,0){\includegraphics[width=\unitlength,page=23]{formulation-exchange.pdf}}%
    \put(0.7334762,0.93176292){\color[rgb]{0,0,0}\makebox(0,0)[b]{\smash{{\ts B: 19}}}}%
    \put(0.73347624,0.89418077){\color[rgb]{0,0,0}\makebox(0,0)[b]{\smash{{\ts C: 15}}}}%
    \put(0,0){\includegraphics[width=\unitlength,page=24]{formulation-exchange.pdf}}%
    \put(0.71526946,0.39620455){\color[rgb]{0,0,0}\makebox(0,0)[b]{\smash{{\ts B: 19}}}}%
    \put(0.7152695,0.35862239){\color[rgb]{0,0,0}\makebox(0,0)[b]{\smash{{\ts C: 15}}}}%
    \put(0,0){\includegraphics[width=\unitlength,page=25]{formulation-exchange.pdf}}%
    \put(0.71231218,0.11694813){\color[rgb]{0,0,0}\makebox(0,0)[b]{\smash{{\ts B: 19}}}}%
    \put(0.71231222,0.07936597){\color[rgb]{0,0,0}\makebox(0,0)[b]{\smash{{\ts C: 15}}}}%
    \put(0,0){\includegraphics[width=\unitlength,page=26]{formulation-exchange.pdf}}%
    \put(0.07759887,0.58595979){\color[rgb]{0,0,0}\makebox(0,0)[b]{\smash{A}}}%
    \put(0,0){\includegraphics[width=\unitlength,page=27]{formulation-exchange.pdf}}%
    \put(0.60824803,0.84116715){\color[rgb]{0,0,0}\makebox(0,0)[b]{\smash{B}}}%
    \put(0.59374306,0.45929117){\color[rgb]{0,0,0}\makebox(0,0)[b]{\smash{C}}}%
    \put(0.30678756,0.16203542){\color[rgb]{0,0,0}\makebox(0,0)[b]{\smash{D}}}%
    \put(0.92178103,0.12981015){\color[rgb]{0,0,0}\makebox(0,0)[b]{\smash{E}}}%
    \put(0,0){\includegraphics[width=\unitlength,page=28]{formulation-exchange.pdf}}%
  \end{picture}%
\endgroup%

%% file: fig/formulation-comparison.pdf_tex
\begingroup%
  \makeatletter%
  \providecommand\color[2][]{%
    \errmessage{(Inkscape) Color is used for the text in Inkscape, but the package 'color.sty' is not loaded}%
    \renewcommand\color[2][]{}%
  }%
  \providecommand\transparent[1]{%
    \errmessage{(Inkscape) Transparency is used (non-zero) for the text in Inkscape, but the package 'transparent.sty' is not loaded}%
    \renewcommand\transparent[1]{}%
  }%
  \providecommand\rotatebox[2]{#2}%
  \ifx\svgwidth\undefined%
    \setlength{\unitlength}{151.2bp}%
    \ifx\svgscale\undefined%
      \relax%
    \else%
      \setlength{\unitlength}{\unitlength * \real{\svgscale}}%
    \fi%
  \else%
    \setlength{\unitlength}{\svgwidth}%
  \fi%
  \global\let\svgwidth\undefined%
  \global\let\svgscale\undefined%
  \makeatother%
  \begin{picture}(1,1)%
    \put(0,0){\includegraphics[width=\unitlength,page=1]{formulation-comparison.pdf}}%
    \put(0.07759887,0.58595979){\color[rgb]{0,0,0}\makebox(0,0)[b]{\smash{A}}}%
    \put(0,0){\includegraphics[width=\unitlength,page=2]{formulation-comparison.pdf}}%
    \put(0.60824803,0.84116715){\color[rgb]{0,0,0}\makebox(0,0)[b]{\smash{B}}}%
    \put(0.59374306,0.45929117){\color[rgb]{0,0,0}\makebox(0,0)[b]{\smash{C}}}%
    \put(0.30678756,0.16203542){\color[rgb]{0,0,0}\makebox(0,0)[b]{\smash{D}}}%
    \put(0.92178103,0.12981015){\color[rgb]{0,0,0}\makebox(0,0)[b]{\smash{E}}}%
    \put(0,0){\includegraphics[width=\unitlength,page=3]{formulation-comparison.pdf}}%
    \put(0.84616848,0.84230946){\color[rgb]{0,0,0}\makebox(0,0)[b]{\smash{{\sss [$\cdots$] A} }}}%
    \put(0.84864379,0.74176792){\color[rgb]{0,0,0}\makebox(0,0)[b]{\smash{{\sss [$\cdots$] D} }}}%
    \put(0.84740614,0.69149716){\color[rgb]{0,0,0}\makebox(0,0)[b]{\smash{{\sss [$\cdots$] E} }}}%
    \put(0.84864379,0.79203869){\color[rgb]{0,0,0}\makebox(0,0)[b]{\smash{{\sss [$\cdots$] C} }}}%
    \put(0,0){\includegraphics[width=\unitlength,page=4]{formulation-comparison.pdf}}%
    \put(0.30188023,0.94159143){\color[rgb]{0,0,0}\makebox(0,0)[b]{\smash{{\sss [$\cdots$] B} }}}%
    \put(0.30318869,0.8410499){\color[rgb]{0,0,0}\makebox(0,0)[b]{\smash{{\sss [$\cdots$] D} }}}%
    \put(0.30195103,0.79077913){\color[rgb]{0,0,0}\makebox(0,0)[b]{\smash{{\sss [$\cdots$] E} }}}%
    \put(0.30318869,0.89132067){\color[rgb]{0,0,0}\makebox(0,0)[b]{\smash{{\sss [$\cdots$] C} }}}%
    \put(0,0){\includegraphics[width=\unitlength,page=5]{formulation-comparison.pdf}}%
    \put(0.84031564,0.48326813){\color[rgb]{0,0,0}\makebox(0,0)[b]{\smash{{\sss [$\cdots$] A} }}}%
    \put(0.84196585,0.3827266){\color[rgb]{0,0,0}\makebox(0,0)[b]{\smash{{\sss [$\cdots$] D} }}}%
    \put(0.84072819,0.33245583){\color[rgb]{0,0,0}\makebox(0,0)[b]{\smash{{\sss [$\cdots$] E} }}}%
    \put(0.840799,0.43299737){\color[rgb]{0,0,0}\makebox(0,0)[b]{\smash{{\sss [$\cdots$] B} }}}%
    \put(0,0){\includegraphics[width=\unitlength,page=6]{formulation-comparison.pdf}}%
    \put(0.66406403,0.23700484){\color[rgb]{0,0,0}\makebox(0,0)[b]{\smash{{\sss [$\cdots$] A} }}}%
    \put(0.66571424,0.13646331){\color[rgb]{0,0,0}\makebox(0,0)[b]{\smash{{\sss [$\cdots$] C} }}}%
    \put(0.6656434,0.08619254){\color[rgb]{0,0,0}\makebox(0,0)[b]{\smash{{\sss [$\cdots$] D} }}}%
    \put(0.66454739,0.18673407){\color[rgb]{0,0,0}\makebox(0,0)[b]{\smash{{\sss [$\cdots$] B} }}}%
    \put(0,0){\includegraphics[width=\unitlength,page=7]{formulation-comparison.pdf}}%
    \put(0.32755118,0.50469147){\color[rgb]{0,0,0}\makebox(0,0)[b]{\smash{{\sss [$\cdots$] A} }}}%
    \put(0.3300265,0.40414994){\color[rgb]{0,0,0}\makebox(0,0)[b]{\smash{{\sss [$\cdots$] C} }}}%
    \put(0.32878884,0.35387917){\color[rgb]{0,0,0}\makebox(0,0)[b]{\smash{{\sss [$\cdots$] E} }}}%
    \put(0.32885964,0.45442071){\color[rgb]{0,0,0}\makebox(0,0)[b]{\smash{{\sss [$\cdots$] B} }}}%
    \put(0,0){\includegraphics[width=\unitlength,page=8]{formulation-comparison.pdf}}%
  \end{picture}%
\endgroup%

%% file: fig/signalprint-plot.pdf_tex
\begingroup%
  \makeatletter%
  \providecommand\color[2][]{%
    \errmessage{(Inkscape) Color is used for the text in Inkscape, but the package 'color.sty' is not loaded}%
    \renewcommand\color[2][]{}%
  }%
  \providecommand\transparent[1]{%
    \errmessage{(Inkscape) Transparency is used (non-zero) for the text in Inkscape, but the package 'transparent.sty' is not loaded}%
    \renewcommand\transparent[1]{}%
  }%
  \providecommand\rotatebox[2]{#2}%
  \ifx\svgwidth\undefined%
    \setlength{\unitlength}{115.2bp}%
    \ifx\svgscale\undefined%
      \relax%
    \else%
      \setlength{\unitlength}{\unitlength * \real{\svgscale}}%
    \fi%
  \else%
    \setlength{\unitlength}{\svgwidth}%
  \fi%
  \global\let\svgwidth\undefined%
  \global\let\svgscale\undefined%
  \makeatother%
  \begin{picture}(1,1)%
    \put(0,0){\includegraphics[width=\unitlength]{signalprint-plot.pdf}}%
    \put(0.54557756,0.02748221){\color[rgb]{0,0,0}\rotatebox{0.33675043}{\makebox(0,0)[b]{\smash{R\textsubscript{1} (dBm)}}}}%
    \put(0.0827118,0.54480368){\color[rgb]{0,0,0}\rotatebox{90}{\makebox(0,0)[b]{\smash{R\textsubscript{2} (dBm)}}}}%
    \put(0.15679447,0.7364049){\color[rgb]{0,0,0}\makebox(0,0)[lb]{\smash{A}}}%
    \put(0.37667921,0.90438001){\color[rgb]{0,0,0}\makebox(0,0)[lb]{\smash{B}}}%
    \put(0.29014052,0.58645636){\color[rgb]{0,0,0}\makebox(0,0)[lb]{\smash{C}}}%
    \put(0.41232308,0.13739105){\color[rgb]{0,0,0}\makebox(0,0)[lb]{\smash{D}}}%
    \put(0.89214749,0.60427142){\color[rgb]{0,0,0}\makebox(0,0)[lb]{\smash{E}}}%
    \put(0.2689931,0.31800045){\color[rgb]{0,0,0}\rotatebox{46.33937226}{\makebox(0,0)[b]{\smash{R\textsubscript{1}=R\textsubscript{2}}}}}%
    \put(0.42001139,0.7696856){\color[rgb]{0,0,0}\rotatebox{44.26028758}{\makebox(0,0)[lb]{\smash{{\footnotesize Signalprint}}}}}%
    \put(0.47774071,0.70976831){\color[rgb]{0,0,0}\rotatebox{45}{\makebox(0,0)[lb]{\smash{{\footnotesize Distance}}}}}%
  \end{picture}%
\endgroup%

%% file: fig/signalprint-dists.pdf_tex
\begingroup%
  \makeatletter%
  \providecommand\color[2][]{%
    \errmessage{(Inkscape) Color is used for the text in Inkscape, but the package 'color.sty' is not loaded}%
    \renewcommand\color[2][]{}%
  }%
  \providecommand\transparent[1]{%
    \errmessage{(Inkscape) Transparency is used (non-zero) for the text in Inkscape, but the package 'transparent.sty' is not loaded}%
    \renewcommand\transparent[1]{}%
  }%
  \providecommand\rotatebox[2]{#2}%
  \ifx\svgwidth\undefined%
    \setlength{\unitlength}{115.2bp}%
    \ifx\svgscale\undefined%
      \relax%
    \else%
      \setlength{\unitlength}{\unitlength * \real{\svgscale}}%
    \fi%
  \else%
    \setlength{\unitlength}{\svgwidth}%
  \fi%
  \global\let\svgwidth\undefined%
  \global\let\svgscale\undefined%
  \makeatother%
  \begin{picture}(1,1)%
    \put(0,0){\includegraphics[width=\unitlength]{signalprint-dists.pdf}}%
    \put(0.54711407,0.02688681){\color[rgb]{0,0,0}\rotatebox{0.33675043}{\makebox(0,0)[b]{\smash{Signalprint Distance}}}}%
    \put(0.07863563,0.54273823){\color[rgb]{0,0,0}\rotatebox{90}{\makebox(0,0)[b]{\smash{Density}}}}%
    \put(0.42417614,0.83333339){\color[rgb]{0,0,0}\makebox(0,0)[rb]{\smash{Sybil}}}%
    \put(0.54130639,0.83333339){\color[rgb]{0,0,0}\makebox(0,0)[lb]{\smash{Non-Sybil}}}%
  \end{picture}%
\endgroup%

%% file: separation.tex
\section{Sybil Classification From Untrusted Signalprints}
\label{sec:separation}

\Note{Summarize section: extend signalprint method to deal with
  invalid observations}

In this section we describe two methods to detect Sybil identities
using untrusted RSSI observations.  In both cases, a set of candidate
views containing the true view (with high probability) is
generated. The accepted view is chosen by a view selection policy. The
first method's policy selects the view indicting the most Sybils,
limiting the total number of incorrect classifications. The second
selects the true view, but works only when conforming nodes outnumber
physical attacker nodes.

\subsection{The Limited Power of Falsified Observations}
\label{sec:falsified-power}

Our key observation is that falsified RSSI observations have limited
power.  Although falsifying observations to make Sybil identities look
non-Sybil is easy, it is extremely difficult to make a non-Sybil look
Sybil. To see this, recall that two identities are considered Sybil
only if all observers report the same RSSI difference for their
transmissions. For any pair of identities, the initiator observes a
RSSI difference itself. On one hand, making true Sybils appear
non-Sybil is easy, because randomly chosen values almost certainly
fail to match this difference. Making a non-Sybil look Sybil, however,
requires learning the difference observed by the initiator itself,
which is kept secret. Guessing is impractical due to the
unpredictability of the wireless channels.  Our methods rely on this
difficulty, which is quantified in \autoref{sec:predictability}.

\subsection{Terminology}

\newcommand{\setI}{\ensuremath{I}\xspace}

\newcommand{\setS}{\ensuremath{S}\xspace}
\newcommand{\setNS}{\ensuremath{\mathit{NS}}\xspace}

\newcommand{\setL}{\ensuremath{L}\xspace}
\newcommand{\setT}{\ensuremath{T}\xspace}

\newcommand{\setC}{\ensuremath{C}\xspace}
\newcommand{\setTS}{\ensuremath{\mathit{TS}}\xspace}
\newcommand{\setLS}{\ensuremath{\mathit{LS}}\xspace}
\newcommand{\setLNS}{\ensuremath{\mathit{LNS}}\xspace}

\newcommand{\view}{\ensuremath{V}\xspace}
\newcommand{\true}[1]{\ensuremath{\overline{#1}}\xspace}
\newcommand{\false}[1]{\ensuremath{\widehat{#1}}\xspace}
\newcommand{\spec}[1]{\ensuremath{\widetilde{#1}}\xspace}

\newcommand{\partNS}[1]{\ensuremath{#1_\mathrm{NS}}}
\newcommand{\partS}[1]{\ensuremath{#1_\mathrm{S}}}

\newcommand{\setR}{\ensuremath{R}\xspace}
\newcommand{\generates}{\mapsto}

\setI is the set of participating identities. Each is either Sybil or
non-Sybil and reports either true or false\footnote{A reported RSSI
  observation is considered \emph{false} if signalprints containing it
  misclassify some identities.}  RSSI observations, partitioning the
identities by their Sybilness (sets \setS and \setNS) and the veracity
of their reported observations (sets \setT and \setL).
\begin{center}
  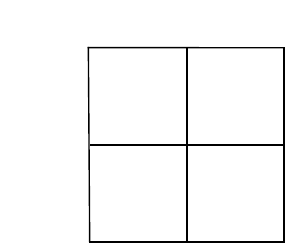
\end{center} Truthtelling, non-Sybil identities are called
\emph{conforming} (set \setC).  Liars and Sybil identities are called
\emph{attacking} (sets \setLS, \setLNS, and \setTS). Our goal is to
distinguish the \setS and \setNS partitions using the reported RSSI
observations without knowing a priori the \setL and \setT partitions.

\begin{definition} An \emph{initiator} is the node performing Sybil
  classification.\footnote{All participants perform classification
    individually, so each is the initiator in its own classification
    session.}  It trusts its own RSSI observations, but no others.
\end{definition}

\begin{definition}
  A \emph{receiver set}, denoted by \setR, is a subset of identities
  ($\setR \subseteq \setI$) whose reported RSSI observations, with the
  initiator's, form signalprints. Those with liars ($\setR \cap \setL
  \neq \emptyset$) produce incorrect classifications and those with
  only truthtellers ($\setR \subseteq \setT$) produce the correct
  classification.
\end{definition}

\begin{definition}
  A \emph{view}, denoted by \view, is a classification of identities
  as Sybil and non-Sybil.  Those classified as Sybil (non-Sybil) are
  said to be Sybil (non-Sybil) \emph{under} \view and are denoted by
  the subset \partS{\view} (\partNS{\view}).  A view \view obtained
  from the signalprints of a receiver set \setR is \emph{generated} by
  \setR, denoted by $\setR \generates \view$ (read: \setR generates
  \view), and can be written $\view(\setR)$. Identities in $\setR$ are
  considered non-Sybil, i.e., $\setR \subseteq \partNS{\view}(R)$.  A
  \emph{true view}, denoted by \true{\view}, correctly labels all
  identities, i.e., $\partS{\true{\view}} = \setS$ and
  $\partNS{\true{\view}} = \setNS$.  Similarly, a \emph{false view},
  denoted by \false{\view}, incorrectly labels some identities, i.e.,
  $\partS{\false{\view}} \neq \setS$ and $\partNS{\false{\view}} \neq
  \setNS$.
\end{definition}

\begin{definition}
  Incorrectly labeling non-Sybil identities as Sybil is called
  \emph{collapsing}.
\end{definition}

\begin{assumption}\label{assumption:noise-free}
  To clearly illustrate the impact of intentionally-falsified
  observations, we first assume that true RSSI observations are
  noise-free and thus always generate the true view. In
  \autoref{sec:handle-noise}, we extend the method to handle
  real-world observations containing, for example, random noise and
  discretization error.
\end{assumption}

\newcommand{\tws}[2]{#1\hspace*{1em}#2}
\begin{table*}[t]
  \centering
  \setarrstr{1}
  \caption{Definitions of Terms and Symbols}
  \label{tbl:terminology}
  \begin{tabular}{lll@{}}
    \toprule
    & \bfseries Definition & \bfseries Notes \\
    \midrule
    \multicolumn{3}{@{}l}{Sets of Identities} \\
    \setI & all participating identities & \\
    \addlinespace[0.5\defaultaddspace]
    \setNS & all non-Sybil identities & \multirow{2}{*}{$\setI = \{\setNS|\setS\}$} \\
    \setS & all Sybil identities & \\
    \addlinespace[0.5\defaultaddspace]
    \setT & all truthful identities & \multirow{2}{*}{$\setI = \{\setT|\setL\}$} \\
    \setL & all lying identities & \\
    \addlinespace[0.5\defaultaddspace]
    \setC & all conforming, or truthful, non-Sybil, identities & $\setNS = \{\setC|\setLNS\}$ \\
    \setLNS & all lying, non-Sybil identities & $\setS = \{\setTS|\setLS\}$ \\
    \setTS & all truthful, Sybil identities & $\setT = \{\setC|\setTS\}$ \\
    \setLS & all lying, Sybil identities & $\setL = \{\setLNS|\setLS\}$ \\
    \addlinespace[0.5\defaultaddspace]
    \partNS{\view} & all identities labeled non-Sybil by view \view & \multirow{2}{*}{$\setI = \{\partNS{\view}|\partS{\view}\}$} \\
    \partS{\view} & all identities labeled Sybil by view \view & \\
    \addlinespace[0.5\defaultaddspace]
    \tws{\setR}{(\emph{receiver set})} & identities used to form signalprints & \\
    \midrule
    \multicolumn{3}{@{}l}{Views} \\
    \tws{\view}{(\emph{view})} & a Sybil--non-Sybil labeling of \setI & \\
    \tws{\true{\view}}{(\emph{true view})} & a view that correctly labels all identities & $\partNS{\true{\view}} = \setNS$ and $\partS{\true{\view}} = \setS$ \\
    \tws{\false{\view}}{(\emph{false view})} & a view that incorrectly labels some identities & $\partNS{\false{\view}} \neq \setNS$ and $\partS{\false{\view}} \neq \setS$ \\
    $\view(\setR)$ & the view generated by receiver set \setR & \\
    \midrule
    \multicolumn{3}{@{}l}{Terms} \\
    \tws{\emph{generates}}{($\setR \generates \view$)} & a receiver set generates a view & \\
    \emph{initiator} & node performing the Sybil classification \\
    \emph{collapse} & classify a non-Sybil identity as Sybil & \\
    \bottomrule
  \end{tabular}
\end{table*}

\subsection{Approach Overview}

\newcommand{\setA}{\ensuremath{A}\xspace}
\newcommand{\setB}{\ensuremath{B}\xspace}

A general separation method does not exist, because different
scenarios can lead to the same reported RSSI observations. To
illustrate, consider identities $\setI = \{\setA | \setB\}$ reporting
observations such that
\begin{align*}
&\setR \subseteq \setA \generates \view^{1} = \{\partNS{\view^{1}}=\setA | \partS{\view^{1}}=\setB \}\textrm{ and} \\
&\setR \subseteq \setB \generates \view^{2} = \{\partNS{\view^{2}}=\setB | \partS{\view^{2}}=\setA \}
\end{align*}
and two different scenarios $x$ and $y$ such that
\begin{align*}
&\textrm{in $x$, } \{\setT^{x} = \setA | \setL^{x} = \setB \} = \setI \textrm{ and} \\
&\textrm{in $y$, } \{\setT^{y} = \setB | \setL^{y} = \setA \} = \setI.
\end{align*}
$\setR \subseteq T \generates \true{\view}$, so $\view^{1}$ and
$\view^{2}$ are both true views, the former in scenario $x$ and the
latter in scenario $y$. Consequently, no method can always choose the
correct view.

We instead develop two different approaches.  The first method, the
\emph{maximum Sybil policy}, simply bounds the number of misclassified
identities by selecting the view reporting the most Sybils. This
selected view must indite at least as many as the true view, bounding
the accepted Sybils by the number of collapsed conforming
identities. Collapsing is difficult, limiting the incorrect
classifications.

The second method, the \emph{view consistency policy}, allows complete
separation, but requires that the following conditions be met.
\begin{itemize}
\item All view correctly classify some conforming identities, because
  collapsing identities is difficult.
\item Conforming identities outnumber lying, non-Sybils (a major
  motivating factor for the Sybil attack).
\end{itemize}
This approach follows from the idea that true observations are
trivially self-consistent, while lies often contradict themselves. We
develop a notion of consistency that allows separation of true and
false observations.

\subsection{Maximum Sybil Policy: Select the View Claiming the Most
  Sybil Identities}
\label{sec:maximum-sybil}

In this section, we prove that the maximum Sybil policy---selecting
the view claiming the most Sybil identities---produces a
classification with bounded error. The number of incorrectly-accepted
Sybil identities is bounded by the number of collapsed conforming
identities.

\begin{lemma}\label{lemma:selected-view-sybil}
  The selected view $\view$ claims at least as many Sybil identities
  as actually exist, i.e., $|\partS{\view}| \geq |\setS|$.
\end{lemma}
\begin{IEEEproof}
  Since the true view $\true{\view}$ claiming $|\setS|$ Sybils always
  exists, the selected view can claim no less.
\end{IEEEproof}

\begin{theorem}\label{thm:max-sybil}
  The selected view $\view$ misclassifies no more Sybil identities
  than it collapses conforming identities, i.e., $|\partNS{\view} \cap
  \setS| \leq |\partS{\view} \cap \setNS|$.
\end{theorem}
\begin{IEEEproof}
  Intuitively, claiming the minimum $|\setS|$ Sybil identities
  requires that each misclassified Sybil be compensated for by a
  collapsed non-Sybil identity. Formally, combining $|\partS{\view}
  \cup \partNS{\view}| = |\setS \cup \setNS|$ with
  \autoref{lemma:selected-view-sybil} yields $|(\partS{\view}
  \cup \partNS{\view}) \cap \setS| \leq |(\setS \cup \setNS)
  \cap \partS{\view}|$. Removing the common $\partS{\view} \cap \setS$
  from both sides gives $|\partNS{\view} \cap \setS| \leq
  |\partS{\view} \cap \setNS|$.
\end{IEEEproof}

\autoref{thm:max-sybil} bounds the misclassifications by the
attacker's collapsing power, $|\partS{\view} \cap \setNS|$. Although
$|\partS{\view} \cap \setNS|$ is small (see
\autoref{sec:predictability}), one Sybil is still accepted for each
conforming identity collapsed. The next few sections develop a second
method that allows accurate classification, but only when conforming
nodes outnumber attackers.

\subsection{View Consistency Policy: Selecting \true{\view} if
  $\setLNS=\emptyset$}
\label{sec:view-consistency}

Our view consistency policy stems from the intuition that lies often
contradict themselves. It is introduced here using the following
unrealistic assumption, which we remove in \autoref{sec:policy1}.

\begin{restrict} \label{restrict:liars} All liars are Sybil, i.e.,
  $LNS = \emptyset$, and thus all non-Sybil identities are truthful,
  i.e., $\setNS \subseteq \setT$.
\end{restrict} 

\autoref{restrict:liars} endows the true view with a useful property:
all receiver sets comprising the non-Sybil identities under the true
view will generate the true view.  We formalize this consistency as
follows.

\begin{definition}
  A view is \emph{view-consistent} if and only if all receiver sets
  comprising a subset of the non-Sybil identities under that view
  generate the same view, i.e., \view is view-consistent iff $\forall \setR
  \in 2^{\partNS{\view}} : \setR \generates \view$.
\end{definition}

\begin{lemma}\label{lemma:true-view-consistent}
  Under \autoref{restrict:liars}, the true view is view-consistent,
  i.e., $\forall \setR \in 2^{\partNS{\true{\view}}} : \setR
  \generates \true{\view}$.
\end{lemma}
\begin{IEEEproof}
  Consider the true view \true{\view}.  By definition,
  $\partNS{\true{\view}} = NS$. By \autoref{restrict:liars}, $\setNS
  \subseteq \setT$ and thus, $\partNS{\true{\view}} \subseteq T$.
  $\forall \setR \in 2^{\setT} \generates \true{\view}$, so $\forall
  \setR \in 2^{\partNS{\true{\view}}} : \setR \generates
  \true{\view}$.
\end{IEEEproof}

Were all false views not consistent, then consistency could be used to
identify the true view.  A fully omniscient attacker could
theoretically generate a false, consistent view by collapsing all
conforming identities. However, the practical difficulty of collapsing
identities prevents this.  We formalize this as follows.
\begin{condition}\label{cond:one-conforming}
  All receiver sets correctly classify at least one conforming
  identity, i.e., $\forall \setR \in 2^{\setI}: \partNS{\view}(\setR)
  \cap \setC \neq \emptyset$.
\end{condition}
\begin{IEEEproof}[Justification]
  Collapsing conforming identities requires knowing the
  hard-to-predict initiator's RSSI
  observations. \autoref{sec:predictability} quantifies the
  probability that our required conditions hold.
  \renewcommand{\IEEEQED}{}
\end{IEEEproof}

\begin{lemma}\label{lemma:false-view-inconsistent}
  Under \autoref{cond:one-conforming}, a view generated by a receiver
  set containing a liar is not view-consistent, i.e., $\setR \cap
  \setL \neq \emptyset$ implies $\view(\setR)$ is not view-consistent.
\end{lemma}
\begin{IEEEproof}
  Consider such a receiver set $\setR$ with $\setR \cap \setL \neq
  \emptyset$.  By \autoref{cond:one-conforming}, $r
  \triangleq \partNS{\view}(\setR) \cap \setC$ is not empty and since
  $r \subseteq \setC \subseteq \setT$, $r \mapsto \true{\view}$.  By
  the definition of a liar, $\view(\setR) \neq \true{\view}$ and thus
  $\setR$ is not consistent.
\end{IEEEproof}

\begin{theorem}\label{thm:true-consistent}
  Under \autoref{restrict:liars} and \autoref{cond:one-conforming} and
  assuming $C \neq \emptyset$, exactly one consistent view is
  generated across all receiver sets and that view is the true view.
\end{theorem}
\begin{IEEEproof}
  By \autoref{lemma:true-view-consistent} and
  \autoref{lemma:false-view-inconsistent}, only the true view is
  consistent, so we need only show that at least one receiver set
  generates the true view.  $\setC \neq \emptyset$ and thus $\setR =
  \setC \generates \true{\view}$.
\end{IEEEproof}

This result suggests a method to identify the true view---select the
only consistent view.  \autoref{restrict:liars} does not hold in
practice, so we develop methods to relax it.

\subsection{Achieving Consistency by Eliminating $\setLNS$}
\label{sec:policy1}

Consider a scenario with some non-Sybil liars.  The true view would be
consistent were the non-Sybil liars excluded.  Similarly, a false view
could be consistent were the correctly classified conforming
identities excluded. If the latter outnumber the former, this yields a
useful property: the consistent view for the largest subset of
identities, i.e., that with the fewest excluded, is the true view, as
we now formalize and prove.

\begin{condition}\label{cond:more-conforming}
  The number of conforming identities is strictly greater than the
  number of non-Sybil liars, i.e., $|\setC| > |\setLNS|$.
\end{condition}
\begin{IEEEproof}[Justification] 
  This is assumed by networks whose protocols require a majority of the
  nodes to behave.  In others, it may hold for economic
  reasons---deploying as many nodes as the conforming participants is
  expensive.  \renewcommand{\IEEEQED}{}
\end{IEEEproof}

\begin{condition}\label{cond:collapse-lns}
  Each receiver set either correctly classifies at least $|\setLNS| +
  1$ conforming identities as non-Sybil or the resulting view, when
  all correctly classified conforming identities are excluded, is not
  consistent, i.e., $\forall \setR \in 2^{\setI}:
  (|\partNS{\view}(\setR) \cap \setC| \geq |\setLNS| + 1) \vee
  (\exists Q \in 2^{\partNS{\view}(R) \setminus \setC} : V(Q) \neq
  V(R))$. Note that this implies \autoref{cond:more-conforming}.
\end{condition}
\begin{IEEEproof}[Justification]
  This is an extension of \autoref{cond:one-conforming}.
  \autoref{sec:predictability} quantifies the probability that it
  holds.  \renewcommand{\IEEEQED}{}
\end{IEEEproof}

\begin{lemma}\label{lemma:largest-subset}
  Under \autoref{cond:more-conforming} and
  \autoref{cond:collapse-lns}, the largest subset of $\setI$
  permitting a consistent view is $\setI \setminus \setLNS$.
\end{lemma}
\begin{IEEEproof}
  $\setI \setminus \setLNS$ permits a consistent view, per
  \autoref{lemma:true-view-consistent}.  Let $E_\setR
  \triangleq \partNS{\false{\view}}(R) \cap \setC$ be the set of correctly
  classified conforming nodes for a lying receiver set $R$, i.e.,
  $\setR \cap \setL \neq \emptyset$.  $\setI \setminus E_\setR$ is the
  largest subset possibly permitting a consistent view under $\setR$.
  By \autoref{cond:collapse-lns}, $\forall \setR : |E_\setR| \geq
  |\setLNS| + 1$.
\end{IEEEproof}

\begin{theorem}\label{thm:policy1}
  Under \autoref{cond:more-conforming} and
  \autoref{cond:collapse-lns}, the largest subset of $\setI$
  permitting a consistent view permits just one consistent view, the
  true view.
\end{theorem}
\begin{IEEEproof}
  This follows directly from \autoref{lemma:largest-subset} and
  \autoref{thm:true-consistent}.
\end{IEEEproof}

In the next section, we extend the approach to handle the noise
inherent in real-world signalprints.

\subsection{Extending Consistency to Handle Noise}
\label{sec:handle-noise}

Noise prevents true signalprints from always generating the true view.
Observing from prior work that the misclassifications are bounded
(e.g., more than 99\% of Sybils detected with fewer than 5\% of
conforming identities collapsed~\cite{xiao09sep,demirbas06jun}), we
extend the notion of consistency as follows.

\begin{definition}
  Let $\gamma_n$ be the maximum fraction\footnote{$\gamma_n$ is not
    the probability that an individual identity is misclassified, but
    an upper bound on the total fraction misclassified.} of non-Sybil
  identities misclassified by a size-$n$ receiver set.  Prior work
  suggests $\gamma_4 = 0.05$ is appropriate (for $|\setC| >
  20$)~\cite{xiao09sep,demirbas06jun}.
\end{definition}

\begin{definition}
  A view is \emph{$\gamma_n$-consistent} if and only if all size-$n$
  receiver sets that are subsets of the non-Sybil identities under
  that view generate a \emph{$\gamma_n$-similar} view. Two views
  $\view^1$ and $\view^2$ are $\gamma_n$-similar if and only if
  \[
  \left( \frac{|\partNS{\view^1}
      \cap \partNS{\view^2}|}{|\partNS{\view^1}
      \setminus \partNS{\view^2}|} > \frac{1-2\gamma_n}{\gamma_n}
  \right) \bigwedge \left( \frac{|\partNS{\view^1}
      \cap \partNS{\view^2}|}{|\partNS{\view^2}
      \setminus \partNS{\view^1}|} > \frac{1-2\gamma_n}{\gamma_n}
  \right)
  \]
  This statement captures the intuitive notion that $\partNS{\view^1}$
  and $\partNS{\view^2}$ should contain the same identities up to
  differences expected under the $\gamma_n$ bound. A view is
  \emph{$\gamma_n$-true} if it is $\gamma_n$-similar to the true view.
\end{definition}

\begin{lemma}
  Under \autoref{restrict:liars}, the view generated by any truthful
  receiver set of size $n$ is $\gamma_n$-consistent.\footnote{This
    assumes that the false negative bound is negligible.  If it is
    not, a similar notion of $\gamma$,$\sigma$-consistency, where
    $\sigma$ is the false negative bound, can be used.  In practice
    $\sigma$ is quite small~\cite{demirbas06jun,xiao09sep}, so simple
    $\gamma_n$-consistency is fine.}
\end{lemma}
\begin{IEEEproof}
  Consider two views $\view^1$ and $\view^2$ generated by conforming
  receiver sets.  Each correctly classifies at least $(1-\gamma_n)$ of
  the non-Sybil identities, so $|\partNS{\view^1}
  \cap \partNS{\view^2}| \geq (1-2\gamma_n)|\setNS|$. Each
  misclassifies at most $\gamma_n$ of the non-Sybil identities, so
  $|\partNS{\view^1} \setminus \partNS{\view^2}| \leq
  \gamma_n|\setNS|$ and similar for $\partNS{\view^2}
  \setminus \partNS{\view^1}$. The ratio of these bounds is the
  result.
\end{IEEEproof}

Substituting $\gamma$-consistency for pure consistency,
\autoref{cond:collapse-lns} still holds with high (albeit different)
probability, quantified in \autoref{sec:predictability}. An analogue
of \autoref{thm:policy1} follows.
\begin{theorem}\label{thm:policy1-gamma}
  Under \autoref{cond:collapse-lns}, the $\gamma_n$-consistent view of
  the largest subset of $\setI$ permitting such a view is
  $\gamma_n$-true.
\end{theorem}

In \autoref{sec:algorithm} we describe an efficient algorithm to
identify the largest subset permitting a $\gamma$-consistent view and
thus the correct (up to errors expected due to signalprint noise)
Sybil classification.

%% file: fig/sets-nolabel.pdf_tex
\begingroup%
  \makeatletter%
  \providecommand\color[2][]{%
    \errmessage{(Inkscape) Color is used for the text in Inkscape, but the package 'color.sty' is not loaded}%
    \renewcommand\color[2][]{}%
  }%
  \providecommand\transparent[1]{%
    \errmessage{(Inkscape) Transparency is used (non-zero) for the text in Inkscape, but the package 'transparent.sty' is not loaded}%
    \renewcommand\transparent[1]{}%
  }%
  \providecommand\rotatebox[2]{#2}%
  \ifx\svgwidth\undefined%
    \setlength{\unitlength}{86.4bp}%
    \ifx\svgscale\undefined%
      \relax%
    \else%
      \setlength{\unitlength}{\unitlength * \real{\svgscale}}%
    \fi%
  \else%
    \setlength{\unitlength}{\svgwidth}%
  \fi%
  \global\let\svgwidth\undefined%
  \global\let\svgscale\undefined%
  \makeatother%
  \begin{picture}(1,0.83333333)%
    \put(0,0){\includegraphics[width=\unitlength]{sets-nolabel.pdf}}%
    \put(0.46286498,0.47249414){\color[rgb]{0,0,0}\makebox(0,0)[b]{\smash{\setLS}}}%
    \put(0.78805556,0.47249414){\color[rgb]{0,0,0}\makebox(0,0)[b]{\smash{\setLNS}}}%
    \put(0.46286498,0.14835281){\color[rgb]{0,0,0}\makebox(0,0)[b]{\smash{\setTS}}}%
    \put(0.78732314,0.14835281){\color[rgb]{0,0,0}\makebox(0,0)[b]{\smash{\setC}}}%
    \put(0.46286498,0.72867282){\color[rgb]{0,0,0}\makebox(0,0)[b]{\smash{\setS}}}%
    \put(0.7880557,0.72867282){\color[rgb]{0,0,0}\makebox(0,0)[b]{\smash{\setNS}}}%
    \put(0.26885633,0.47249414){\color[rgb]{0,0,0}\makebox(0,0)[rb]{\smash{\setL}}}%
    \put(0.26712022,0.14835281){\color[rgb]{0,0,0}\makebox(0,0)[rb]{\smash{\setT}}}%
  \end{picture}%
\endgroup%

%% file: algorithm.tex
\section{Efficient Implementation of the\\Selection Policies}
\label{sec:algorithm}

Both the maximum Sybil and view consistency policies offer ways to
select a view, either the one claiming the most Sybils or the largest
one that is $\gamma_n$-true, but brute-force examination of
$2^{|\setI|}$ receiver sets is infeasible. Instead, we describe
$\bigO(|\setI|^3)$ algorithms for both policies. In summary, both
start by generating $\bigO(|\setI|)$ candidate views
(\autoref{alg:build-receiver-sets}). For the maximum Sybil policy, the
one claiming the most Sybil identities is trivially identified. For
the view consistency policy, \autoref{alg:find-consistent-subset} is
used to identify largest $\gamma_n$-consistent view.

\begin{algorithm}
  \caption{Choose the receiver sets to consider}
  \label{alg:build-receiver-sets}
  \fontsize{9.5}{10.5}\selectfont
  \begin{algorithmic}[1]
    \Require $i_0$ is the identity running the procedure
    \Require $n$ is the desired receiver set size
    \State $S \gets \emptyset$
    \ForAll {$i \in \setI$}
        \State $\setR \gets \{i_0,i\}$
        \For {$cnt = 3 \to n$}
            \State $\setR \gets \setR \cup \{\textrm{RandElement}(\partNS{\view}(\setR))\}$
        \EndFor
        \State $S \gets S \cup \{R\}$ 
    \EndFor
    \State \Return $S$ \Comment{\parbox[t]{.65\linewidth}{with high probability, $S$ contains a truthful receiver set}}
  \end{algorithmic}
\end{algorithm}

\begin{figure}
  \centering
  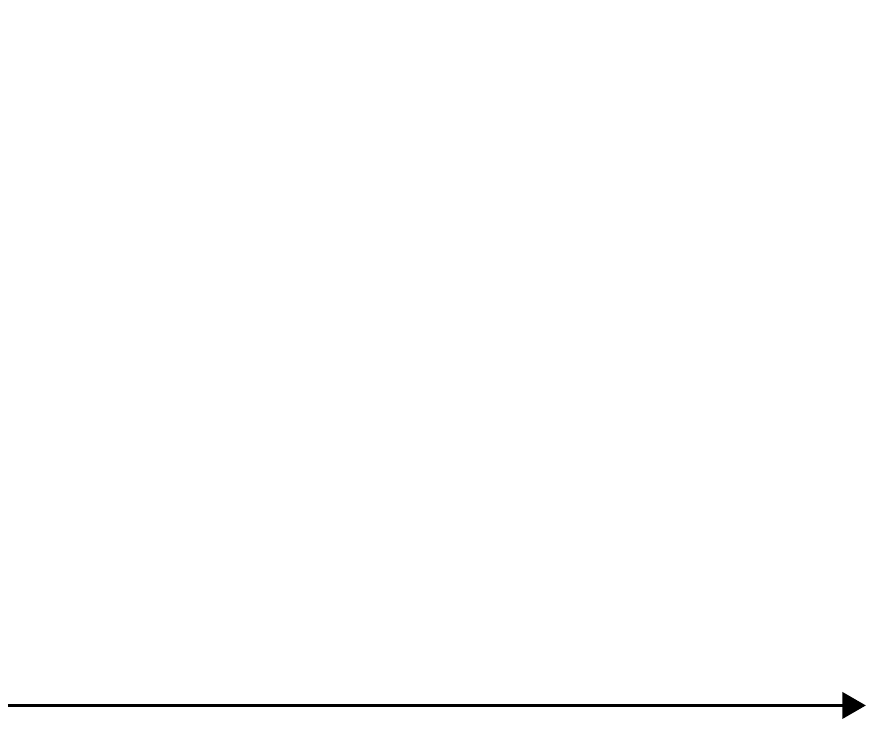
  \caption{Illustration of \autoref{alg:build-receiver-sets}. All
    $|\setI|$ size-2 receiver sets are increased to size-4 by
    iteratively adding a random identity from those labeled non-Sybil
    by the current set. With high probability, at least one of the
    final sets will contain only conforming identities.  }
  \label{fig:build-receiver-sets}
\end{figure}

\subsection{Candidate Receiver Set Selection}

\begin{figure}
  \centering
  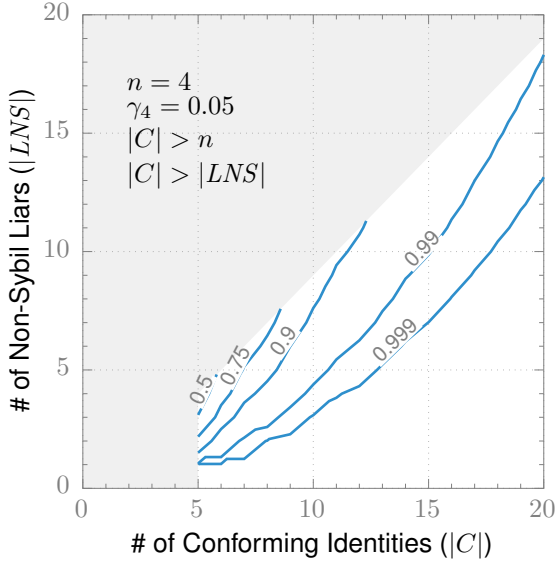
  \sbox0{\footnotemark}
  \caption{Contours of probability that at least one of the receiver
    sets from \autoref{alg:build-receiver-sets} is
    conforming\protect.\usebox0}
  \label{fig:build-receiver-set-probability}
\end{figure}

The only requirement for candidate receiver set selection is that at
least one of the candidates must be truthful.
\autoref{alg:build-receiver-sets} selects $|\setI|,$ size-$n$ (we
suggest $n=4$) receiver sets of which at least one is truthful with
high probability.  As illustrated in
\autoref{fig:build-receiver-sets}, the algorithm starts with all
$|\setI|$ size-2 receiver sets (lines 2--3) and builds each up to the
full size-$n$ by iteratively (line 4) adding a randomly selected
identity from those indicated to be conforming at the prior lower
dimensionality (line 5).  At least $|\setC|$ of the initial size-2
receiver sets are conforming and after increasing to size-$n$, at
least one is still conforming with high probability (graphed in
\autoref{fig:build-receiver-set-probability}):
\[1-\left(1 -
  \prod_{m=2}^{n-1}\frac{(1-\gamma_m)\cdot|\setC|-(m-1)}{|\setLNS|+(1-\gamma_m)\cdot|\setC|-(m-1)}
\right)^{|\setC|}
\]
The signalprint threshold for this process is chosen to eliminate
(nearly) all false negatives, as having a few false positives does not
matter.  The complexity of a straightforward implementation is
$\bigO(|\setI|^3)$.  \autoref{sec:discussion} further discusses the
runtime.

\footnotetext{For small $|\setC|$ and relatively large $|\setLNS|$ the
  probability can be increased by building $2\cdot|\setI|$ or $3\cdot|\setI|$ or
  more receiver sets instead. We omit further details due to lack of
  space.}

\subsection{Finding the Largest $\gamma_n$-Consistent View}

\begin{algorithm}
  \caption{Find receiver set permitting the largest $\gamma_n$-consistent subset}
  \label{alg:find-consistent-subset}
  \fontsize{9.5}{10.5}\selectfont
  \begin{algorithmic}[1]
    \Require $S$ is the set of receivers sets generated by \autoref{alg:build-receiver-sets}
    \Require $\partNS{\view}(\setR)$ for each $\setR \in $ \{size-2 receiver sets\} computed by \autoref{alg:build-receiver-sets}
    \Require $s$ is the initiator running the algorithm
    \State $(C,\setR_\textrm{max}) \gets (\infty,\textrm{null})$
    \ForAll {$\setR \in \setS$}
        \State Compute RSSI ratio for each Sybil set in $\partS{\view}(\setR)$
        \State $c \gets 0$
        \ForAll {$i \in \partNS{\view}(\setR)$}
            \State $e \gets 0$
            \State $n \gets$ number of identities whose RSSI ratios reported by $i$ do not match that for $\setR$ 
            \If {$\frac{|\partNS{\view}(\setR)| + n}{n} < \frac{1-2\gamma_n}{\gamma_n}$} 
                  \State $e \gets 1$
            \EndIf
            \If {$\view(\setR)$ and $\view(\{i,s\})$ are not $\gamma_2$-similar}
                  \State $e \gets 1$
            \EndIf
            \If {$e = 1$}
                  \State $c \gets c + 1$ \Comment {exclude $i$}                        
            \EndIf
        \EndFor
        \If {$c < C$}
            \State $(C,\setR_\textrm{max}) \gets (c,\setR)$ \Comment{new largest $\gamma$-consistent subset found}
        \EndIf
    \EndFor
    \State \Return $R_\textrm{max}$
  \end{algorithmic}
\end{algorithm}

Given the $|\setI|$ candidate receiver sets, the next task is
identifying the one generating a $\gamma_n$-true view, which, pursuant
to \autoref{thm:policy1-gamma}, is that permitting the largest subset
of $\setI$ to be $\gamma_n$-consistent. Checking consistency by
examining all $2^{|\partNS{\view}|}$ receiver sets is infeasible, so
we make several observations leading to the $\bigO(|\setI|^3)$
\autoref{alg:find-consistent-subset}. For each candidate receiver set
(line 2), we determine how many identities must be excluded for the
view to be $\gamma_n$-consistent (lines 3--17).  That excluding the
fewest is $\gamma_n$-true and the desired classification (line 22).

The crux of the algorithm is lines 3--17, which use the following
observations to efficiently determine which identities must be
excluded.
\begin{enumerate}
\item Adding an identity to a receiver set can change the view in one
  direction only---an identity can go from Sybil to non-Sybil, but not
  vice versa---because uncorrelated RSSI vectors cannot become
  correlated by increasing the dimension.\footnote{This is not true for
    low dimension receiver sets affected by noise, but is for the
    size-$(n>4)$ sets considered here.}
\item For identities $a$ and $b$, $\setR \cup \{a\} \generates
  \view(\setR)$ and $\setR \cup \{b\} \generates \view(\setR)$ implies
  $\setR \cup \{a,b\} \generates \view(\setR)$ because $a$ and $b$
  must have the same RSSI ratios for the Sybils as $\setR$.
\end{enumerate}
From these observations, we determine the excluded identities by
computing, for each identity in $\partS{\view}(\setR)$, the RSSI ratio
with an arbitrary sibling (line 3) and comparing against those
reported by potential non-Sybils in $\partNS{\view}(\setR)$ (line
7). If the number not matching is too large (line 8), the view is not
$\gamma_n$-consistent and the identity is excluded (line 15). Further,
if the receiver set consisting of just the supposedly non-Sybil
identity and the initiator is not $\gamma_2$-similar to $\setR$ (line
11).

\subsection{Runtime in the Absence of Liars}

In a typical situation with no liars, the consistency algorithm can
detect the Sybils in $\bigO(|\setI|^2)$ time.  Since all identities
are truthful, any chosen receiver set will be $\gamma_n$-consistent
with no exclusions---clearly the largest possible---and thus the other
$|\setI|-1$ also-truthful receiver sets need not be checked.  With
lying attackers present, the overall runtime is $\bigO(|\setI|^3)$, as
each algorithm takes $\bigO(|\setI|^3)$ time.

%% file: fig/build-receiver-sets.pdf_tex
\begingroup%
  \makeatletter%
  \providecommand\color[2][]{%
    \errmessage{(Inkscape) Color is used for the text in Inkscape, but the package 'color.sty' is not loaded}%
    \renewcommand\color[2][]{}%
  }%
  \providecommand\transparent[1]{%
    \errmessage{(Inkscape) Transparency is used (non-zero) for the text in Inkscape, but the package 'transparent.sty' is not loaded}%
    \renewcommand\transparent[1]{}%
  }%
  \providecommand\rotatebox[2]{#2}%
  \ifx\svgwidth\undefined%
    \setlength{\unitlength}{251.28000488bp}%
    \ifx\svgscale\undefined%
      \relax%
    \else%
      \setlength{\unitlength}{\unitlength * \real{\svgscale}}%
    \fi%
  \else%
    \setlength{\unitlength}{\svgwidth}%
  \fi%
  \global\let\svgwidth\undefined%
  \global\let\svgscale\undefined%
  \makeatother%
  \begin{tikzpicture}[x=\unitlength,y=\unitlength,remember picture]%
    \node[anchor=south west,inner sep=0] at (0,0) {\includegraphics[width=\unitlength,page=1]{build-receiver-sets.pdf}};%
    \node[anchor=center] at (0.50910538,0.0077095) {\color[rgb]{0,0,0}\smash{Algorithm Progression}};%
    \node[anchor=east] at (0.16640397,0.83011561) {\color[rgb]{0,0,0}\smash{$\setR_1$}};%
    \node[anchor=center] at (0.27961893,0.79376413) {\color[rgb]{0,0,0}\rotatebox{-90}{\smash{$\generates$}}};%
    \node[anchor=center] at (0.28870612,0.83011561) {\color[rgb]{0,0,0}\smash{$(i_1,i_0)$}};%
    \node[anchor=center] at (0.25116656,0.73444321) {\color[rgb]{0,0,0}\smash{S}};%
    \node[anchor=center] at (0.31963713,0.73444321) {\color[rgb]{0,0,0}\smash{NS}};%
    \node[anchor=center] at (0.25722332,0.69461621) {\color[rgb]{0,0,0}\smash{$i_3$}};%
    \node[anchor=center] at (0.32018895,0.64748486) {\color[rgb]{0,0,0}\smash{\circled[n15]{$i_5$}}};%
    \node[anchor=center] at (0.25722332,0.6005253) {\color[rgb]{0,0,0}\smash{$i_6$}};%
    \node[anchor=center] at (0.32018895,0.69444443) {\color[rgb]{0,0,0}\smash{$i_2$}};%
    \node[anchor=center] at (0.25722332,0.64757075) {\color[rgb]{0,0,0}\smash{$i_4$}};%
    \node[anchor=center] at (0.32018895,0.6005253) {\color[rgb]{0,0,0}\smash{$i_8$}};%
    \node[anchor=south west,inner sep=0] at (0,0) {\includegraphics[width=\unitlength,page=2]{build-receiver-sets.pdf}};%
    \node[anchor=center] at (0.25544339,0.55658166) {\color[rgb]{0,0,0}\rotatebox{-90}{\smash{…}}};%
    \node[anchor=center] at (0.318409,0.55658166) {\color[rgb]{0,0,0}\rotatebox{-90}{\smash{…}}};%
    \node[anchor=south west,inner sep=0] at (0,0) {\includegraphics[width=\unitlength,page=3]{build-receiver-sets.pdf}};%
    \node[anchor=east] at (0.16640397,0.37810671) {\color[rgb]{0,0,0}\smash{$\setR_{|I|}$}};%
    \node[anchor=center] at (0.27961893,0.34175524) {\color[rgb]{0,0,0}\rotatebox{-90}{\smash{$\generates$}}};%
    \node[anchor=center] at (0.28870612,0.37810671) {\color[rgb]{0,0,0}\smash{$(i_{|I|},i_0)$}};%
    \node[anchor=center] at (0.25116656,0.2824343) {\color[rgb]{0,0,0}\smash{S}};%
    \node[anchor=center] at (0.31963713,0.2824343) {\color[rgb]{0,0,0}\smash{NS}};%
    \node[anchor=center] at (0.25722332,0.24260731) {\color[rgb]{0,0,0}\smash{$i_1$}};%
    \node[anchor=center] at (0.32018895,0.19547597) {\color[rgb]{0,0,0}\smash{$i_3$}};%
    \node[anchor=center] at (0.25722332,0.1485164) {\color[rgb]{0,0,0}\smash{$i_5$}};%
    \node[anchor=center] at (0.32018895,0.24243553) {\color[rgb]{0,0,0}\smash{$i_2$}};%
    \node[anchor=center] at (0.25722332,0.19556185) {\color[rgb]{0,0,0}\smash{$i_4$}};%
    \node[anchor=center] at (0.32018895,0.1485164) {\color[rgb]{0,0,0}\smash{\circled[n16]{$i_6$}}};%
    \node[anchor=south west,inner sep=0] at (0,0) {\includegraphics[width=\unitlength,page=4]{build-receiver-sets.pdf}};%
    \node[anchor=center] at (0.25544339,0.10457274) {\color[rgb]{0,0,0}\rotatebox{-90}{\smash{…}}};%
    \node[anchor=center] at (0.318409,0.10457274) {\color[rgb]{0,0,0}\rotatebox{-90}{\smash{…}}};%
    \node[anchor=south west,inner sep=0] at (0,0) {\includegraphics[width=\unitlength,page=5]{build-receiver-sets.pdf}};%
    \node[anchor=east] at (0.21195186,0.83030216) {\color[rgb]{0,0,0}\smash{$\triangleq$}};%
    \node[anchor=east] at (0.21195186,0.37810671) {\color[rgb]{0,0,0}\smash{$\triangleq$}};%
    \node[anchor=east] at (0.21195186,0.69480275) {\color[rgb]{0,0,0}\smash{$\triangleq$}};%
    \node[anchor=east] at (0.16640397,0.69480275) {\color[rgb]{0,0,0}\smash{$\view(\setR_1)$}};%
    \node[anchor=east] at (0.16640397,0.24279385) {\color[rgb]{0,0,0}\smash{$\view(\setR_{|I|})$}};%
    \node[anchor=east] at (0.21195186,0.24279385) {\color[rgb]{0,0,0}\smash{$\triangleq$}};%
    \node[anchor=center] at (0.28711959,0.46383718) {\color[rgb]{0,0,0}\rotatebox{-90}{\smash{…}}};%
    \node[anchor=center] at (0.12987226,0.54555653) {\color[rgb]{0,0,0}\rotatebox{-90}{\smash{…}}};%
    \node[anchor=center] at (0.56167441,0.79376413) {\color[rgb]{0,0,0}\rotatebox{-90}{\smash{$\generates$}}};%
    \node[anchor=center] at (0.57095502,0.83011561) {\color[rgb]{0,0,0}\smash{$($\circled[n25]{$i_5$}$,i_1,i_0)$}};%
    \node[anchor=center] at (0.53351211,0.73444321) {\color[rgb]{0,0,0}\smash{S}};%
    \node[anchor=center] at (0.60198269,0.73444321) {\color[rgb]{0,0,0}\smash{NS}};%
    \node[anchor=center] at (0.53956887,0.69461621) {\color[rgb]{0,0,0}\smash{$i_1$}};%
    \node[anchor=center] at (0.6025345,0.64748486) {\color[rgb]{0,0,0}\smash{$i_8$}};%
    \node[anchor=center] at (0.53956887,0.6005253) {\color[rgb]{0,0,0}\smash{$i_6$}};%
    \node[anchor=center] at (0.6025345,0.69444443) {\color[rgb]{0,0,0}\smash{\circled[n23a]{$i_3$}}};%
    \node[anchor=center] at (0.53956887,0.64757075) {\color[rgb]{0,0,0}\smash{$i_4$}};%
    \node[anchor=center] at (0.6025345,0.6005253) {\color[rgb]{0,0,0}\smash{$i_9$}};%
    \node[anchor=south west,inner sep=0] at (0,0) {\includegraphics[width=\unitlength,page=6]{build-receiver-sets.pdf}};%
    \node[anchor=center] at (0.53778895,0.55658166) {\color[rgb]{0,0,0}\rotatebox{-90}{\smash{…}}};%
    \node[anchor=center] at (0.60075455,0.55658166) {\color[rgb]{0,0,0}\rotatebox{-90}{\smash{…}}};%
    \node[anchor=south west,inner sep=0] at (0,0) {\includegraphics[width=\unitlength,page=7]{build-receiver-sets.pdf}};%
    \node[anchor=center] at (0.56167441,0.34175524) {\color[rgb]{0,0,0}\rotatebox{-90}{\smash{$\generates$}}};%
    \node[anchor=center] at (0.57095502,0.37810671) {\color[rgb]{0,0,0}\smash{$($\circled[n26]{$i_6$}$,i_{|I|},i_0)$}};%
    \node[anchor=center] at (0.53351211,0.2824343) {\color[rgb]{0,0,0}\smash{S}};%
    \node[anchor=center] at (0.60198269,0.2824343) {\color[rgb]{0,0,0}\smash{NS}};%
    \node[anchor=center] at (0.53956887,0.24260731) {\color[rgb]{0,0,0}\smash{$i_1$}};%
    \node[anchor=center] at (0.6025345,0.19547597) {\color[rgb]{0,0,0}\smash{\circled[n23b]{$i_3$}}};%
    \node[anchor=center] at (0.53956887,0.1485164) {\color[rgb]{0,0,0}\smash{$i_5$}};%
    \node[anchor=center] at (0.6025345,0.24243553) {\color[rgb]{0,0,0}\smash{$i_2$}};%
    \node[anchor=center] at (0.53956887,0.19556185) {\color[rgb]{0,0,0}\smash{$i_4$}};%
    \node[anchor=center] at (0.6025345,0.1485164) {\color[rgb]{0,0,0}\smash{$i_7$}};%
    \node[anchor=south west,inner sep=0] at (0,0) {\includegraphics[width=\unitlength,page=8]{build-receiver-sets.pdf}};%
    \node[anchor=center] at (0.53778895,0.10457274) {\color[rgb]{0,0,0}\rotatebox{-90}{\smash{…}}};%
    \node[anchor=center] at (0.60075455,0.10457274) {\color[rgb]{0,0,0}\rotatebox{-90}{\smash{…}}};%
    \node[anchor=south west,inner sep=0] at (0,0) {\includegraphics[width=\unitlength,page=9]{build-receiver-sets.pdf}};%
    \node[anchor=center] at (0.56917507,0.46383718) {\color[rgb]{0,0,0}\rotatebox{-90}{\smash{…}}};%
    \node[anchor=center] at (0.84372989,0.79376413) {\color[rgb]{0,0,0}\rotatebox{-90}{\smash{$\generates$}}};%
    \node[anchor=center] at (0.8530105,0.83011561) {\color[rgb]{0,0,0}\smash{$($\circled[n33a]{$i_3$}$,i_5,i_1,i_0)$}};%
    \node[anchor=center] at (0.81566432,0.73444321) {\color[rgb]{0,0,0}\smash{S}};%
    \node[anchor=center] at (0.88413489,0.73444321) {\color[rgb]{0,0,0}\smash{NS}};%
    \node[anchor=center] at (0.82172108,0.69461621) {\color[rgb]{0,0,0}\smash{$i_2$}};%
    \node[anchor=center] at (0.88468671,0.64748486) {\color[rgb]{0,0,0}\smash{$i_9$}};%
    \node[anchor=center] at (0.82172108,0.6005253) {\color[rgb]{0,0,0}\smash{$i_6$}};%
    \node[anchor=center] at (0.88468671,0.69444443) {\color[rgb]{0,0,0}\smash{$i_8$}};%
    \node[anchor=center] at (0.82172108,0.64757075) {\color[rgb]{0,0,0}\smash{$i_4$}};%
    \node[anchor=center] at (0.88468671,0.6005253) {\color[rgb]{0,0,0}\smash{$i_{11}$}};%
    \node[anchor=south west,inner sep=0] at (0,0) {\includegraphics[width=\unitlength,page=10]{build-receiver-sets.pdf}};%
    \node[anchor=center] at (0.81994115,0.55658166) {\color[rgb]{0,0,0}\rotatebox{-90}{\smash{…}}};%
    \node[anchor=center] at (0.88290676,0.55658166) {\color[rgb]{0,0,0}\rotatebox{-90}{\smash{…}}};%
    \node[anchor=south west,inner sep=0] at (0,0) {\includegraphics[width=\unitlength,page=11]{build-receiver-sets.pdf}};%
    \node[anchor=center] at (0.84372989,0.34175524) {\color[rgb]{0,0,0}\rotatebox{-90}{\smash{$\generates$}}};%
    \node[anchor=center] at (0.8530105,0.37810671) {\color[rgb]{0,0,0}\smash{$($\circled[n33b]{$i_3$}$,i_6,i_{|I|},i_0)$}};%
    \node[anchor=center] at (0.81566432,0.2824343) {\color[rgb]{0,0,0}\smash{S}};%
    \node[anchor=center] at (0.88413489,0.2824343) {\color[rgb]{0,0,0}\smash{NS}};%
    \node[anchor=center] at (0.82172108,0.24260731) {\color[rgb]{0,0,0}\smash{$i_1$}};%
    \node[anchor=center] at (0.88468671,0.19547597) {\color[rgb]{0,0,0}\smash{$i_7$}};%
    \node[anchor=center] at (0.82172108,0.1485164) {\color[rgb]{0,0,0}\smash{$i_5$}};%
    \node[anchor=center] at (0.88468671,0.24243553) {\color[rgb]{0,0,0}\smash{$i_2$}};%
    \node[anchor=center] at (0.82172108,0.19556185) {\color[rgb]{0,0,0}\smash{$i_4$}};%
    \node[anchor=center] at (0.88468671,0.1485164) {\color[rgb]{0,0,0}\smash{$i_8$}};%
    \node[anchor=south west,inner sep=0] at (0,0) {\includegraphics[width=\unitlength,page=12]{build-receiver-sets.pdf}};%
    \node[anchor=center] at (0.81994115,0.10457274) {\color[rgb]{0,0,0}\rotatebox{-90}{\smash{…}}};%
    \node[anchor=center] at (0.88290676,0.10457274) {\color[rgb]{0,0,0}\rotatebox{-90}{\smash{…}}};%
    \node[anchor=south west,inner sep=0] at (0,0) {\includegraphics[width=\unitlength,page=13]{build-receiver-sets.pdf}};%
    \node[anchor=center] at (0.85123055,0.46383718) {\color[rgb]{0,0,0}\rotatebox{-90}{\smash{…}}};%
    \InputIfFileExists{fig/build-receiver-sets.pdf_tikz}{}{}%
  \end{tikzpicture}%
\endgroup%

%% file: fig/build-receiver-set-probability-contours.pdf_tex
\begingroup
  \makeatletter
  \providecommand\color[2][]{%
    \GenericError{(gnuplot) \space\space\space\@spaces}{%
      Package color not loaded in conjunction with
      terminal option `colourtext'%
    }{See the gnuplot documentation for explanation.%
    }{Either use 'blacktext' in gnuplot or load the package
      color.sty in LaTeX.}%
    \renewcommand\color[2][]{}%
  }%
  \providecommand\includegraphics[2][]{%
    \GenericError{(gnuplot) \space\space\space\@spaces}{%
      Package graphicx or graphics not loaded%
    }{See the gnuplot documentation for explanation.%
    }{The gnuplot epslatex terminal needs graphicx.sty or graphics.sty.}%
    \renewcommand\includegraphics[2][]{}%
  }%
  \providecommand\rotatebox[2]{#2}%
  \@ifundefined{ifGPcolor}{%
    \newif\ifGPcolor
    \GPcolortrue
  }{}%
  \@ifundefined{ifGPblacktext}{%
    \newif\ifGPblacktext
    \GPblacktextfalse
  }{}%
  \let\gplgaddtomacro\g@addto@macro
  \gdef\gplbacktext{}%
  \gdef\gplfronttext{}%
  \makeatother
  \ifGPblacktext
    \def\colorrgb#1{}%
    \def\colorgray#1{}%
  \else
    \ifGPcolor
      \def\colorrgb#1{\color[rgb]{#1}}%
      \def\colorgray#1{\color[gray]{#1}}%
      \expandafter\def\csname LTw\endcsname{\color{white}}%
      \expandafter\def\csname LTb\endcsname{\color{black}}%
      \expandafter\def\csname LTa\endcsname{\color{black}}%
      \expandafter\def\csname LT0\endcsname{\color[rgb]{1,0,0}}%
      \expandafter\def\csname LT1\endcsname{\color[rgb]{0,1,0}}%
      \expandafter\def\csname LT2\endcsname{\color[rgb]{0,0,1}}%
      \expandafter\def\csname LT3\endcsname{\color[rgb]{1,0,1}}%
      \expandafter\def\csname LT4\endcsname{\color[rgb]{0,1,1}}%
      \expandafter\def\csname LT5\endcsname{\color[rgb]{1,1,0}}%
      \expandafter\def\csname LT6\endcsname{\color[rgb]{0,0,0}}%
      \expandafter\def\csname LT7\endcsname{\color[rgb]{1,0.3,0}}%
      \expandafter\def\csname LT8\endcsname{\color[rgb]{0.5,0.5,0.5}}%
    \else
      \def\colorrgb#1{\color{black}}%
      \def\colorgray#1{\color[gray]{#1}}%
      \expandafter\def\csname LTw\endcsname{\color{white}}%
      \expandafter\def\csname LTb\endcsname{\color{black}}%
      \expandafter\def\csname LTa\endcsname{\color{black}}%
      \expandafter\def\csname LT0\endcsname{\color{black}}%
      \expandafter\def\csname LT1\endcsname{\color{black}}%
      \expandafter\def\csname LT2\endcsname{\color{black}}%
      \expandafter\def\csname LT3\endcsname{\color{black}}%
      \expandafter\def\csname LT4\endcsname{\color{black}}%
      \expandafter\def\csname LT5\endcsname{\color{black}}%
      \expandafter\def\csname LT6\endcsname{\color{black}}%
      \expandafter\def\csname LT7\endcsname{\color{black}}%
      \expandafter\def\csname LT8\endcsname{\color{black}}%
    \fi
  \fi
    \setlength{\unitlength}{0.0500bp}%
    \ifx\gptboxheight\undefined%
      \newlength{\gptboxheight}%
      \newlength{\gptboxwidth}%
      \newsavebox{\gptboxtext}%
    \fi%
    \setlength{\fboxrule}{0.5pt}%
    \setlength{\fboxsep}{1pt}%
\begin{picture}(4320.00,4320.00)%
    \gplgaddtomacro\gplbacktext{%
      \colorrgb{0.50,0.50,0.50}%
      \put(489,569){\makebox(0,0)[r]{\strut{}$0$}}%
      \colorrgb{0.50,0.50,0.50}%
      \put(489,1462){\makebox(0,0)[r]{\strut{}$5$}}%
      \colorrgb{0.50,0.50,0.50}%
      \put(489,2355){\makebox(0,0)[r]{\strut{}$10$}}%
      \colorrgb{0.50,0.50,0.50}%
      \put(489,3248){\makebox(0,0)[r]{\strut{}$15$}}%
      \colorrgb{0.50,0.50,0.50}%
      \put(489,4141){\makebox(0,0)[r]{\strut{}$20$}}%
      \colorrgb{0.50,0.50,0.50}%
      \put(578,391){\makebox(0,0){\strut{}$0$}}%
      \colorrgb{0.50,0.50,0.50}%
      \put(1447,391){\makebox(0,0){\strut{}$5$}}%
      \colorrgb{0.50,0.50,0.50}%
      \put(2315,391){\makebox(0,0){\strut{}$10$}}%
      \colorrgb{0.50,0.50,0.50}%
      \put(3184,391){\makebox(0,0){\strut{}$15$}}%
      \colorrgb{0.50,0.50,0.50}%
      \put(4052,391){\makebox(0,0){\strut{}$20$}}%
    }%
    \gplgaddtomacro\gplfronttext{%
      \csname LTb\endcsname%
      \put(133,2355){\rotatebox{-270}{\makebox(0,0){\strut{}\# of Non-Sybil Liars ($|\setLNS|$)}}}%
      \csname LTb\endcsname%
      \put(2315,124){\makebox(0,0){\strut{}\# of Conforming Identities ($|\setC|$)}}%
      \colorrgb{0.50,0.50,0.50}%
      \put(2142,1641){\makebox(0,0){\rotatebox{55.243536}{\colorbox{white}{\smash{\fns{0.9}}}}}}%
      \colorrgb{0.50,0.50,0.50}%
      \put(1447,838){\makebox(0,0){\strut{}}}%
      \colorrgb{0.50,0.50,0.50}%
      \put(1548,926){\makebox(0,0){\strut{}}}%
      \colorrgb{0.50,0.50,0.50}%
      \put(1620,1016){\makebox(0,0){\strut{}}}%
      \colorrgb{0.50,0.50,0.50}%
      \put(1718,1105){\makebox(0,0){\strut{}}}%
      \colorrgb{0.50,0.50,0.50}%
      \put(1794,1215){\makebox(0,0){\strut{}}}%
      \colorrgb{0.50,0.50,0.50}%
      \put(1875,1283){\makebox(0,0){\strut{}}}%
      \colorrgb{0.50,0.50,0.50}%
      \put(1968,1377){\makebox(0,0){\strut{}}}%
      \colorrgb{0.50,0.50,0.50}%
      \put(2039,1462){\makebox(0,0){\strut{}}}%
      \colorrgb{0.50,0.50,0.50}%
      \put(2141,1640){\makebox(0,0){\strut{}}}%
      \colorrgb{0.50,0.50,0.50}%
      \put(2142,1641){\makebox(0,0){\strut{}}}%
      \colorrgb{0.50,0.50,0.50}%
      \put(2142,1641){\makebox(0,0){\strut{}}}%
      \colorrgb{0.50,0.50,0.50}%
      \put(2262,1819){\makebox(0,0){\strut{}}}%
      \colorrgb{0.50,0.50,0.50}%
      \put(2315,1929){\makebox(0,0){\strut{}}}%
      \colorrgb{0.50,0.50,0.50}%
      \put(2361,1998){\makebox(0,0){\strut{}}}%
      \colorrgb{0.50,0.50,0.50}%
      \put(2406,2092){\makebox(0,0){\strut{}}}%
      \colorrgb{0.50,0.50,0.50}%
      \put(2456,2176){\makebox(0,0){\strut{}}}%
      \colorrgb{0.50,0.50,0.50}%
      \put(2489,2252){\makebox(0,0){\strut{}}}%
      \colorrgb{0.50,0.50,0.50}%
      \put(2573,2355){\makebox(0,0){\strut{}}}%
      \colorrgb{0.50,0.50,0.50}%
      \put(2662,2483){\makebox(0,0){\strut{}}}%
      \colorrgb{0.50,0.50,0.50}%
      \put(2691,2534){\makebox(0,0){\strut{}}}%
      \colorrgb{0.50,0.50,0.50}%
      \put(2714,2586){\makebox(0,0){\strut{}}}%
      \colorrgb{0.50,0.50,0.50}%
      \put(3184,2331){\makebox(0,0){\rotatebox{46.460492}{\colorbox{white}{\smash{\fns{0.99}}}}}}%
      \colorrgb{0.50,0.50,0.50}%
      \put(1447,757){\makebox(0,0){\strut{}}}%
      \colorrgb{0.50,0.50,0.50}%
      \put(1502,805){\makebox(0,0){\strut{}}}%
      \colorrgb{0.50,0.50,0.50}%
      \put(1620,805){\makebox(0,0){\strut{}}}%
      \colorrgb{0.50,0.50,0.50}%
      \put(1756,926){\makebox(0,0){\strut{}}}%
      \colorrgb{0.50,0.50,0.50}%
      \put(1794,951){\makebox(0,0){\strut{}}}%
      \colorrgb{0.50,0.50,0.50}%
      \put(1878,1012){\makebox(0,0){\strut{}}}%
      \colorrgb{0.50,0.50,0.50}%
      \put(1968,1032){\makebox(0,0){\strut{}}}%
      \colorrgb{0.50,0.50,0.50}%
      \put(2052,1105){\makebox(0,0){\strut{}}}%
      \colorrgb{0.50,0.50,0.50}%
      \put(2141,1181){\makebox(0,0){\strut{}}}%
      \colorrgb{0.50,0.50,0.50}%
      \put(2253,1283){\makebox(0,0){\strut{}}}%
      \colorrgb{0.50,0.50,0.50}%
      \put(2315,1352){\makebox(0,0){\strut{}}}%
      \colorrgb{0.50,0.50,0.50}%
      \put(2427,1462){\makebox(0,0){\strut{}}}%
      \colorrgb{0.50,0.50,0.50}%
      \put(2489,1541){\makebox(0,0){\strut{}}}%
      \colorrgb{0.50,0.50,0.50}%
      \put(2609,1641){\makebox(0,0){\strut{}}}%
      \colorrgb{0.50,0.50,0.50}%
      \put(2662,1685){\makebox(0,0){\strut{}}}%
      \colorrgb{0.50,0.50,0.50}%
      \put(2781,1819){\makebox(0,0){\strut{}}}%
      \colorrgb{0.50,0.50,0.50}%
      \put(2836,1905){\makebox(0,0){\strut{}}}%
      \colorrgb{0.50,0.50,0.50}%
      \put(2915,1998){\makebox(0,0){\strut{}}}%
      \colorrgb{0.50,0.50,0.50}%
      \put(3010,2150){\makebox(0,0){\strut{}}}%
      \colorrgb{0.50,0.50,0.50}%
      \put(3037,2176){\makebox(0,0){\strut{}}}%
      \colorrgb{0.50,0.50,0.50}%
      \put(3184,2331){\makebox(0,0){\strut{}}}%
      \colorrgb{0.50,0.50,0.50}%
      \put(3202,2355){\makebox(0,0){\strut{}}}%
      \colorrgb{0.50,0.50,0.50}%
      \put(3229,2402){\makebox(0,0){\strut{}}}%
      \colorrgb{0.50,0.50,0.50}%
      \put(3324,2534){\makebox(0,0){\strut{}}}%
      \colorrgb{0.50,0.50,0.50}%
      \put(3357,2597){\makebox(0,0){\strut{}}}%
      \colorrgb{0.50,0.50,0.50}%
      \put(3439,2712){\makebox(0,0){\strut{}}}%
      \colorrgb{0.50,0.50,0.50}%
      \put(3531,2890){\makebox(0,0){\strut{}}}%
      \colorrgb{0.50,0.50,0.50}%
      \put(3532,2891){\makebox(0,0){\strut{}}}%
      \colorrgb{0.50,0.50,0.50}%
      \put(3532,2892){\makebox(0,0){\strut{}}}%
      \colorrgb{0.50,0.50,0.50}%
      \put(3648,3069){\makebox(0,0){\strut{}}}%
      \colorrgb{0.50,0.50,0.50}%
      \put(3705,3187){\makebox(0,0){\strut{}}}%
      \colorrgb{0.50,0.50,0.50}%
      \put(3744,3248){\makebox(0,0){\strut{}}}%
      \colorrgb{0.50,0.50,0.50}%
      \put(3783,3329){\makebox(0,0){\strut{}}}%
      \colorrgb{0.50,0.50,0.50}%
      \put(3843,3427){\makebox(0,0){\strut{}}}%
      \colorrgb{0.50,0.50,0.50}%
      \put(3878,3505){\makebox(0,0){\strut{}}}%
      \colorrgb{0.50,0.50,0.50}%
      \put(3940,3605){\makebox(0,0){\strut{}}}%
      \colorrgb{0.50,0.50,0.50}%
      \put(3993,3723){\makebox(0,0){\strut{}}}%
      \colorrgb{0.50,0.50,0.50}%
      \put(4028,3784){\makebox(0,0){\strut{}}}%
      \colorrgb{0.50,0.50,0.50}%
      \put(4052,3839){\makebox(0,0){\strut{}}}%
      \colorrgb{0.50,0.50,0.50}%
      \put(1529,1283){\makebox(0,0){\rotatebox{64.670266}{\colorbox{white}{\smash{\fns{0.5}}}}}}%
      \colorrgb{0.50,0.50,0.50}%
      \put(1447,1123){\makebox(0,0){\strut{}}}%
      \colorrgb{0.50,0.50,0.50}%
      \put(1529,1283){\makebox(0,0){\strut{}}}%
      \colorrgb{0.50,0.50,0.50}%
      \put(1587,1428){\makebox(0,0){\strut{}}}%
      \colorrgb{0.50,0.50,0.50}%
      \put(2978,1641){\makebox(0,0){\rotatebox{44.227537}{\colorbox{white}{\smash{\fns{0.999}}}}}}%
      \colorrgb{0.50,0.50,0.50}%
      \put(1447,749){\makebox(0,0){\strut{}}}%
      \colorrgb{0.50,0.50,0.50}%
      \put(1452,753){\makebox(0,0){\strut{}}}%
      \colorrgb{0.50,0.50,0.50}%
      \put(1620,753){\makebox(0,0){\strut{}}}%
      \colorrgb{0.50,0.50,0.50}%
      \put(1663,791){\makebox(0,0){\strut{}}}%
      \colorrgb{0.50,0.50,0.50}%
      \put(1794,791){\makebox(0,0){\strut{}}}%
      \colorrgb{0.50,0.50,0.50}%
      \put(1956,926){\makebox(0,0){\strut{}}}%
      \colorrgb{0.50,0.50,0.50}%
      \put(1968,929){\makebox(0,0){\strut{}}}%
      \colorrgb{0.50,0.50,0.50}%
      \put(1982,941){\makebox(0,0){\strut{}}}%
      \colorrgb{0.50,0.50,0.50}%
      \put(2141,977){\makebox(0,0){\strut{}}}%
      \colorrgb{0.50,0.50,0.50}%
      \put(2289,1105){\makebox(0,0){\strut{}}}%
      \colorrgb{0.50,0.50,0.50}%
      \put(2315,1121){\makebox(0,0){\strut{}}}%
      \colorrgb{0.50,0.50,0.50}%
      \put(2434,1227){\makebox(0,0){\strut{}}}%
      \colorrgb{0.50,0.50,0.50}%
      \put(2489,1252){\makebox(0,0){\strut{}}}%
      \colorrgb{0.50,0.50,0.50}%
      \put(2530,1283){\makebox(0,0){\strut{}}}%
      \colorrgb{0.50,0.50,0.50}%
      \put(2662,1340){\makebox(0,0){\strut{}}}%
      \colorrgb{0.50,0.50,0.50}%
      \put(2795,1462){\makebox(0,0){\strut{}}}%
      \colorrgb{0.50,0.50,0.50}%
      \put(2836,1503){\makebox(0,0){\strut{}}}%
      \colorrgb{0.50,0.50,0.50}%
      \put(2978,1641){\makebox(0,0){\strut{}}}%
      \colorrgb{0.50,0.50,0.50}%
      \put(3010,1677){\makebox(0,0){\strut{}}}%
      \colorrgb{0.50,0.50,0.50}%
      \put(3180,1819){\makebox(0,0){\strut{}}}%
      \colorrgb{0.50,0.50,0.50}%
      \put(3184,1822){\makebox(0,0){\strut{}}}%
      \colorrgb{0.50,0.50,0.50}%
      \put(3347,1998){\makebox(0,0){\strut{}}}%
      \colorrgb{0.50,0.50,0.50}%
      \put(3357,2012){\makebox(0,0){\strut{}}}%
      \colorrgb{0.50,0.50,0.50}%
      \put(3505,2176){\makebox(0,0){\strut{}}}%
      \colorrgb{0.50,0.50,0.50}%
      \put(3531,2214){\makebox(0,0){\strut{}}}%
      \colorrgb{0.50,0.50,0.50}%
      \put(3654,2355){\makebox(0,0){\strut{}}}%
      \colorrgb{0.50,0.50,0.50}%
      \put(3705,2431){\makebox(0,0){\strut{}}}%
      \colorrgb{0.50,0.50,0.50}%
      \put(3792,2534){\makebox(0,0){\strut{}}}%
      \colorrgb{0.50,0.50,0.50}%
      \put(3878,2668){\makebox(0,0){\strut{}}}%
      \colorrgb{0.50,0.50,0.50}%
      \put(3915,2712){\makebox(0,0){\strut{}}}%
      \colorrgb{0.50,0.50,0.50}%
      \put(3981,2818){\makebox(0,0){\strut{}}}%
      \colorrgb{0.50,0.50,0.50}%
      \put(4038,2891){\makebox(0,0){\strut{}}}%
      \colorrgb{0.50,0.50,0.50}%
      \put(4052,2916){\makebox(0,0){\strut{}}}%
      \colorrgb{0.50,0.50,0.50}%
      \put(1783,1462){\makebox(0,0){\rotatebox{62.579925}{\colorbox{white}{\smash{\fns{0.75}}}}}}%
      \colorrgb{0.50,0.50,0.50}%
      \put(1447,959){\makebox(0,0){\strut{}}}%
      \colorrgb{0.50,0.50,0.50}%
      \put(1575,1105){\makebox(0,0){\strut{}}}%
      \colorrgb{0.50,0.50,0.50}%
      \put(1620,1196){\makebox(0,0){\strut{}}}%
      \colorrgb{0.50,0.50,0.50}%
      \put(1689,1283){\makebox(0,0){\strut{}}}%
      \colorrgb{0.50,0.50,0.50}%
      \put(1766,1434){\makebox(0,0){\strut{}}}%
      \colorrgb{0.50,0.50,0.50}%
      \put(1783,1462){\makebox(0,0){\strut{}}}%
      \colorrgb{0.50,0.50,0.50}%
      \put(1794,1488){\makebox(0,0){\strut{}}}%
      \colorrgb{0.50,0.50,0.50}%
      \put(1916,1641){\makebox(0,0){\strut{}}}%
      \colorrgb{0.50,0.50,0.50}%
      \put(1968,1720){\makebox(0,0){\strut{}}}%
      \colorrgb{0.50,0.50,0.50}%
      \put(2024,1819){\makebox(0,0){\strut{}}}%
      \colorrgb{0.50,0.50,0.50}%
      \put(2069,1923){\makebox(0,0){\strut{}}}%
      \colorrgb{0.00,0.00,0.00}%
      \put(1447,3248){\makebox(0,0){\strut{}\shortstack[l]{$n = 4$\\$\gamma_4=0.05$\\$|\setC| > n$\\$|\setC| > |\setLNS|$}}}%
    }%
    \gplbacktext
    \put(0,0){\includegraphics{build-receiver-set-probability-contours}}%
    \gplfronttext
  \end{picture}%
\endgroup

%% file: predictability.tex
\section{Classification Performance Against Optimal Attackers}
\label{sec:predictability}

Both view selection policies depend directly on the unpredictability
of RSSIs, because collapsing identities requires knowing the
observations of the initiator, as explained in
\autoref{sec:falsified-power}. An intelligent attacker can attempt
educated guesses, resulting in some successful collapses. In this
section, we evaluate the two selection policies against optimal
attackers, as defined in \autoref{sec:optimal-max-sybil} and
\autoref{sec:optimal-consistency}.

\subsection{RSSI Unpredictability}
\label{sec:rssi-unpredictability}

Accurate guessing of RSSIs is difficult because the wireless channel
varies significantly with small displacements in location and
orientation (\emph{spatial variation}) and environmental changes over
time (\emph{temporal variation})~\cite{rappaport02,hashemi92mayII}.
Pre-characterization could account for spatial variation, but would be
prohibitively expensive at the needed spatial and orientation
granularity (\SI{6.25}{\centi\meter}~\cite{rappaport91may} and
\SI{3}{\degree} for our test devices).

\begin{figure}
  \centering
  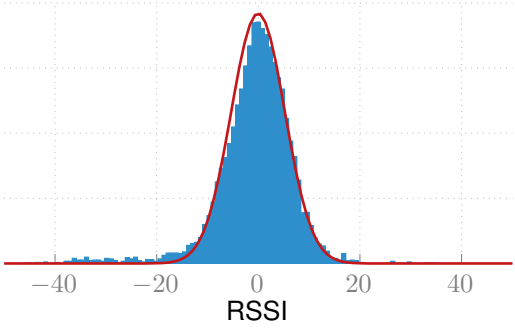
  \caption{Distribution of RSSI variations in real-world deployment.}
  \label{fig:rssi-distribution}
\end{figure}

We empirically determined the RSSI variation for human-carried
smartphones by deploying experimental phones to eleven graduate
students in two adjacent offices and measuring fixed-distance,
pairwise RSSIs for fifteen hours.  The observed distribution of
deviations\footnote{For each pair of transceivers, we subtracted the
  mean of all their measurements to get the deviations and aggregated
  all the pairwise deviations.}, shown in
\autoref{fig:rssi-distribution}, is roughly normal with a standard
deviation of \SI{7.3}{\dBm}, in line with other real-world
measurements for spatial and orientation variations
(\SIrange{4}{12}{\dBm} and \SI{5.3}{\dBm}~\cite{rappaport02}). We use
this distribution to model the attacker uncertainty of RSSIs.

\subsection{Optimal Attacker Strategy---Maximum Sybil Policy}
\label{sec:optimal-max-sybil}

\autoref{thm:max-sybil} shows that the performance of the maximum
Sybil policy is inversely related to the number of collapsed non-Sybil
identities. Therefore, the optimal attacker tries to collapse as many
as possible.  We give two observations about this goal.
\begin{enumerate}
\item More distinct guesses increase the probability of success, so an
  optimal attacker partitions its (mostly Sybil) identities, with each
  group making a different guess.
\item Smaller group size increases the number of groups, but decreases
  the probability the group is considered---recall that
  \autoref{alg:build-receiver-sets} generates only $|\setI|$ of the
  possible $2^{|\setI|}$ candidate receiver sets. 
\end{enumerate}
Consequently, there is an optimal group size that maximizes the total
number of groups (guesses) outputted by
\autoref{alg:build-receiver-sets}, which we obtained via Monte Carlo
simulations. We model the initiator's RSSI observation as a random
vector whose elements are drawn i.i.d from the Gaussian distribution in
\autoref{fig:rssi-distribution}. Given the total number of guesses,
the best choices are the vectors with the highest joint
probabilities.  The performance against this strategy is discussed in
\autoref{sec:performance-analysis}.

\subsection{Optimal Attacker Strategy---View Consistency Policy}
\label{sec:optimal-consistency}

\begin{figure}
  \centering
  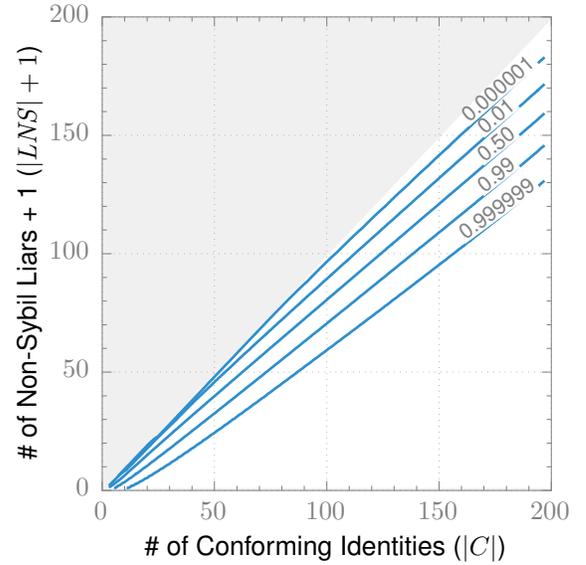
  \caption{Contours of a lower bound on the probability that
    \autoref{cond:collapse-lns} holds under an optimal attacker
    strategy with the attacker's knowledge of RSSIs modeled as a
    normal distribution with standard deviation \SI{7.3}{\dBm}.}
  \label{fig:true-view-selection-probability}
\end{figure}

The view consistency policy depends on \autoref{cond:collapse-lns}
holding, i.e., all consistent views must correctly classify at least
$|\setLNS| + 1$ conforming identities.  In this section we quantify
the probability that it holds against an optimal attacker.  To break
\autoref{cond:collapse-lns}, an attacker must generate a consistent
view that collapses at least $|\setC|-|\setLNS|$ conforming
identities.  We give three observations about the optimal attacker
strategy for this goal.
\begin{enumerate}
\item Collapsing $|\setC| - |\setLNS|$ identities is easiest with
  larger $|\setLNS|$.  Thus, the optimal attacker uses only one
  physical node to claim Sybils---the others just lie.
\item For a particular false view to be consistent, all supposedly
  non-Sybil identities must indict the same identities, e.g., have the
  same RSSI guesses for the collapsed conforming identities. The
  optimal attacker must divide its (mostly Sybil) identities into
  groups, each using a different set of guesses.
\item More groups increases the probability of success, but decreases
  the number of Sybils actually accepted, as each group is smaller.
\end{enumerate}
We assume the optimal attacker wishes to maximize the probability of
success and thus uses minimum-sized groups (three identities, for
size-4 signalprints).

For each group, the attacker must guess RSSI values for the conforming
identities with the goal of collapsing at least $s \triangleq
|\setC|-|\setLNS|$ of them.  There are $\binom{|\setC|}{s}$ such sets
and the optimal attacker guesses values that maximize the probability
of at least one (across all groups) being correct.  The first group is
easy; the $|\setC|$ guesses are simply the most likely values, i.e.,
the expected values for the conforming identities' RSSIs, under the
uncertainty distribution.

For the next (and subsequent) groups, the optimal attacker should pick
the next most likely RSSI values for each of the $\binom{|\setC|}{s}$
sets. However, the sets share elements (only $|\setC|$ RSSIs are
actually guessed), so the attacker must determine the most probable
sets that are compatible.  This is a non-trivial problem, so for our
analysis, we assume that all sets are compatible (e.g., one group can
guess \SI{-78}{\dBm} and \SI{-49}{\dBm} RSSIs for nodes $a$ and $b$,
but \SI{-82}{\dBm} and \SI{-54}{\dBm} RSSIs for nodes $a$ and $c$).
This is clearly impossible, but leads to a conservative lower bound on
the probability that the attacker fails---a feasible, optimal strategy
is less likely to succeed.

\autoref{fig:true-view-selection-probability} shows contours of this
lower bound on the probability that \autoref{cond:collapse-lns} holds
as a function of $|\setC|$ and $|\setLNS|$.  When the conforming nodes
outnumber the attacker nodes by at least 1.5$\times$ ---the expected
case in real networks---the condition holds with very high
probability. In practice, it will hold with even higher probability,
as this is a lower bound.

\subsection{Performance Comparison of Both Policies}
\label{sec:performance-analysis}

We compare the performance of the two policies against the optimal
attackers using the \emph{final Sybil ratio}, the fraction of accepted
identities that are Sybil. Optimal attacker group sizes are chosen to
maximize this metric.  We model the attacker's knowledge of initiator
RSSIs as a normal distribution with standard deviation \SI{7.3}{\dBm},
which conservatively assumes fine-grained temporal and spatial
characterization (see \autoref{sec:rssi-unpredictability}). We expect
real-world attackers to have less knowledge, leading to even better
classification performance.

Our procedure for generating candidate receiver sets
(\autoref{alg:build-receiver-sets}) works best when conforming nodes
outnumber physical attackers. This condition should normally hold in
real-world networks (it is the major motivation for a Sybil attack),
so for both policies, we report results assuming that it does.

\begin{figure}
  \centering
  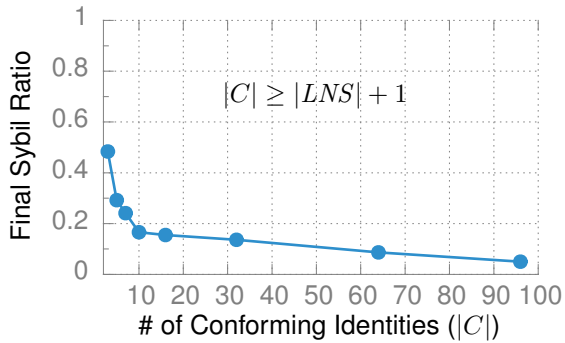
  \caption{The final Sybil ratio, i.e., fraction of accepted
    identities that are Sybil, produced by the maximum Sybil policy
    against an optimal attacker strategy.}
  \label{fig:sybil-ratio-maximum-sybil}
\end{figure}

\autoref{fig:sybil-ratio-maximum-sybil} graphs the final Sybil ratio
of the maximum Sybil policy, which roughly corresponds to the collapse
ratio $\frac{|\view{\setS} \cap \setNS|}{|\setC|}$, a direct measure
of the attacker's collapsing power. The performance is irrelevant to
the number of physical attackers. The Sybil ratio decreases to
0.05-0.2 when $|\setC|>10$. When $|C| < 10$, the Sybil ratio is high
(0.2--0.5), despite elimination of most Sybil identities
(92\%--99\%). This behavior is due to the ease of guessing the
low-dimension RSSI observation vectors.

\begin{figure}
  \centering
  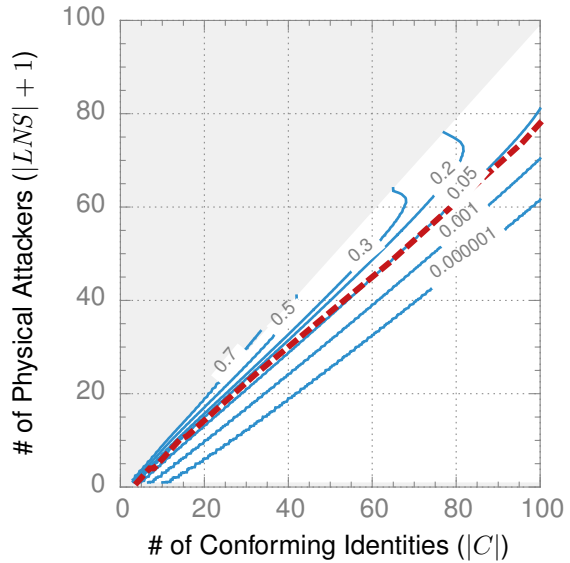
  \caption{Contours showing the final Sybil ratio for the view
    consistency policy against an optimal attacker strategy. The
    dashed line corresponds to situations where this policy performs
    equally to the maximum Sybil policy for same numbers of conforming
    nodes.}
  \label{fig:sybil-ratio-consistency-contours}
\end{figure}

\autoref{fig:sybil-ratio-consistency-contours} shows the performance
of the consistency policy. Performance increases rapidly with the
ratio of conforming nodes to physical attackers---recall the attacker
needs to collapse $|\setC| - |\setLNS|$ to break
\autoref{cond:collapse-lns}. For example, the final Sybil ratio drops
below $10^{-6}$ when $\frac{|\setC|}{|\setLNS| + 1} \geq 1.6$.  As the
collapse rate is usually below $0.2$ (see
\autoref{fig:sybil-ratio-maximum-sybil} when $|\setC|>10$), we observe
good performance when $|\setC| - |\setLNS| \geq 0.2|C|$ (below the
0.05 contour). The dashed line (roughly
$\frac{|\setC|}{|\setLNS|+1}=$1.2) indicates the points at which both
policies perform equally. Below it, the consistency policy performs
better than the maximum Sybil policy and above it does worse.

The view consistency policy is superior when conforming nodes are
expected to outnumber attacker nodes by at least $1.2\times$, the
common case in urban environments. The max Sybil policy remains viable
when the number of physical attackers is comparable to (or even larger
than) that of the conforming nodes. We suggest users of the Mason test
consider their deployment- and application-specific knowledge when
choosing a policy.

%% file: fig/rssi_distribution.pdf_tex
\begingroup
  \makeatletter
  \providecommand\color[2][]{%
    \GenericError{(gnuplot) \space\space\space\@spaces}{%
      Package color not loaded in conjunction with
      terminal option `colourtext'%
    }{See the gnuplot documentation for explanation.%
    }{Either use 'blacktext' in gnuplot or load the package
      color.sty in LaTeX.}%
    \renewcommand\color[2][]{}%
  }%
  \providecommand\includegraphics[2][]{%
    \GenericError{(gnuplot) \space\space\space\@spaces}{%
      Package graphicx or graphics not loaded%
    }{See the gnuplot documentation for explanation.%
    }{The gnuplot epslatex terminal needs graphicx.sty or graphics.sty.}%
    \renewcommand\includegraphics[2][]{}%
  }%
  \providecommand\rotatebox[2]{#2}%
  \@ifundefined{ifGPcolor}{%
    \newif\ifGPcolor
    \GPcolortrue
  }{}%
  \@ifundefined{ifGPblacktext}{%
    \newif\ifGPblacktext
    \GPblacktextfalse
  }{}%
  \let\gplgaddtomacro\g@addto@macro
  \gdef\gplbacktext{}%
  \gdef\gplfronttext{}%
  \makeatother
  \ifGPblacktext
    \def\colorrgb#1{}%
    \def\colorgray#1{}%
  \else
    \ifGPcolor
      \def\colorrgb#1{\color[rgb]{#1}}%
      \def\colorgray#1{\color[gray]{#1}}%
      \expandafter\def\csname LTw\endcsname{\color{white}}%
      \expandafter\def\csname LTb\endcsname{\color{black}}%
      \expandafter\def\csname LTa\endcsname{\color{black}}%
      \expandafter\def\csname LT0\endcsname{\color[rgb]{1,0,0}}%
      \expandafter\def\csname LT1\endcsname{\color[rgb]{0,1,0}}%
      \expandafter\def\csname LT2\endcsname{\color[rgb]{0,0,1}}%
      \expandafter\def\csname LT3\endcsname{\color[rgb]{1,0,1}}%
      \expandafter\def\csname LT4\endcsname{\color[rgb]{0,1,1}}%
      \expandafter\def\csname LT5\endcsname{\color[rgb]{1,1,0}}%
      \expandafter\def\csname LT6\endcsname{\color[rgb]{0,0,0}}%
      \expandafter\def\csname LT7\endcsname{\color[rgb]{1,0.3,0}}%
      \expandafter\def\csname LT8\endcsname{\color[rgb]{0.5,0.5,0.5}}%
    \else
      \def\colorrgb#1{\color{black}}%
      \def\colorgray#1{\color[gray]{#1}}%
      \expandafter\def\csname LTw\endcsname{\color{white}}%
      \expandafter\def\csname LTb\endcsname{\color{black}}%
      \expandafter\def\csname LTa\endcsname{\color{black}}%
      \expandafter\def\csname LT0\endcsname{\color{black}}%
      \expandafter\def\csname LT1\endcsname{\color{black}}%
      \expandafter\def\csname LT2\endcsname{\color{black}}%
      \expandafter\def\csname LT3\endcsname{\color{black}}%
      \expandafter\def\csname LT4\endcsname{\color{black}}%
      \expandafter\def\csname LT5\endcsname{\color{black}}%
      \expandafter\def\csname LT6\endcsname{\color{black}}%
      \expandafter\def\csname LT7\endcsname{\color{black}}%
      \expandafter\def\csname LT8\endcsname{\color{black}}%
    \fi
  \fi
    \setlength{\unitlength}{0.0500bp}%
    \ifx\gptboxheight\undefined%
      \newlength{\gptboxheight}%
      \newlength{\gptboxwidth}%
      \newsavebox{\gptboxtext}%
    \fi%
    \setlength{\fboxrule}{0.5pt}%
    \setlength{\fboxsep}{1pt}%
\begin{picture}(4320.00,2660.00)%
    \gplgaddtomacro\gplbacktext{%
      \colorrgb{0.50,0.50,0.50}%
      \put(133,516){\makebox(0,0)[r]{\strut{}}}%
      \colorrgb{0.50,0.50,0.50}%
      \put(133,1007){\makebox(0,0)[r]{\strut{}}}%
      \colorrgb{0.50,0.50,0.50}%
      \put(133,1499){\makebox(0,0)[r]{\strut{}}}%
      \colorrgb{0.50,0.50,0.50}%
      \put(133,1990){\makebox(0,0)[r]{\strut{}}}%
      \colorrgb{0.50,0.50,0.50}%
      \put(133,2481){\makebox(0,0)[r]{\strut{}}}%
      \colorrgb{0.50,0.50,0.50}%
      \put(605,338){\makebox(0,0){\strut{}$-40$}}%
      \colorrgb{0.50,0.50,0.50}%
      \put(1371,338){\makebox(0,0){\strut{}$-20$}}%
      \colorrgb{0.50,0.50,0.50}%
      \put(2137,338){\makebox(0,0){\strut{}$0$}}%
      \colorrgb{0.50,0.50,0.50}%
      \put(2903,338){\makebox(0,0){\strut{}$20$}}%
      \colorrgb{0.50,0.50,0.50}%
      \put(3669,338){\makebox(0,0){\strut{}$40$}}%
    }%
    \gplgaddtomacro\gplfronttext{%
      \csname LTb\endcsname%
      \put(2137,124){\makebox(0,0){\strut{}RSSI}}%
    }%
    \gplbacktext
    \put(0,0){\includegraphics{rssi_distribution}}%
    \gplfronttext
  \end{picture}%
\endgroup

%% file: fig/sybil-ratio-maximum-sybil.pdf_tex
\begingroup
  \makeatletter
  \providecommand\color[2][]{%
    \GenericError{(gnuplot) \space\space\space\@spaces}{%
      Package color not loaded in conjunction with
      terminal option `colourtext'%
    }{See the gnuplot documentation for explanation.%
    }{Either use 'blacktext' in gnuplot or load the package
      color.sty in LaTeX.}%
    \renewcommand\color[2][]{}%
  }%
  \providecommand\includegraphics[2][]{%
    \GenericError{(gnuplot) \space\space\space\@spaces}{%
      Package graphicx or graphics not loaded%
    }{See the gnuplot documentation for explanation.%
    }{The gnuplot epslatex terminal needs graphicx.sty or graphics.sty.}%
    \renewcommand\includegraphics[2][]{}%
  }%
  \providecommand\rotatebox[2]{#2}%
  \@ifundefined{ifGPcolor}{%
    \newif\ifGPcolor
    \GPcolortrue
  }{}%
  \@ifundefined{ifGPblacktext}{%
    \newif\ifGPblacktext
    \GPblacktextfalse
  }{}%
  \let\gplgaddtomacro\g@addto@macro
  \gdef\gplbacktext{}%
  \gdef\gplfronttext{}%
  \makeatother
  \ifGPblacktext
    \def\colorrgb#1{}%
    \def\colorgray#1{}%
  \else
    \ifGPcolor
      \def\colorrgb#1{\color[rgb]{#1}}%
      \def\colorgray#1{\color[gray]{#1}}%
      \expandafter\def\csname LTw\endcsname{\color{white}}%
      \expandafter\def\csname LTb\endcsname{\color{black}}%
      \expandafter\def\csname LTa\endcsname{\color{black}}%
      \expandafter\def\csname LT0\endcsname{\color[rgb]{1,0,0}}%
      \expandafter\def\csname LT1\endcsname{\color[rgb]{0,1,0}}%
      \expandafter\def\csname LT2\endcsname{\color[rgb]{0,0,1}}%
      \expandafter\def\csname LT3\endcsname{\color[rgb]{1,0,1}}%
      \expandafter\def\csname LT4\endcsname{\color[rgb]{0,1,1}}%
      \expandafter\def\csname LT5\endcsname{\color[rgb]{1,1,0}}%
      \expandafter\def\csname LT6\endcsname{\color[rgb]{0,0,0}}%
      \expandafter\def\csname LT7\endcsname{\color[rgb]{1,0.3,0}}%
      \expandafter\def\csname LT8\endcsname{\color[rgb]{0.5,0.5,0.5}}%
    \else
      \def\colorrgb#1{\color{black}}%
      \def\colorgray#1{\color[gray]{#1}}%
      \expandafter\def\csname LTw\endcsname{\color{white}}%
      \expandafter\def\csname LTb\endcsname{\color{black}}%
      \expandafter\def\csname LTa\endcsname{\color{black}}%
      \expandafter\def\csname LT0\endcsname{\color{black}}%
      \expandafter\def\csname LT1\endcsname{\color{black}}%
      \expandafter\def\csname LT2\endcsname{\color{black}}%
      \expandafter\def\csname LT3\endcsname{\color{black}}%
      \expandafter\def\csname LT4\endcsname{\color{black}}%
      \expandafter\def\csname LT5\endcsname{\color{black}}%
      \expandafter\def\csname LT6\endcsname{\color{black}}%
      \expandafter\def\csname LT7\endcsname{\color{black}}%
      \expandafter\def\csname LT8\endcsname{\color{black}}%
    \fi
  \fi
  \setlength{\unitlength}{0.0500bp}%
  \begin{picture}(4320.00,2660.00)%
    \gplgaddtomacro\gplbacktext{%
      \colorrgb{0.50,0.50,0.50}%
      \put(634,539){\makebox(0,0)[r]{\strut{} 0}}%
      \colorrgb{0.50,0.50,0.50}%
      \put(634,922){\makebox(0,0)[r]{\strut{} 0.2}}%
      \colorrgb{0.50,0.50,0.50}%
      \put(634,1305){\makebox(0,0)[r]{\strut{} 0.4}}%
      \colorrgb{0.50,0.50,0.50}%
      \put(634,1689){\makebox(0,0)[r]{\strut{} 0.6}}%
      \colorrgb{0.50,0.50,0.50}%
      \put(634,2072){\makebox(0,0)[r]{\strut{} 0.8}}%
      \colorrgb{0.50,0.50,0.50}%
      \put(634,2455){\makebox(0,0)[r]{\strut{} 1}}%
      \colorrgb{0.50,0.50,0.50}%
      \put(1004,353){\makebox(0,0){\strut{} 10}}%
      \colorrgb{0.50,0.50,0.50}%
      \put(1338,353){\makebox(0,0){\strut{} 20}}%
      \colorrgb{0.50,0.50,0.50}%
      \put(1672,353){\makebox(0,0){\strut{} 30}}%
      \colorrgb{0.50,0.50,0.50}%
      \put(2007,353){\makebox(0,0){\strut{} 40}}%
      \colorrgb{0.50,0.50,0.50}%
      \put(2341,353){\makebox(0,0){\strut{} 50}}%
      \colorrgb{0.50,0.50,0.50}%
      \put(2675,353){\makebox(0,0){\strut{} 60}}%
      \colorrgb{0.50,0.50,0.50}%
      \put(3010,353){\makebox(0,0){\strut{} 70}}%
      \colorrgb{0.50,0.50,0.50}%
      \put(3344,353){\makebox(0,0){\strut{} 80}}%
      \colorrgb{0.50,0.50,0.50}%
      \put(3679,353){\makebox(0,0){\strut{} 90}}%
      \colorrgb{0.50,0.50,0.50}%
      \put(4013,353){\makebox(0,0){\strut{} 100}}%
      \csname LTb\endcsname%
      \put(144,1497){\rotatebox{-270}{\makebox(0,0){\strut{}Final Sybil Ratio}}}%
      \csname LTb\endcsname%
      \put(2374,130){\makebox(0,0){\strut{}\# of Conforming Identities ($|\setC|$)}}%
    }%
    \gplgaddtomacro\gplfronttext{%
      \colorrgb{0.00,0.00,0.00}%
      \put(2341,1880){\makebox(0,0){\strut{}\shortstack[l]{$|\setC| \ge |\setLNS|+1$}}}%
    }%
    \gplbacktext
    \put(0,0){\includegraphics{sybil-ratio-maximum-sybil}}%
    \gplfronttext
  \end{picture}%
\endgroup

%% file: motion.tex
\section{Detecting Mobile Attackers}
\label{sec:motion}

\begin{figure}
  \centering
  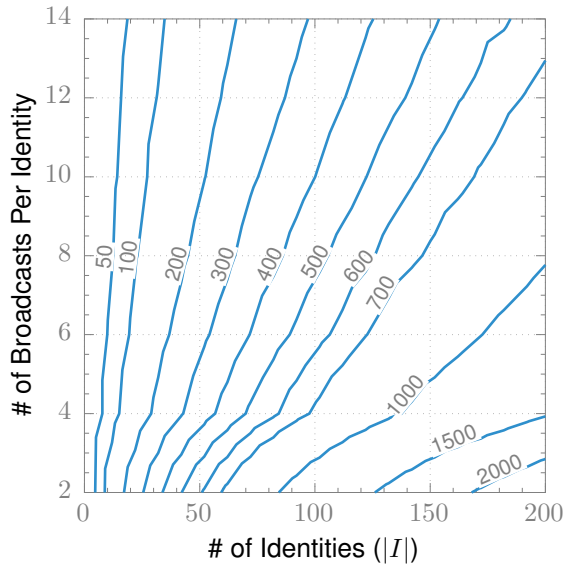
  \caption{Contours showing the response time (in ms, 99$^\textrm{th}$
    percentile) to precisely switch between two positions required to
    defeat the challenge-response moving node detection.}
  \label{fig:challenge-response-time-contours}
\end{figure}

A mobile attacker can defeat signalprint comparison by changing
locations or orientations between transmissions to associate distinct
signalprints with each Sybil identity.  Fortunately, in the networks
we consider, most conforming nodes (e.g., human-carried smartphones
and laptops) are stationary over most short time-spans
(\SIrange{1}{2}{\minute}), due to human mobility habits.  Thus,
multiple transmissions should have the same signalprint, as noted, but
not pursued, by Xiao \etal~\cite{xiao09sep}. From this observation, we
develop a protocol to detect moving attackers.

Again, the lack of trusted observations is troublesome. Instead of
comparing signalprints, we compare the initiator's observations: all
transmissions from a conforming node should have the same RSSI. As
shown in \autoref{sec:evaluation}, this simple criterion yields
acceptable detection.

The protocol collection phase (\autoref{fig:formulation-hello}) is
extended to request multiple probe packets (e.g., 14) from each
identity in a pseudo-random order (see
\autoref{sec:idcollection}). During the classification phase
(\autoref{fig:formulation-comparison}) each participant rejects all
identities with large RSSI variation across its transmissions
(specifically, a standard deviation larger than \SI{2.5}{\dBm}). In
essence, an attacker is challenged to quickly and precisely switch
between the multiple positions associated with its Sybil identities
(\SI{6.25}{\centi\meter} location precision according to coherence
length theory~\cite{rappaport91may} and \SI{3}{\degree} orientation
precision according to our measurements).

\autoref{fig:challenge-response-time-contours} plots the required
response time for an attacker to pass the challenge.  Given human
reaction times~\cite{humanbenchmark}, reliably mounting such an attack
would require specialized hardware---precise electromechanical control
or beam steering antenna arrays---that is outside our attack model and
substantially more expensive to deploy than compromised commodity
devices.

%% file: fig/challenge-response-time-contours.pdf_tex
\begingroup
  \makeatletter
  \providecommand\color[2][]{%
    \GenericError{(gnuplot) \space\space\space\@spaces}{%
      Package color not loaded in conjunction with
      terminal option `colourtext'%
    }{See the gnuplot documentation for explanation.%
    }{Either use 'blacktext' in gnuplot or load the package
      color.sty in LaTeX.}%
    \renewcommand\color[2][]{}%
  }%
  \providecommand\includegraphics[2][]{%
    \GenericError{(gnuplot) \space\space\space\@spaces}{%
      Package graphicx or graphics not loaded%
    }{See the gnuplot documentation for explanation.%
    }{The gnuplot epslatex terminal needs graphicx.sty or graphics.sty.}%
    \renewcommand\includegraphics[2][]{}%
  }%
  \providecommand\rotatebox[2]{#2}%
  \@ifundefined{ifGPcolor}{%
    \newif\ifGPcolor
    \GPcolortrue
  }{}%
  \@ifundefined{ifGPblacktext}{%
    \newif\ifGPblacktext
    \GPblacktextfalse
  }{}%
  \let\gplgaddtomacro\g@addto@macro
  \gdef\gplbacktext{}%
  \gdef\gplfronttext{}%
  \makeatother
  \ifGPblacktext
    \def\colorrgb#1{}%
    \def\colorgray#1{}%
  \else
    \ifGPcolor
      \def\colorrgb#1{\color[rgb]{#1}}%
      \def\colorgray#1{\color[gray]{#1}}%
      \expandafter\def\csname LTw\endcsname{\color{white}}%
      \expandafter\def\csname LTb\endcsname{\color{black}}%
      \expandafter\def\csname LTa\endcsname{\color{black}}%
      \expandafter\def\csname LT0\endcsname{\color[rgb]{1,0,0}}%
      \expandafter\def\csname LT1\endcsname{\color[rgb]{0,1,0}}%
      \expandafter\def\csname LT2\endcsname{\color[rgb]{0,0,1}}%
      \expandafter\def\csname LT3\endcsname{\color[rgb]{1,0,1}}%
      \expandafter\def\csname LT4\endcsname{\color[rgb]{0,1,1}}%
      \expandafter\def\csname LT5\endcsname{\color[rgb]{1,1,0}}%
      \expandafter\def\csname LT6\endcsname{\color[rgb]{0,0,0}}%
      \expandafter\def\csname LT7\endcsname{\color[rgb]{1,0.3,0}}%
      \expandafter\def\csname LT8\endcsname{\color[rgb]{0.5,0.5,0.5}}%
    \else
      \def\colorrgb#1{\color{black}}%
      \def\colorgray#1{\color[gray]{#1}}%
      \expandafter\def\csname LTw\endcsname{\color{white}}%
      \expandafter\def\csname LTb\endcsname{\color{black}}%
      \expandafter\def\csname LTa\endcsname{\color{black}}%
      \expandafter\def\csname LT0\endcsname{\color{black}}%
      \expandafter\def\csname LT1\endcsname{\color{black}}%
      \expandafter\def\csname LT2\endcsname{\color{black}}%
      \expandafter\def\csname LT3\endcsname{\color{black}}%
      \expandafter\def\csname LT4\endcsname{\color{black}}%
      \expandafter\def\csname LT5\endcsname{\color{black}}%
      \expandafter\def\csname LT6\endcsname{\color{black}}%
      \expandafter\def\csname LT7\endcsname{\color{black}}%
      \expandafter\def\csname LT8\endcsname{\color{black}}%
    \fi
  \fi
    \setlength{\unitlength}{0.0500bp}%
    \ifx\gptboxheight\undefined%
      \newlength{\gptboxheight}%
      \newlength{\gptboxwidth}%
      \newsavebox{\gptboxtext}%
    \fi%
    \setlength{\fboxrule}{0.5pt}%
    \setlength{\fboxsep}{1pt}%
\begin{picture}(4320.00,4320.00)%
    \gplgaddtomacro\gplbacktext{%
      \colorrgb{0.50,0.50,0.50}%
      \put(489,569){\makebox(0,0)[r]{\strut{}$2$}}%
      \colorrgb{0.50,0.50,0.50}%
      \put(489,1164){\makebox(0,0)[r]{\strut{}$4$}}%
      \colorrgb{0.50,0.50,0.50}%
      \put(489,1760){\makebox(0,0)[r]{\strut{}$6$}}%
      \colorrgb{0.50,0.50,0.50}%
      \put(489,2355){\makebox(0,0)[r]{\strut{}$8$}}%
      \colorrgb{0.50,0.50,0.50}%
      \put(489,2950){\makebox(0,0)[r]{\strut{}$10$}}%
      \colorrgb{0.50,0.50,0.50}%
      \put(489,3546){\makebox(0,0)[r]{\strut{}$12$}}%
      \colorrgb{0.50,0.50,0.50}%
      \put(489,4141){\makebox(0,0)[r]{\strut{}$14$}}%
      \colorrgb{0.50,0.50,0.50}%
      \put(578,391){\makebox(0,0){\strut{}$0$}}%
      \colorrgb{0.50,0.50,0.50}%
      \put(1447,391){\makebox(0,0){\strut{}$50$}}%
      \colorrgb{0.50,0.50,0.50}%
      \put(2315,391){\makebox(0,0){\strut{}$100$}}%
      \colorrgb{0.50,0.50,0.50}%
      \put(3184,391){\makebox(0,0){\strut{}$150$}}%
      \colorrgb{0.50,0.50,0.50}%
      \put(4052,391){\makebox(0,0){\strut{}$200$}}%
    }%
    \gplgaddtomacro\gplfronttext{%
      \csname LTb\endcsname%
      \put(133,2355){\rotatebox{-270}{\makebox(0,0){\strut{}\# of Broadcasts Per Identity}}}%
      \csname LTb\endcsname%
      \put(2315,124){\makebox(0,0){\strut{}\# of Identities ($|\setI|$)}}%
      \colorrgb{0.50,0.50,0.50}%
      \put(2030,2260){\makebox(0,0){\rotatebox{64.983107}{\colorbox{white}{\smash{\fns{400}}}}}}%
      \colorrgb{0.50,0.50,0.50}%
      \put(1166,569){\makebox(0,0){\strut{}}}%
      \colorrgb{0.50,0.50,0.50}%
      \put(1246,780){\makebox(0,0){\strut{}}}%
      \colorrgb{0.50,0.50,0.50}%
      \put(1308,855){\makebox(0,0){\strut{}}}%
      \colorrgb{0.50,0.50,0.50}%
      \put(1365,966){\makebox(0,0){\strut{}}}%
      \colorrgb{0.50,0.50,0.50}%
      \put(1481,1082){\makebox(0,0){\strut{}}}%
      \colorrgb{0.50,0.50,0.50}%
      \put(1497,1109){\makebox(0,0){\strut{}}}%
      \colorrgb{0.50,0.50,0.50}%
      \put(1563,1164){\makebox(0,0){\strut{}}}%
      \colorrgb{0.50,0.50,0.50}%
      \put(1611,1315){\makebox(0,0){\strut{}}}%
      \colorrgb{0.50,0.50,0.50}%
      \put(1655,1400){\makebox(0,0){\strut{}}}%
      \colorrgb{0.50,0.50,0.50}%
      \put(1713,1561){\makebox(0,0){\strut{}}}%
      \colorrgb{0.50,0.50,0.50}%
      \put(1823,1760){\makebox(0,0){\strut{}}}%
      \colorrgb{0.50,0.50,0.50}%
      \put(1826,1770){\makebox(0,0){\strut{}}}%
      \colorrgb{0.50,0.50,0.50}%
      \put(1829,1778){\makebox(0,0){\strut{}}}%
      \colorrgb{0.50,0.50,0.50}%
      \put(1917,2053){\makebox(0,0){\strut{}}}%
      \colorrgb{0.50,0.50,0.50}%
      \put(2002,2199){\makebox(0,0){\strut{}}}%
      \colorrgb{0.50,0.50,0.50}%
      \put(2030,2260){\makebox(0,0){\strut{}}}%
      \colorrgb{0.50,0.50,0.50}%
      \put(2073,2355){\makebox(0,0){\strut{}}}%
      \colorrgb{0.50,0.50,0.50}%
      \put(2125,2531){\makebox(0,0){\strut{}}}%
      \colorrgb{0.50,0.50,0.50}%
      \put(2176,2644){\makebox(0,0){\strut{}}}%
      \colorrgb{0.50,0.50,0.50}%
      \put(2228,2773){\makebox(0,0){\strut{}}}%
      \colorrgb{0.50,0.50,0.50}%
      \put(2317,2950){\makebox(0,0){\strut{}}}%
      \colorrgb{0.50,0.50,0.50}%
      \put(2334,3003){\makebox(0,0){\strut{}}}%
      \colorrgb{0.50,0.50,0.50}%
      \put(2350,3050){\makebox(0,0){\strut{}}}%
      \colorrgb{0.50,0.50,0.50}%
      \put(2427,3281){\makebox(0,0){\strut{}}}%
      \colorrgb{0.50,0.50,0.50}%
      \put(2523,3502){\makebox(0,0){\strut{}}}%
      \colorrgb{0.50,0.50,0.50}%
      \put(2530,3522){\makebox(0,0){\strut{}}}%
      \colorrgb{0.50,0.50,0.50}%
      \put(2543,3546){\makebox(0,0){\strut{}}}%
      \colorrgb{0.50,0.50,0.50}%
      \put(2626,3790){\makebox(0,0){\strut{}}}%
      \colorrgb{0.50,0.50,0.50}%
      \put(2697,3999){\makebox(0,0){\strut{}}}%
      \colorrgb{0.50,0.50,0.50}%
      \put(2718,4070){\makebox(0,0){\strut{}}}%
      \colorrgb{0.50,0.50,0.50}%
      \put(2755,4141){\makebox(0,0){\strut{}}}%
      \colorrgb{0.50,0.50,0.50}%
      \put(969,2324){\makebox(0,0){\rotatebox{82.673593}{\colorbox{white}{\smash{\fns{100}}}}}}%
      \colorrgb{0.50,0.50,0.50}%
      \put(731,569){\makebox(0,0){\strut{}}}%
      \colorrgb{0.50,0.50,0.50}%
      \put(734,749){\makebox(0,0){\strut{}}}%
      \colorrgb{0.50,0.50,0.50}%
      \put(786,946){\makebox(0,0){\strut{}}}%
      \colorrgb{0.50,0.50,0.50}%
      \put(808,1092){\makebox(0,0){\strut{}}}%
      \colorrgb{0.50,0.50,0.50}%
      \put(840,1164){\makebox(0,0){\strut{}}}%
      \colorrgb{0.50,0.50,0.50}%
      \put(864,1495){\makebox(0,0){\strut{}}}%
      \colorrgb{0.50,0.50,0.50}%
      \put(915,1760){\makebox(0,0){\strut{}}}%
      \colorrgb{0.50,0.50,0.50}%
      \put(923,1886){\makebox(0,0){\strut{}}}%
      \colorrgb{0.50,0.50,0.50}%
      \put(960,2223){\makebox(0,0){\strut{}}}%
      \colorrgb{0.50,0.50,0.50}%
      \put(969,2324){\makebox(0,0){\strut{}}}%
      \colorrgb{0.50,0.50,0.50}%
      \put(977,2355){\makebox(0,0){\strut{}}}%
      \colorrgb{0.50,0.50,0.50}%
      \put(1024,2732){\makebox(0,0){\strut{}}}%
      \colorrgb{0.50,0.50,0.50}%
      \put(1051,2950){\makebox(0,0){\strut{}}}%
      \colorrgb{0.50,0.50,0.50}%
      \put(1061,3198){\makebox(0,0){\strut{}}}%
      \colorrgb{0.50,0.50,0.50}%
      \put(1122,3546){\makebox(0,0){\strut{}}}%
      \colorrgb{0.50,0.50,0.50}%
      \put(1126,3574){\makebox(0,0){\strut{}}}%
      \colorrgb{0.50,0.50,0.50}%
      \put(1134,3645){\makebox(0,0){\strut{}}}%
      \colorrgb{0.50,0.50,0.50}%
      \put(1170,4017){\makebox(0,0){\strut{}}}%
      \colorrgb{0.50,0.50,0.50}%
      \put(1182,4141){\makebox(0,0){\strut{}}}%
      \colorrgb{0.50,0.50,0.50}%
      \put(3045,1255){\makebox(0,0){\rotatebox{50.381726}{\colorbox{white}{\smash{\fns{1000}}}}}}%
      \colorrgb{0.50,0.50,0.50}%
      \put(2043,569){\makebox(0,0){\strut{}}}%
      \colorrgb{0.50,0.50,0.50}%
      \put(2145,676){\makebox(0,0){\strut{}}}%
      \colorrgb{0.50,0.50,0.50}%
      \put(2176,696){\makebox(0,0){\strut{}}}%
      \colorrgb{0.50,0.50,0.50}%
      \put(2283,798){\makebox(0,0){\strut{}}}%
      \colorrgb{0.50,0.50,0.50}%
      \put(2350,832){\makebox(0,0){\strut{}}}%
      \colorrgb{0.50,0.50,0.50}%
      \put(2427,898){\makebox(0,0){\strut{}}}%
      \colorrgb{0.50,0.50,0.50}%
      \put(2523,948){\makebox(0,0){\strut{}}}%
      \colorrgb{0.50,0.50,0.50}%
      \put(2575,988){\makebox(0,0){\strut{}}}%
      \colorrgb{0.50,0.50,0.50}%
      \put(2697,1041){\makebox(0,0){\strut{}}}%
      \colorrgb{0.50,0.50,0.50}%
      \put(2727,1063){\makebox(0,0){\strut{}}}%
      \colorrgb{0.50,0.50,0.50}%
      \put(2871,1125){\makebox(0,0){\strut{}}}%
      \colorrgb{0.50,0.50,0.50}%
      \put(2880,1132){\makebox(0,0){\strut{}}}%
      \colorrgb{0.50,0.50,0.50}%
      \put(2967,1164){\makebox(0,0){\strut{}}}%
      \colorrgb{0.50,0.50,0.50}%
      \put(3024,1233){\makebox(0,0){\strut{}}}%
      \colorrgb{0.50,0.50,0.50}%
      \put(3045,1255){\makebox(0,0){\strut{}}}%
      \colorrgb{0.50,0.50,0.50}%
      \put(3152,1392){\makebox(0,0){\strut{}}}%
      \colorrgb{0.50,0.50,0.50}%
      \put(3218,1437){\makebox(0,0){\strut{}}}%
      \colorrgb{0.50,0.50,0.50}%
      \put(3291,1512){\makebox(0,0){\strut{}}}%
      \colorrgb{0.50,0.50,0.50}%
      \put(3392,1594){\makebox(0,0){\strut{}}}%
      \colorrgb{0.50,0.50,0.50}%
      \put(3430,1629){\makebox(0,0){\strut{}}}%
      \colorrgb{0.50,0.50,0.50}%
      \put(3566,1750){\makebox(0,0){\strut{}}}%
      \colorrgb{0.50,0.50,0.50}%
      \put(3568,1752){\makebox(0,0){\strut{}}}%
      \colorrgb{0.50,0.50,0.50}%
      \put(3577,1760){\makebox(0,0){\strut{}}}%
      \colorrgb{0.50,0.50,0.50}%
      \put(3697,1906){\makebox(0,0){\strut{}}}%
      \colorrgb{0.50,0.50,0.50}%
      \put(3739,1945){\makebox(0,0){\strut{}}}%
      \colorrgb{0.50,0.50,0.50}%
      \put(3831,2040){\makebox(0,0){\strut{}}}%
      \colorrgb{0.50,0.50,0.50}%
      \put(3913,2124){\makebox(0,0){\strut{}}}%
      \colorrgb{0.50,0.50,0.50}%
      \put(3962,2186){\makebox(0,0){\strut{}}}%
      \colorrgb{0.50,0.50,0.50}%
      \put(3392,912){\makebox(0,0){\rotatebox{21.647332}{\colorbox{white}{\smash{\fns{1500}}}}}}%
      \colorrgb{0.50,0.50,0.50}%
      \put(2767,569){\makebox(0,0){\strut{}}}%
      \colorrgb{0.50,0.50,0.50}%
      \put(2853,631){\makebox(0,0){\strut{}}}%
      \colorrgb{0.50,0.50,0.50}%
      \put(2871,639){\makebox(0,0){\strut{}}}%
      \colorrgb{0.50,0.50,0.50}%
      \put(2999,725){\makebox(0,0){\strut{}}}%
      \colorrgb{0.50,0.50,0.50}%
      \put(3045,742){\makebox(0,0){\strut{}}}%
      \colorrgb{0.50,0.50,0.50}%
      \put(3149,807){\makebox(0,0){\strut{}}}%
      \colorrgb{0.50,0.50,0.50}%
      \put(3218,834){\makebox(0,0){\strut{}}}%
      \colorrgb{0.50,0.50,0.50}%
      \put(3301,882){\makebox(0,0){\strut{}}}%
      \colorrgb{0.50,0.50,0.50}%
      \put(3392,912){\makebox(0,0){\strut{}}}%
      \colorrgb{0.50,0.50,0.50}%
      \put(3456,945){\makebox(0,0){\strut{}}}%
      \colorrgb{0.50,0.50,0.50}%
      \put(3566,976){\makebox(0,0){\strut{}}}%
      \colorrgb{0.50,0.50,0.50}%
      \put(3613,1001){\makebox(0,0){\strut{}}}%
      \colorrgb{0.50,0.50,0.50}%
      \put(3739,1038){\makebox(0,0){\strut{}}}%
      \colorrgb{0.50,0.50,0.50}%
      \put(3772,1053){\makebox(0,0){\strut{}}}%
      \colorrgb{0.50,0.50,0.50}%
      \put(3913,1100){\makebox(0,0){\strut{}}}%
      \colorrgb{0.50,0.50,0.50}%
      \put(3930,1107){\makebox(0,0){\strut{}}}%
      \colorrgb{0.50,0.50,0.50}%
      \put(1681,2266){\makebox(0,0){\rotatebox{71.565051}{\colorbox{white}{\smash{\fns{300}}}}}}%
      \colorrgb{0.50,0.50,0.50}%
      \put(1021,569){\makebox(0,0){\strut{}}}%
      \colorrgb{0.50,0.50,0.50}%
      \put(1074,773){\makebox(0,0){\strut{}}}%
      \colorrgb{0.50,0.50,0.50}%
      \put(1134,867){\makebox(0,0){\strut{}}}%
      \colorrgb{0.50,0.50,0.50}%
      \put(1183,996){\makebox(0,0){\strut{}}}%
      \colorrgb{0.50,0.50,0.50}%
      \put(1308,1147){\makebox(0,0){\strut{}}}%
      \colorrgb{0.50,0.50,0.50}%
      \put(1311,1153){\makebox(0,0){\strut{}}}%
      \colorrgb{0.50,0.50,0.50}%
      \put(1322,1164){\makebox(0,0){\strut{}}}%
      \colorrgb{0.50,0.50,0.50}%
      \put(1398,1449){\makebox(0,0){\strut{}}}%
      \colorrgb{0.50,0.50,0.50}%
      \put(1481,1668){\makebox(0,0){\strut{}}}%
      \colorrgb{0.50,0.50,0.50}%
      \put(1495,1712){\makebox(0,0){\strut{}}}%
      \colorrgb{0.50,0.50,0.50}%
      \put(1520,1760){\makebox(0,0){\strut{}}}%
      \colorrgb{0.50,0.50,0.50}%
      \put(1587,1991){\makebox(0,0){\strut{}}}%
      \colorrgb{0.50,0.50,0.50}%
      \put(1655,2176){\makebox(0,0){\strut{}}}%
      \colorrgb{0.50,0.50,0.50}%
      \put(1681,2266){\makebox(0,0){\strut{}}}%
      \colorrgb{0.50,0.50,0.50}%
      \put(1713,2355){\makebox(0,0){\strut{}}}%
      \colorrgb{0.50,0.50,0.50}%
      \put(1764,2578){\makebox(0,0){\strut{}}}%
      \colorrgb{0.50,0.50,0.50}%
      \put(1829,2802){\makebox(0,0){\strut{}}}%
      \colorrgb{0.50,0.50,0.50}%
      \put(1852,2871){\makebox(0,0){\strut{}}}%
      \colorrgb{0.50,0.50,0.50}%
      \put(1887,2950){\makebox(0,0){\strut{}}}%
      \colorrgb{0.50,0.50,0.50}%
      \put(1949,3134){\makebox(0,0){\strut{}}}%
      \colorrgb{0.50,0.50,0.50}%
      \put(2002,3291){\makebox(0,0){\strut{}}}%
      \colorrgb{0.50,0.50,0.50}%
      \put(2040,3418){\makebox(0,0){\strut{}}}%
      \colorrgb{0.50,0.50,0.50}%
      \put(2089,3546){\makebox(0,0){\strut{}}}%
      \colorrgb{0.50,0.50,0.50}%
      \put(2129,3708){\makebox(0,0){\strut{}}}%
      \colorrgb{0.50,0.50,0.50}%
      \put(2176,3843){\makebox(0,0){\strut{}}}%
      \colorrgb{0.50,0.50,0.50}%
      \put(2216,4004){\makebox(0,0){\strut{}}}%
      \colorrgb{0.50,0.50,0.50}%
      \put(2263,4141){\makebox(0,0){\strut{}}}%
      \colorrgb{0.50,0.50,0.50}%
      \put(3739,686){\makebox(0,0){\rotatebox{23.468545}{\colorbox{white}{\smash{\fns{2000}}}}}}%
      \colorrgb{0.50,0.50,0.50}%
      \put(3496,569){\makebox(0,0){\strut{}}}%
      \colorrgb{0.50,0.50,0.50}%
      \put(3556,601){\makebox(0,0){\strut{}}}%
      \colorrgb{0.50,0.50,0.50}%
      \put(3566,603){\makebox(0,0){\strut{}}}%
      \colorrgb{0.50,0.50,0.50}%
      \put(3708,677){\makebox(0,0){\strut{}}}%
      \colorrgb{0.50,0.50,0.50}%
      \put(3739,686){\makebox(0,0){\strut{}}}%
      \colorrgb{0.50,0.50,0.50}%
      \put(3862,745){\makebox(0,0){\strut{}}}%
      \colorrgb{0.50,0.50,0.50}%
      \put(3913,762){\makebox(0,0){\strut{}}}%
      \colorrgb{0.50,0.50,0.50}%
      \put(4016,810){\makebox(0,0){\strut{}}}%
      \colorrgb{0.50,0.50,0.50}%
      \put(1331,2273){\makebox(0,0){\rotatebox{75.291696}{\colorbox{white}{\smash{\fns{200}}}}}}%
      \colorrgb{0.50,0.50,0.50}%
      \put(876,569){\makebox(0,0){\strut{}}}%
      \colorrgb{0.50,0.50,0.50}%
      \put(904,761){\makebox(0,0){\strut{}}}%
      \colorrgb{0.50,0.50,0.50}%
      \put(960,889){\makebox(0,0){\strut{}}}%
      \colorrgb{0.50,0.50,0.50}%
      \put(998,1034){\makebox(0,0){\strut{}}}%
      \colorrgb{0.50,0.50,0.50}%
      \put(1081,1164){\makebox(0,0){\strut{}}}%
      \colorrgb{0.50,0.50,0.50}%
      \put(1098,1286){\makebox(0,0){\strut{}}}%
      \colorrgb{0.50,0.50,0.50}%
      \put(1134,1437){\makebox(0,0){\strut{}}}%
      \colorrgb{0.50,0.50,0.50}%
      \put(1171,1631){\makebox(0,0){\strut{}}}%
      \colorrgb{0.50,0.50,0.50}%
      \put(1217,1760){\makebox(0,0){\strut{}}}%
      \colorrgb{0.50,0.50,0.50}%
      \put(1254,1945){\makebox(0,0){\strut{}}}%
      \colorrgb{0.50,0.50,0.50}%
      \put(1308,2157){\makebox(0,0){\strut{}}}%
      \colorrgb{0.50,0.50,0.50}%
      \put(1331,2273){\makebox(0,0){\strut{}}}%
      \colorrgb{0.50,0.50,0.50}%
      \put(1358,2355){\makebox(0,0){\strut{}}}%
      \colorrgb{0.50,0.50,0.50}%
      \put(1406,2615){\makebox(0,0){\strut{}}}%
      \colorrgb{0.50,0.50,0.50}%
      \put(1481,2917){\makebox(0,0){\strut{}}}%
      \colorrgb{0.50,0.50,0.50}%
      \put(1485,2936){\makebox(0,0){\strut{}}}%
      \colorrgb{0.50,0.50,0.50}%
      \put(1490,2950){\makebox(0,0){\strut{}}}%
      \colorrgb{0.50,0.50,0.50}%
      \put(1550,3311){\makebox(0,0){\strut{}}}%
      \colorrgb{0.50,0.50,0.50}%
      \put(1607,3546){\makebox(0,0){\strut{}}}%
      \colorrgb{0.50,0.50,0.50}%
      \put(1623,3656){\makebox(0,0){\strut{}}}%
      \colorrgb{0.50,0.50,0.50}%
      \put(1655,3794){\makebox(0,0){\strut{}}}%
      \colorrgb{0.50,0.50,0.50}%
      \put(1695,4002){\makebox(0,0){\strut{}}}%
      \colorrgb{0.50,0.50,0.50}%
      \put(1722,4141){\makebox(0,0){\strut{}}}%
      \colorrgb{0.50,0.50,0.50}%
      \put(792,2337){\makebox(0,0){\rotatebox{85.710847}{\colorbox{white}{\smash{\fns{50}}}}}}%
      \colorrgb{0.50,0.50,0.50}%
      \put(659,569){\makebox(0,0){\strut{}}}%
      \colorrgb{0.50,0.50,0.50}%
      \put(665,985){\makebox(0,0){\strut{}}}%
      \colorrgb{0.50,0.50,0.50}%
      \put(713,1164){\makebox(0,0){\strut{}}}%
      \colorrgb{0.50,0.50,0.50}%
      \put(713,1417){\makebox(0,0){\strut{}}}%
      \colorrgb{0.50,0.50,0.50}%
      \put(750,1760){\makebox(0,0){\strut{}}}%
      \colorrgb{0.50,0.50,0.50}%
      \put(754,1870){\makebox(0,0){\strut{}}}%
      \colorrgb{0.50,0.50,0.50}%
      \put(786,2256){\makebox(0,0){\strut{}}}%
      \colorrgb{0.50,0.50,0.50}%
      \put(792,2337){\makebox(0,0){\strut{}}}%
      \colorrgb{0.50,0.50,0.50}%
      \put(794,2355){\makebox(0,0){\strut{}}}%
      \colorrgb{0.50,0.50,0.50}%
      \put(812,2862){\makebox(0,0){\strut{}}}%
      \colorrgb{0.50,0.50,0.50}%
      \put(825,2950){\makebox(0,0){\strut{}}}%
      \colorrgb{0.50,0.50,0.50}%
      \put(844,3347){\makebox(0,0){\strut{}}}%
      \colorrgb{0.50,0.50,0.50}%
      \put(854,3546){\makebox(0,0){\strut{}}}%
      \colorrgb{0.50,0.50,0.50}%
      \put(869,3858){\makebox(0,0){\strut{}}}%
      \colorrgb{0.50,0.50,0.50}%
      \put(902,4141){\makebox(0,0){\strut{}}}%
      \colorrgb{0.50,0.50,0.50}%
      \put(2373,2276){\makebox(0,0){\rotatebox{61.927513}{\colorbox{white}{\smash{\fns{500}}}}}}%
      \colorrgb{0.50,0.50,0.50}%
      \put(1311,569){\makebox(0,0){\strut{}}}%
      \colorrgb{0.50,0.50,0.50}%
      \put(1420,779){\makebox(0,0){\strut{}}}%
      \colorrgb{0.50,0.50,0.50}%
      \put(1481,840){\makebox(0,0){\strut{}}}%
      \colorrgb{0.50,0.50,0.50}%
      \put(1545,946){\makebox(0,0){\strut{}}}%
      \colorrgb{0.50,0.50,0.50}%
      \put(1655,1039){\makebox(0,0){\strut{}}}%
      \colorrgb{0.50,0.50,0.50}%
      \put(1681,1075){\makebox(0,0){\strut{}}}%
      \colorrgb{0.50,0.50,0.50}%
      \put(1793,1164){\makebox(0,0){\strut{}}}%
      \colorrgb{0.50,0.50,0.50}%
      \put(1813,1219){\makebox(0,0){\strut{}}}%
      \colorrgb{0.50,0.50,0.50}%
      \put(1829,1248){\makebox(0,0){\strut{}}}%
      \colorrgb{0.50,0.50,0.50}%
      \put(1923,1436){\makebox(0,0){\strut{}}}%
      \colorrgb{0.50,0.50,0.50}%
      \put(2002,1571){\makebox(0,0){\strut{}}}%
      \colorrgb{0.50,0.50,0.50}%
      \put(2036,1645){\makebox(0,0){\strut{}}}%
      \colorrgb{0.50,0.50,0.50}%
      \put(2125,1760){\makebox(0,0){\strut{}}}%
      \colorrgb{0.50,0.50,0.50}%
      \put(2156,1829){\makebox(0,0){\strut{}}}%
      \colorrgb{0.50,0.50,0.50}%
      \put(2176,1873){\makebox(0,0){\strut{}}}%
      \colorrgb{0.50,0.50,0.50}%
      \put(2258,2074){\makebox(0,0){\strut{}}}%
      \colorrgb{0.50,0.50,0.50}%
      \put(2350,2231){\makebox(0,0){\strut{}}}%
      \colorrgb{0.50,0.50,0.50}%
      \put(2373,2276){\makebox(0,0){\strut{}}}%
      \colorrgb{0.50,0.50,0.50}%
      \put(2414,2355){\makebox(0,0){\strut{}}}%
      \colorrgb{0.50,0.50,0.50}%
      \put(2477,2516){\makebox(0,0){\strut{}}}%
      \colorrgb{0.50,0.50,0.50}%
      \put(2523,2596){\makebox(0,0){\strut{}}}%
      \colorrgb{0.50,0.50,0.50}%
      \put(2592,2714){\makebox(0,0){\strut{}}}%
      \colorrgb{0.50,0.50,0.50}%
      \put(2697,2936){\makebox(0,0){\strut{}}}%
      \colorrgb{0.50,0.50,0.50}%
      \put(2700,2942){\makebox(0,0){\strut{}}}%
      \colorrgb{0.50,0.50,0.50}%
      \put(2704,2950){\makebox(0,0){\strut{}}}%
      \colorrgb{0.50,0.50,0.50}%
      \put(2801,3191){\makebox(0,0){\strut{}}}%
      \colorrgb{0.50,0.50,0.50}%
      \put(2871,3331){\makebox(0,0){\strut{}}}%
      \colorrgb{0.50,0.50,0.50}%
      \put(2910,3410){\makebox(0,0){\strut{}}}%
      \colorrgb{0.50,0.50,0.50}%
      \put(2996,3546){\makebox(0,0){\strut{}}}%
      \colorrgb{0.50,0.50,0.50}%
      \put(3025,3612){\makebox(0,0){\strut{}}}%
      \colorrgb{0.50,0.50,0.50}%
      \put(3045,3656){\makebox(0,0){\strut{}}}%
      \colorrgb{0.50,0.50,0.50}%
      \put(3124,3868){\makebox(0,0){\strut{}}}%
      \colorrgb{0.50,0.50,0.50}%
      \put(3218,4062){\makebox(0,0){\strut{}}}%
      \colorrgb{0.50,0.50,0.50}%
      \put(3232,4094){\makebox(0,0){\strut{}}}%
      \colorrgb{0.50,0.50,0.50}%
      \put(3251,4141){\makebox(0,0){\strut{}}}%
      \colorrgb{0.50,0.50,0.50}%
      \put(2697,2236){\makebox(0,0){\rotatebox{51.529536}{\colorbox{white}{\smash{\fns{600}}}}}}%
      \colorrgb{0.50,0.50,0.50}%
      \put(1463,569){\makebox(0,0){\strut{}}}%
      \colorrgb{0.50,0.50,0.50}%
      \put(1475,591){\makebox(0,0){\strut{}}}%
      \colorrgb{0.50,0.50,0.50}%
      \put(1481,598){\makebox(0,0){\strut{}}}%
      \colorrgb{0.50,0.50,0.50}%
      \put(1592,784){\makebox(0,0){\strut{}}}%
      \colorrgb{0.50,0.50,0.50}%
      \put(1655,836){\makebox(0,0){\strut{}}}%
      \colorrgb{0.50,0.50,0.50}%
      \put(1723,931){\makebox(0,0){\strut{}}}%
      \colorrgb{0.50,0.50,0.50}%
      \put(1829,1016){\makebox(0,0){\strut{}}}%
      \colorrgb{0.50,0.50,0.50}%
      \put(1861,1055){\makebox(0,0){\strut{}}}%
      \colorrgb{0.50,0.50,0.50}%
      \put(2002,1137){\makebox(0,0){\strut{}}}%
      \colorrgb{0.50,0.50,0.50}%
      \put(2008,1143){\makebox(0,0){\strut{}}}%
      \colorrgb{0.50,0.50,0.50}%
      \put(2042,1164){\makebox(0,0){\strut{}}}%
      \colorrgb{0.50,0.50,0.50}%
      \put(2123,1345){\makebox(0,0){\strut{}}}%
      \colorrgb{0.50,0.50,0.50}%
      \put(2176,1412){\makebox(0,0){\strut{}}}%
      \colorrgb{0.50,0.50,0.50}%
      \put(2244,1528){\makebox(0,0){\strut{}}}%
      \colorrgb{0.50,0.50,0.50}%
      \put(2350,1668){\makebox(0,0){\strut{}}}%
      \colorrgb{0.50,0.50,0.50}%
      \put(2368,1699){\makebox(0,0){\strut{}}}%
      \colorrgb{0.50,0.50,0.50}%
      \put(2427,1760){\makebox(0,0){\strut{}}}%
      \colorrgb{0.50,0.50,0.50}%
      \put(2489,1879){\makebox(0,0){\strut{}}}%
      \colorrgb{0.50,0.50,0.50}%
      \put(2523,1946){\makebox(0,0){\strut{}}}%
      \colorrgb{0.50,0.50,0.50}%
      \put(2594,2112){\makebox(0,0){\strut{}}}%
      \colorrgb{0.50,0.50,0.50}%
      \put(2697,2236){\makebox(0,0){\strut{}}}%
      \colorrgb{0.50,0.50,0.50}%
      \put(2720,2276){\makebox(0,0){\strut{}}}%
      \colorrgb{0.50,0.50,0.50}%
      \put(2767,2355){\makebox(0,0){\strut{}}}%
      \colorrgb{0.50,0.50,0.50}%
      \put(2826,2510){\makebox(0,0){\strut{}}}%
      \colorrgb{0.50,0.50,0.50}%
      \put(2871,2593){\makebox(0,0){\strut{}}}%
      \colorrgb{0.50,0.50,0.50}%
      \put(2942,2707){\makebox(0,0){\strut{}}}%
      \colorrgb{0.50,0.50,0.50}%
      \put(3045,2871){\makebox(0,0){\strut{}}}%
      \colorrgb{0.50,0.50,0.50}%
      \put(3060,2899){\makebox(0,0){\strut{}}}%
      \colorrgb{0.50,0.50,0.50}%
      \put(3094,2950){\makebox(0,0){\strut{}}}%
      \colorrgb{0.50,0.50,0.50}%
      \put(3175,3099){\makebox(0,0){\strut{}}}%
      \colorrgb{0.50,0.50,0.50}%
      \put(3218,3179){\makebox(0,0){\strut{}}}%
      \colorrgb{0.50,0.50,0.50}%
      \put(3291,3295){\makebox(0,0){\strut{}}}%
      \colorrgb{0.50,0.50,0.50}%
      \put(3392,3496){\makebox(0,0){\strut{}}}%
      \colorrgb{0.50,0.50,0.50}%
      \put(3401,3516){\makebox(0,0){\strut{}}}%
      \colorrgb{0.50,0.50,0.50}%
      \put(3427,3546){\makebox(0,0){\strut{}}}%
      \colorrgb{0.50,0.50,0.50}%
      \put(3519,3704){\makebox(0,0){\strut{}}}%
      \colorrgb{0.50,0.50,0.50}%
      \put(3566,3810){\makebox(0,0){\strut{}}}%
      \colorrgb{0.50,0.50,0.50}%
      \put(3617,3966){\makebox(0,0){\strut{}}}%
      \colorrgb{0.50,0.50,0.50}%
      \put(3739,4056){\makebox(0,0){\strut{}}}%
      \colorrgb{0.50,0.50,0.50}%
      \put(3757,4081){\makebox(0,0){\strut{}}}%
      \colorrgb{0.50,0.50,0.50}%
      \put(3789,4141){\makebox(0,0){\strut{}}}%
      \colorrgb{0.50,0.50,0.50}%
      \put(2871,2037){\makebox(0,0){\rotatebox{60.314163}{\colorbox{white}{\smash{\fns{700}}}}}}%
      \colorrgb{0.50,0.50,0.50}%
      \put(1609,569){\makebox(0,0){\strut{}}}%
      \colorrgb{0.50,0.50,0.50}%
      \put(1640,621){\makebox(0,0){\strut{}}}%
      \colorrgb{0.50,0.50,0.50}%
      \put(1655,634){\makebox(0,0){\strut{}}}%
      \colorrgb{0.50,0.50,0.50}%
      \put(1765,788){\makebox(0,0){\strut{}}}%
      \colorrgb{0.50,0.50,0.50}%
      \put(1829,838){\makebox(0,0){\strut{}}}%
      \colorrgb{0.50,0.50,0.50}%
      \put(1899,924){\makebox(0,0){\strut{}}}%
      \colorrgb{0.50,0.50,0.50}%
      \put(2002,984){\makebox(0,0){\strut{}}}%
      \colorrgb{0.50,0.50,0.50}%
      \put(2043,1027){\makebox(0,0){\strut{}}}%
      \colorrgb{0.50,0.50,0.50}%
      \put(2176,1109){\makebox(0,0){\strut{}}}%
      \colorrgb{0.50,0.50,0.50}%
      \put(2189,1121){\makebox(0,0){\strut{}}}%
      \colorrgb{0.50,0.50,0.50}%
      \put(2273,1164){\makebox(0,0){\strut{}}}%
      \colorrgb{0.50,0.50,0.50}%
      \put(2324,1253){\makebox(0,0){\strut{}}}%
      \colorrgb{0.50,0.50,0.50}%
      \put(2350,1286){\makebox(0,0){\strut{}}}%
      \colorrgb{0.50,0.50,0.50}%
      \put(2442,1444){\makebox(0,0){\strut{}}}%
      \colorrgb{0.50,0.50,0.50}%
      \put(2523,1528){\makebox(0,0){\strut{}}}%
      \colorrgb{0.50,0.50,0.50}%
      \put(2572,1594){\makebox(0,0){\strut{}}}%
      \colorrgb{0.50,0.50,0.50}%
      \put(2697,1747){\makebox(0,0){\strut{}}}%
      \colorrgb{0.50,0.50,0.50}%
      \put(2700,1751){\makebox(0,0){\strut{}}}%
      \colorrgb{0.50,0.50,0.50}%
      \put(2709,1760){\makebox(0,0){\strut{}}}%
      \colorrgb{0.50,0.50,0.50}%
      \put(2817,1945){\makebox(0,0){\strut{}}}%
      \colorrgb{0.50,0.50,0.50}%
      \put(2871,2037){\makebox(0,0){\strut{}}}%
      \colorrgb{0.50,0.50,0.50}%
      \put(2931,2150){\makebox(0,0){\strut{}}}%
      \colorrgb{0.50,0.50,0.50}%
      \put(3045,2264){\makebox(0,0){\strut{}}}%
      \colorrgb{0.50,0.50,0.50}%
      \put(3064,2287){\makebox(0,0){\strut{}}}%
      \colorrgb{0.50,0.50,0.50}%
      \put(3124,2355){\makebox(0,0){\strut{}}}%
      \colorrgb{0.50,0.50,0.50}%
      \put(3186,2467){\makebox(0,0){\strut{}}}%
      \colorrgb{0.50,0.50,0.50}%
      \put(3218,2516){\makebox(0,0){\strut{}}}%
      \colorrgb{0.50,0.50,0.50}%
      \put(3296,2683){\makebox(0,0){\strut{}}}%
      \colorrgb{0.50,0.50,0.50}%
      \put(3392,2782){\makebox(0,0){\strut{}}}%
      \colorrgb{0.50,0.50,0.50}%
      \put(3428,2825){\makebox(0,0){\strut{}}}%
      \colorrgb{0.50,0.50,0.50}%
      \put(3515,2950){\makebox(0,0){\strut{}}}%
      \colorrgb{0.50,0.50,0.50}%
      \put(3545,3020){\makebox(0,0){\strut{}}}%
      \colorrgb{0.50,0.50,0.50}%
      \put(3566,3043){\makebox(0,0){\strut{}}}%
      \colorrgb{0.50,0.50,0.50}%
      \put(3666,3203){\makebox(0,0){\strut{}}}%
      \colorrgb{0.50,0.50,0.50}%
      \put(3739,3347){\makebox(0,0){\strut{}}}%
      \colorrgb{0.50,0.50,0.50}%
      \put(3775,3425){\makebox(0,0){\strut{}}}%
      \colorrgb{0.50,0.50,0.50}%
      \put(3874,3546){\makebox(0,0){\strut{}}}%
      \colorrgb{0.50,0.50,0.50}%
      \put(3901,3585){\makebox(0,0){\strut{}}}%
      \colorrgb{0.50,0.50,0.50}%
      \put(3913,3607){\makebox(0,0){\strut{}}}%
      \colorrgb{0.50,0.50,0.50}%
      \put(4020,3775){\makebox(0,0){\strut{}}}%
    }%
    \gplbacktext
    \put(0,0){\includegraphics{challenge-response-time-contours}}%
    \gplfronttext
  \end{picture}%
\endgroup

%% file: mason.tex
\section{The Mason Test}
\label{sec:mason}

This section describes the full Mason test protocol, an implementation
of the concepts introduced in the previous sections.  Those results
impose four main requirements on the protocol.
\begin{enumerate}
\item Conforming neighbors must be able to participate. That is,
  selective jamming of conforming identities must be detectable.

\item Probe packets must be transmitted in a pseudo-random order.
  Further, each participant must be able to verify that no group of
  identities controlled the order.

\item Moving identities are rejected, so to save energy and time,
  conforming nodes that are moving when the protocol begins should not
  participate.

\item Attackers must not know the RSSI observations of conforming
  identities when computing lies.

\end{enumerate}

We assume a known upper bound on the number of conforming neighbors,
i.e., those within the one-hop transmission range. In most
applications, a bound in the hundreds (we use 400 in our experiments)
will be acceptable.  If more identities attempt to participate, the
protocol aborts and no classification is made.  This appears to open a
denial-of-service attack. However, we do not attempt to prevent,
instead only detect, DOS attacks, because one such attack---simply
jamming the wireless channel---is unpreventable (with commodity
hardware). See \autoref{sec:discussion} for more discussion.

The rest of this section describes the two components of the protocol:
collection of RSSI observations and Sybil classification. We assume
one identity, the initiator, starts the protocol and leads the
collection, but all identities still individually and safely perform
Sybil classification.

\subsection{Collection of RSSI Observations}
\label{sec:idcollection}

\Note{Describe the protocol step by step.}

\textbf{Phase I: Identity Collection}. The first phase gathers
participating neighbors, ensuring that no conforming identities are
jammed by attackers.  The initiator sends a REQUEST message stating
its identity, e.g., a public key.  All stationary neighbors respond
with their identities via HELLO-I messages, ACKed by the
initiator. Unacknowledged HELLO-Is are re-transmitted. The process
terminates when the channel is idle---indicating all HELLO-I's were
received and ACKed.  If the channel does not go idle before a timeout
(e.g., 15 seconds), an attacker may be selectively jamming some
HELLO-Is and the protocol aborts.  The protocol also aborts if too
many identities join, e.g., 400.

\textbf{Phase II: Randomized Broadcast Request}: The second phase is
the challenge-response protocol to collect RSSI observations for
motion detection and Sybil classification.  First, each identity
contributes a random value; all are hashed together to produce a seed
to generate the random sequence of broadcast requests issued by the
initiator.  Specifically, it sends a TRANSMIT message to the next
participant in the random sequence, who must quickly broadcast a
signed HELLO-II, e.g., within \SI{10}{\milli\second} in our
implementation\footnote{\SI{10}{\milli\second} is larger than the
  typical roundtrip time for 802.11b with packets handled in interrupt
  context for low-latency responses. These packets can be signed ahead
  of time and cached---signatures do not need to be computed in the
  \SI{10}{\milli\second} interval.}.  Each participant records the
RSSIs of the HELLO-II messages it hears. Some identities will not hear
each other; this is acceptable because the initiator needs
observations from only three other conforming identities.  $|\setI|
\times s$ requests are issued, where $s$ is large enough to ensure a
short minimum duration between consecutive requests for any two pairs
of nodes, e.g., 14 in our tests.  An identity that fails to respond in
time might be an attacker attempting to change positions and is
rejected.

\textbf{Phase III: RSSI Observations Report}.  In the third phase, the
RSSI observations are shared.  First, each identity broadcasts a hash
of its observations. Then the actual values are shared. Those not
matching the respective hash are rejected, preventing attackers from
using the reported values to fabricate plausible observations. The
same mechanism from Phase~1 is used to detect selective jamming.

\subsection{Sybil Classification}
\label{sec:classification}

\begin{figure}
  \centering
  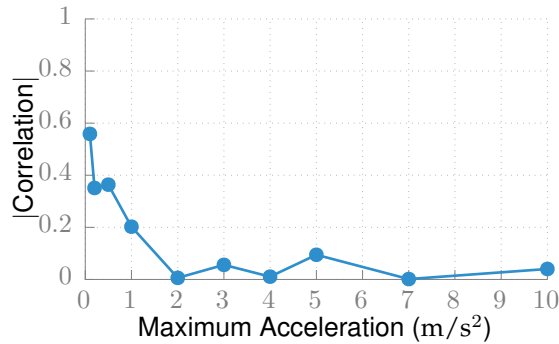
  \caption{RSSI correlation as a function of the maximum device
    acceleration between observations.}
  \label{fig:rssi-acceleration-correlation}
\end{figure}

Each participant performs Sybil classification individually. First,
the identity verifies that its observations were not potentially
predictable from those reported in prior rounds, possibly violating
\autoref{cond:collapse-lns}. Correlation in RSSI values between
observations decreases with motion between observations, as shown by
our experiments (\autoref{fig:rssi-acceleration-correlation}). Thus, a
node only performs Sybil classification if it has strong evidence the
current observations are uncorrelated with prior ones\footnote{Note
  that although we did not encounter this case in our experiments, it
  is conceivable that some devices will return to the same location
  and orientation after motion.}, i.e., it has observed an
acceleration of at least \SI{2}{\meter\per\second\squared}.

Classification is a simple application of the methods of
\autoref{sec:motion} and \autoref{sec:algorithm}. Each identity with
an RSSI variance across its multiple broadcasts higher than a
threshold is rejected. Then, \autoref{alg:build-receiver-sets} and
\autoref{alg:find-consistent-subset} are used to identify a
$\gamma$-true Sybil classification over the remaining, stationary
identities.

%% file: fig/rssi-acceleration-correlation.pdf_tex
\begingroup
  \makeatletter
  \providecommand\color[2][]{%
    \GenericError{(gnuplot) \space\space\space\@spaces}{%
      Package color not loaded in conjunction with
      terminal option `colourtext'%
    }{See the gnuplot documentation for explanation.%
    }{Either use 'blacktext' in gnuplot or load the package
      color.sty in LaTeX.}%
    \renewcommand\color[2][]{}%
  }%
  \providecommand\includegraphics[2][]{%
    \GenericError{(gnuplot) \space\space\space\@spaces}{%
      Package graphicx or graphics not loaded%
    }{See the gnuplot documentation for explanation.%
    }{The gnuplot epslatex terminal needs graphicx.sty or graphics.sty.}%
    \renewcommand\includegraphics[2][]{}%
  }%
  \providecommand\rotatebox[2]{#2}%
  \@ifundefined{ifGPcolor}{%
    \newif\ifGPcolor
    \GPcolortrue
  }{}%
  \@ifundefined{ifGPblacktext}{%
    \newif\ifGPblacktext
    \GPblacktextfalse
  }{}%
  \let\gplgaddtomacro\g@addto@macro
  \gdef\gplbacktext{}%
  \gdef\gplfronttext{}%
  \makeatother
  \ifGPblacktext
    \def\colorrgb#1{}%
    \def\colorgray#1{}%
  \else
    \ifGPcolor
      \def\colorrgb#1{\color[rgb]{#1}}%
      \def\colorgray#1{\color[gray]{#1}}%
      \expandafter\def\csname LTw\endcsname{\color{white}}%
      \expandafter\def\csname LTb\endcsname{\color{black}}%
      \expandafter\def\csname LTa\endcsname{\color{black}}%
      \expandafter\def\csname LT0\endcsname{\color[rgb]{1,0,0}}%
      \expandafter\def\csname LT1\endcsname{\color[rgb]{0,1,0}}%
      \expandafter\def\csname LT2\endcsname{\color[rgb]{0,0,1}}%
      \expandafter\def\csname LT3\endcsname{\color[rgb]{1,0,1}}%
      \expandafter\def\csname LT4\endcsname{\color[rgb]{0,1,1}}%
      \expandafter\def\csname LT5\endcsname{\color[rgb]{1,1,0}}%
      \expandafter\def\csname LT6\endcsname{\color[rgb]{0,0,0}}%
      \expandafter\def\csname LT7\endcsname{\color[rgb]{1,0.3,0}}%
      \expandafter\def\csname LT8\endcsname{\color[rgb]{0.5,0.5,0.5}}%
    \else
      \def\colorrgb#1{\color{black}}%
      \def\colorgray#1{\color[gray]{#1}}%
      \expandafter\def\csname LTw\endcsname{\color{white}}%
      \expandafter\def\csname LTb\endcsname{\color{black}}%
      \expandafter\def\csname LTa\endcsname{\color{black}}%
      \expandafter\def\csname LT0\endcsname{\color{black}}%
      \expandafter\def\csname LT1\endcsname{\color{black}}%
      \expandafter\def\csname LT2\endcsname{\color{black}}%
      \expandafter\def\csname LT3\endcsname{\color{black}}%
      \expandafter\def\csname LT4\endcsname{\color{black}}%
      \expandafter\def\csname LT5\endcsname{\color{black}}%
      \expandafter\def\csname LT6\endcsname{\color{black}}%
      \expandafter\def\csname LT7\endcsname{\color{black}}%
      \expandafter\def\csname LT8\endcsname{\color{black}}%
    \fi
  \fi
    \setlength{\unitlength}{0.0500bp}%
    \ifx\gptboxheight\undefined%
      \newlength{\gptboxheight}%
      \newlength{\gptboxwidth}%
      \newsavebox{\gptboxtext}%
    \fi%
    \setlength{\fboxrule}{0.5pt}%
    \setlength{\fboxsep}{1pt}%
\begin{picture}(4320.00,2660.00)%
    \gplgaddtomacro\gplbacktext{%
      \colorrgb{0.50,0.50,0.50}%
      \put(480,516){\makebox(0,0)[r]{\strut{}$0$}}%
      \colorrgb{0.50,0.50,0.50}%
      \put(480,909){\makebox(0,0)[r]{\strut{}$0.2$}}%
      \colorrgb{0.50,0.50,0.50}%
      \put(480,1302){\makebox(0,0)[r]{\strut{}$0.4$}}%
      \colorrgb{0.50,0.50,0.50}%
      \put(480,1695){\makebox(0,0)[r]{\strut{}$0.6$}}%
      \colorrgb{0.50,0.50,0.50}%
      \put(480,2088){\makebox(0,0)[r]{\strut{}$0.8$}}%
      \colorrgb{0.50,0.50,0.50}%
      \put(480,2481){\makebox(0,0)[r]{\strut{}$1$}}%
      \colorrgb{0.50,0.50,0.50}%
      \put(569,338){\makebox(0,0){\strut{}$0$}}%
      \colorrgb{0.50,0.50,0.50}%
      \put(917,338){\makebox(0,0){\strut{}$1$}}%
      \colorrgb{0.50,0.50,0.50}%
      \put(1266,338){\makebox(0,0){\strut{}$2$}}%
      \colorrgb{0.50,0.50,0.50}%
      \put(1614,338){\makebox(0,0){\strut{}$3$}}%
      \colorrgb{0.50,0.50,0.50}%
      \put(1962,338){\makebox(0,0){\strut{}$4$}}%
      \colorrgb{0.50,0.50,0.50}%
      \put(2311,338){\makebox(0,0){\strut{}$5$}}%
      \colorrgb{0.50,0.50,0.50}%
      \put(2659,338){\makebox(0,0){\strut{}$6$}}%
      \colorrgb{0.50,0.50,0.50}%
      \put(3007,338){\makebox(0,0){\strut{}$7$}}%
      \colorrgb{0.50,0.50,0.50}%
      \put(3355,338){\makebox(0,0){\strut{}$8$}}%
      \colorrgb{0.50,0.50,0.50}%
      \put(3704,338){\makebox(0,0){\strut{}$9$}}%
      \colorrgb{0.50,0.50,0.50}%
      \put(4052,338){\makebox(0,0){\strut{}$10$}}%
    }%
    \gplgaddtomacro\gplfronttext{%
      \csname LTb\endcsname%
      \put(133,1498){\rotatebox{-270}{\makebox(0,0){\strut{}$|$Correlation$|$}}}%
      \csname LTb\endcsname%
      \put(2310,124){\makebox(0,0){\strut{}Maximum Acceleration (\si[per-mode=symbol]{\meter\per\second\squared})}}%
    }%
    \gplbacktext
    \put(0,0){\includegraphics{rssi-acceleration-correlation}}%
    \gplfronttext
  \end{picture}%
\endgroup

%% file: evaluation.tex
\section{Prototype and Evaluation}
\label{sec:evaluation}

We implemented the Mason test as a Linux kernel module and tested its
performance on HTC Magic Android smartphones in various operating
environments.  It sits directly above the 802.11 link layer,
responding to requests in interrupt context, to minimize response
latency for the REQUEST--HELLO-II sequence (\SI{12}{\milli\second}
roundtrip time on our hardware). The classification algorithms are
implemented in Python.

\Note{Explain training/testing conventions for following experiments
here.}

Wireless channels are environment-dependent, so we evaluated the Mason
test in four different environments.
\begin{LaTeXdescription}
\item[Office I] Eleven participants in two adjacent offices for
  fifteen hours.
\item[Office II] Eleven participants in two adjacent offices in a
  different building for one hour, to determine whether performance
  varies across similar, but non-identical environments.
\item[Cafeteria] Eleven participants in a crowded cafeteria during
  lunch. This was a rapidly-changing wireless environment due to
  frequent motion of the cafeteria patrons.
\item[Outdoor] Eleven participants meeting in a cold, open, grassy
  courtyard for one hour, capturing the outdoor
  environment. Participants moved frequently to stay warm.
\end{LaTeXdescription}
In each environment, we conducted multiple trials with one Sybil
attacker\footnote{As discussed in \autoref{sec:separation} and
  \autoref{sec:predictability}, additional physical nodes are best
  used as lying, non-Sybils.} generating 4, 20, 40, and 160 Sybil
identities. The gathered traces were split into testing and training
sets.

\begin{table}
  \centering
  \setarrstr{1.2}
  \caption{Thresholds for Signalprint Comparison and Motion Filtering}
  \label{tbl:thresholds}
  \begin{tabular}{@{}llS[table-format=1.2,table-alignment=left]@{}}
    \toprule
    \multicolumn{2}{@{}l}{\bfseries Name} & \bfseries Threshold (\si\dBm) \\
    \midrule
    Signalprint Distance & dimension-2 & 0.85 \\
                         & dimension-3 & 3.6 \\
                         & dimension-4 & 1.2 \\
    \midrule
    \multicolumn{2}{@{}l}{RSSI Standard Deviation} & 2.5 \\
    \bottomrule
  \end{tabular}
\end{table}

\subsection{Selection and Robustness of Thresholds}
\label{sec:thresholds}

\begin{figure}
  \centering
  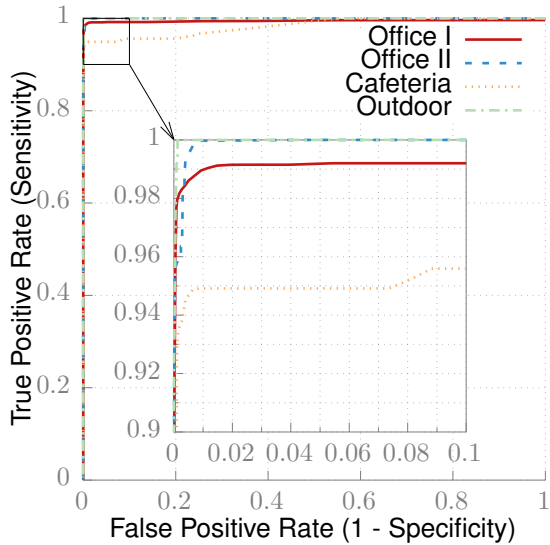
  \caption{ROC curve showing the classification performance of
    signalprint comparison in different environments for varying
    distance thresholds.  Only identities that passed the motion
    filter are considered.  The knees of the curves all correspond to
    the same thresholds, suggesting that the same value can be used in
    all locations.}
  \label{fig:roc}
\end{figure}

\begin{figure*}
  \centering
  \begin{minipage}{0.23\linewidth}
    \centering
    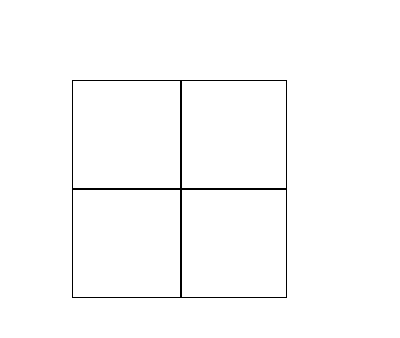
    Office I
  \end{minipage}
  \hfill
  \begin{minipage}{0.23\linewidth}
    \centering
    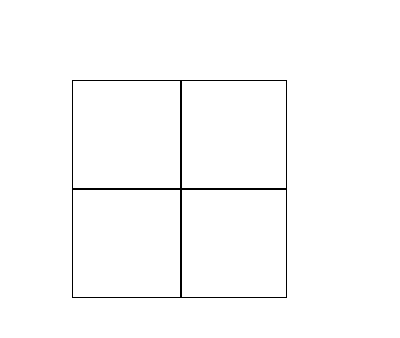
    Office II
  \end{minipage}
  \hfill
  \begin{minipage}{0.23\linewidth}
    \centering
    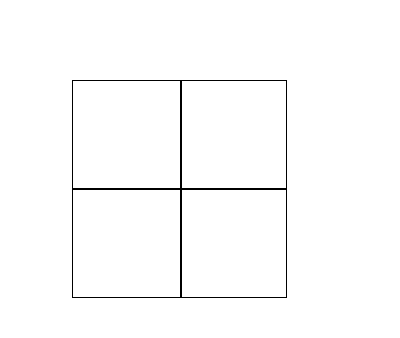
    Cafeteria
  \end{minipage}
  \hfill
  \begin{minipage}{0.23\linewidth}
    \centering
    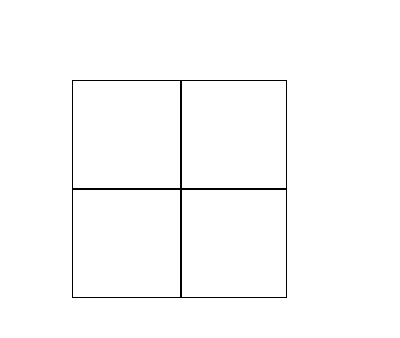
    Outdoor
  \end{minipage}
  \caption{Confusion matrices detailing the classifier performance in
    the four environments. \emph{S} is Sybil and \emph{C} is
    conforming.  Multiple tests were run in each environment, so mean
    percentages are shown instead of absolute counts.}
  \label{fig:confusion}
\end{figure*}

\begin{table}
  \setarrstr{1.2}
  \centering
  \caption{Classification Performance}
  \label{tbl:sensitivity}
  \begin{tabular}{@{}lS[table-format=3.1,table-alignment=left]S[table-format=3.1,table-alignment=left]@{}}
    \toprule
    Environment & {Sensitivity (\si{\percent})} & {Specificity (\si{\percent})} \\
    \midrule
    Office I  &  99.6 & 96.5 \\
    Office II & 100.0 & 87.7 \\
    Cafeteria &  91.4 & 86.6 \\
    Outdoor   &  95.9 & 61.1 \\
    \bottomrule
  \end{tabular}
\end{table}

The training data were used to determine good motion filter and
signalprint distance thresholds, shown in \autoref{tbl:thresholds}.

The motion filter threshold was chosen such that at least $95\%$ of
the conforming participants (averaged over all training rounds) in the
low-motion Office I environment would pass.  This policy ensures that
conforming smartphones, which are usually left mostly stationary,
e.g., on desks, in purses, or in the pockets of seated people, will
usually pass the test.  Devices exhibiting more motion (i.e., a
standard deviation of RSSIs at the initiator larger than
\SI{2.5}{\dBm})---as would be expected from an attacker trying to
defeat signalprint detection---will be rejected.

The signalprint distance thresholds were chosen by evaluating the
signalprint classification performance at various possible
values. \autoref{fig:roc} shows the ROC curves for size-4 receiver
sets (a ``positive'' is an identity classified as Sybil).  Note that
the true positive and false positive rates consider only identities
that passed the motion filter, in order to isolate the effects of the
signalprint distance threshold.  The curves show that a good threshold
has performance in line with prior
work~\cite{xiao09sep,demirbas06jun}, as expected.

In all environments, the knees of the curve correspond to the same
thresholds, suggesting that these values can be used in general,
across environments.  A possible explanation is that despite
environment differences, the signalprint distance distributions for
stationary Sybil siblings are identical.  All results in this paper
use these uniform thresholds, show in \autoref{tbl:thresholds}.

\subsection{Classification Performance}
\label{sec:performance}

The performance of the full Mason test---motion filtering and
signalprint comparison---is detailed by the confusion matrices in
\autoref{fig:confusion}.  Many tests were conducted in each
environment, so average percentages are shown instead of absolute
counts.  To evaluate the performance, we consider two standard
classification metrics derived from these matrices, \emph{sensitivity}
(percentage of Sybil identities correctly identified) and
\emph{specificity} (percentage of conforming identities correctly
identified).

Note that 100\% sensitivity is not necessary. Most protocols that
would use Mason require a majority of the participants to be
conforming.  The total identities are limited (e.g., 400) so rejecting
most of the Sybils and accepting most of the conforming identities is
sufficient to meet this requirement.

\autoref{tbl:sensitivity} shows the performance for all four
environments.  The test performs best in the low-motion indoor
environments, with over 99.5\% sensitivity and over 85\% specificity.
The sensitivity in the cafeteria environment is just 91.4\%, likely
due to the rapid and frequent changes in the wireless environment
resulting from the movement of cafeteria patrons.  In the outdoor
environment, with participants moving, the specificity is 61.1\%
because some conforming identities are rejected by the motion
filter\footnote{These moving participants normally do not
  participate. Including them here makes clears that they are not
  identified as conforming.}.

\begin{figure}
\centering
    \centering
    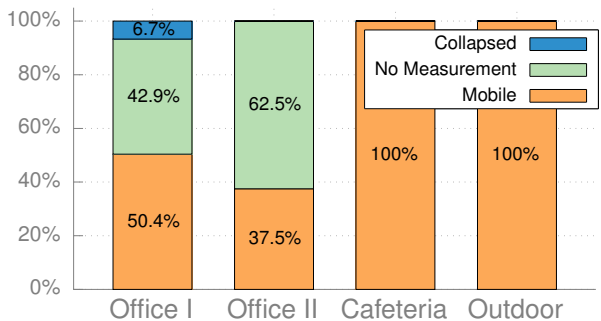
    \caption{Relative frequencies of the three causes of false
      positives.}
    \label{fig:conforming-breakdown}
\end{figure}

An identity is classified as Sybil for three reasons: if it has
similar signalprints with another, the initiator has too few RSSI
reports to form a signalprint, or it fails the motion filter.
\autoref{fig:conforming-breakdown} shows the relative prevalence of
these causes of the false positives.  Interestingly, the first
cause---collapsing---is rare, occurring only in the first office
environment.  Not surprisingly, missing RSSI reports are an issue only
in the environments with significant obstructions, the indoor offices,
accounting for about half of these false positives.  In the open
cafeteria and outdoor environments, all false positives are due to
participant motion.

\subsection{Overhead Evaluation}

\begin{figure*}
\begin{minipage}{0.66\textwidth}
  \centering
  \begin{subfigure}[b]{0.49\textwidth}
    \centering
    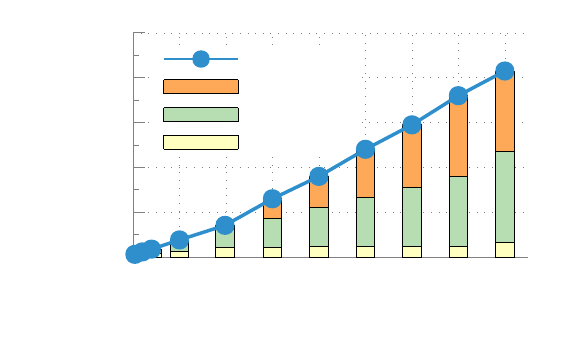
    \caption{Runtime in seconds.}
    \label{fig:collection_time}
  \end{subfigure}
  \hfill
  \begin{subfigure}[b]{0.49\textwidth}
    \centering
    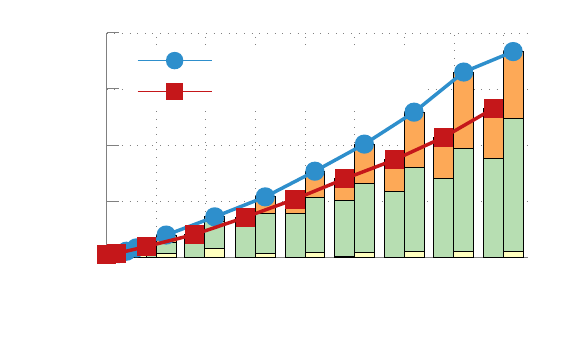
    \caption{Energy consumption in joules.}
    \label{fig:collection_energy}
  \end{subfigure}
  \caption{Overhead of the collection phase. The stacked bars
    partition the cost among the participant collection (HELLO I),
    RSSI measurement (HELLO II), and RSSI observation exchange (RSST)
    steps.}
  \label{fig:overhead_collection_phase}
\end{minipage}
\hfill
\begin{minipage}{0.32\textwidth}
  \centering
  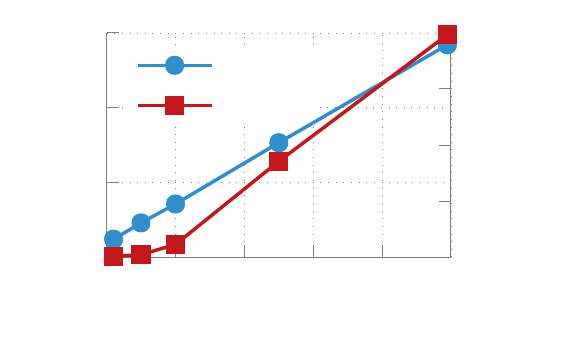
  \caption{Runtime and energy consumption of the classification phase.}
  \label{fig:overhead_classification_phase}
\end{minipage}
\end{figure*}

\autoref{fig:overhead_collection_phase} shows the runtime and energy
overhead for the Mason test collection phase, with the stacked bars
separating the costs by sub-phase. For a protocol not run frequently
(once every hour is often sufficient), the runtimes of 10--90 seconds
are acceptable.  Likewise, smartphone energy consumption is
acceptable, with the extreme \SI{18}{\joule} consumption for 400
identities representing 0.01\% of the \SI{17,500}{\joule} capacity of
a typical smartphone batteries.

\autoref{fig:overhead_classification_phase} show the classification
phase overheads for 2--100 identities.  At a small fraction of the
collection costs, these are also acceptable.  For more than 100
participants, costs become excessive due to the $\bigO(n^3)$ scaling
behavior\footnote{A native-C implementation might scale to 300--400
  identities.}.  Limiting participation to 100 identities may be
necessary for energy-constrained devices, but will generally not
reduce performance because having 100 non-Sybil, one-hop neighbors is
rare.

The periodic accelerometer sampling used to measure motion between
Mason test rounds consumes 5.2\% of battery capacity over a typical
\SI{18}{\hour} work day.

%% file: fig/roc.pdf_tex
\begingroup
  \makeatletter
  \providecommand\color[2][]{%
    \GenericError{(gnuplot) \space\space\space\@spaces}{%
      Package color not loaded in conjunction with
      terminal option `colourtext'%
    }{See the gnuplot documentation for explanation.%
    }{Either use 'blacktext' in gnuplot or load the package
      color.sty in LaTeX.}%
    \renewcommand\color[2][]{}%
  }%
  \providecommand\includegraphics[2][]{%
    \GenericError{(gnuplot) \space\space\space\@spaces}{%
      Package graphicx or graphics not loaded%
    }{See the gnuplot documentation for explanation.%
    }{The gnuplot epslatex terminal needs graphicx.sty or graphics.sty.}%
    \renewcommand\includegraphics[2][]{}%
  }%
  \providecommand\rotatebox[2]{#2}%
  \@ifundefined{ifGPcolor}{%
    \newif\ifGPcolor
    \GPcolortrue
  }{}%
  \@ifundefined{ifGPblacktext}{%
    \newif\ifGPblacktext
    \GPblacktextfalse
  }{}%
  \let\gplgaddtomacro\g@addto@macro
  \gdef\gplbacktext{}%
  \gdef\gplfronttext{}%
  \makeatother
  \ifGPblacktext
    \def\colorrgb#1{}%
    \def\colorgray#1{}%
  \else
    \ifGPcolor
      \def\colorrgb#1{\color[rgb]{#1}}%
      \def\colorgray#1{\color[gray]{#1}}%
      \expandafter\def\csname LTw\endcsname{\color{white}}%
      \expandafter\def\csname LTb\endcsname{\color{black}}%
      \expandafter\def\csname LTa\endcsname{\color{black}}%
      \expandafter\def\csname LT0\endcsname{\color[rgb]{1,0,0}}%
      \expandafter\def\csname LT1\endcsname{\color[rgb]{0,1,0}}%
      \expandafter\def\csname LT2\endcsname{\color[rgb]{0,0,1}}%
      \expandafter\def\csname LT3\endcsname{\color[rgb]{1,0,1}}%
      \expandafter\def\csname LT4\endcsname{\color[rgb]{0,1,1}}%
      \expandafter\def\csname LT5\endcsname{\color[rgb]{1,1,0}}%
      \expandafter\def\csname LT6\endcsname{\color[rgb]{0,0,0}}%
      \expandafter\def\csname LT7\endcsname{\color[rgb]{1,0.3,0}}%
      \expandafter\def\csname LT8\endcsname{\color[rgb]{0.5,0.5,0.5}}%
    \else
      \def\colorrgb#1{\color{black}}%
      \def\colorgray#1{\color[gray]{#1}}%
      \expandafter\def\csname LTw\endcsname{\color{white}}%
      \expandafter\def\csname LTb\endcsname{\color{black}}%
      \expandafter\def\csname LTa\endcsname{\color{black}}%
      \expandafter\def\csname LT0\endcsname{\color{black}}%
      \expandafter\def\csname LT1\endcsname{\color{black}}%
      \expandafter\def\csname LT2\endcsname{\color{black}}%
      \expandafter\def\csname LT3\endcsname{\color{black}}%
      \expandafter\def\csname LT4\endcsname{\color{black}}%
      \expandafter\def\csname LT5\endcsname{\color{black}}%
      \expandafter\def\csname LT6\endcsname{\color{black}}%
      \expandafter\def\csname LT7\endcsname{\color{black}}%
      \expandafter\def\csname LT8\endcsname{\color{black}}%
    \fi
  \fi
    \setlength{\unitlength}{0.0500bp}%
    \ifx\gptboxheight\undefined%
      \newlength{\gptboxheight}%
      \newlength{\gptboxwidth}%
      \newsavebox{\gptboxtext}%
    \fi%
    \setlength{\fboxrule}{0.5pt}%
    \setlength{\fboxsep}{1pt}%
\begin{picture}(4320.00,4320.00)%
    \gplgaddtomacro\gplbacktext{%
      \colorrgb{0.50,0.50,0.50}%
      \put(480,587){\makebox(0,0)[r]{\strut{}$0$}}%
      \colorrgb{0.50,0.50,0.50}%
      \put(480,1284){\makebox(0,0)[r]{\strut{}$0.2$}}%
      \colorrgb{0.50,0.50,0.50}%
      \put(480,1980){\makebox(0,0)[r]{\strut{}$0.4$}}%
      \colorrgb{0.50,0.50,0.50}%
      \put(480,2677){\makebox(0,0)[r]{\strut{}$0.6$}}%
      \colorrgb{0.50,0.50,0.50}%
      \put(480,3373){\makebox(0,0)[r]{\strut{}$0.8$}}%
      \colorrgb{0.50,0.50,0.50}%
      \put(480,4070){\makebox(0,0)[r]{\strut{}$1$}}%
      \colorrgb{0.50,0.50,0.50}%
      \put(569,409){\makebox(0,0){\strut{}$0$}}%
      \colorrgb{0.50,0.50,0.50}%
      \put(1266,409){\makebox(0,0){\strut{}$0.2$}}%
      \colorrgb{0.50,0.50,0.50}%
      \put(1962,409){\makebox(0,0){\strut{}$0.4$}}%
      \colorrgb{0.50,0.50,0.50}%
      \put(2659,409){\makebox(0,0){\strut{}$0.6$}}%
      \colorrgb{0.50,0.50,0.50}%
      \put(3355,409){\makebox(0,0){\strut{}$0.8$}}%
      \colorrgb{0.50,0.50,0.50}%
      \put(4052,409){\makebox(0,0){\strut{}$1$}}%
    }%
    \gplgaddtomacro\gplfronttext{%
      \csname LTb\endcsname%
      \put(133,2328){\rotatebox{-270}{\makebox(0,0){\strut{}True Positive Rate (Sensitivity)}}}%
      \csname LTb\endcsname%
      \put(2310,195){\makebox(0,0){\strut{}False Positive Rate (1 - Specificity)}}%
      \csname LTb\endcsname%
      \put(3358,3910){\makebox(0,0)[r]{\strut{}Office I}}%
      \csname LTb\endcsname%
      \put(3358,3732){\makebox(0,0)[r]{\strut{}Office II}}%
      \csname LTb\endcsname%
      \put(3358,3554){\makebox(0,0)[r]{\strut{}Cafeteria}}%
      \csname LTb\endcsname%
      \put(3358,3376){\makebox(0,0)[r]{\strut{}Outdoor}}%
    }%
    \gplgaddtomacro\gplbacktext{%
      \colorrgb{0.50,0.50,0.50}%
      \put(1163,950){\makebox(0,0)[r]{\strut{}$0.9$}}%
      \colorrgb{0.50,0.50,0.50}%
      \put(1163,1391){\makebox(0,0)[r]{\strut{}$0.92$}}%
      \colorrgb{0.50,0.50,0.50}%
      \put(1163,1831){\makebox(0,0)[r]{\strut{}$0.94$}}%
      \colorrgb{0.50,0.50,0.50}%
      \put(1163,2272){\makebox(0,0)[r]{\strut{}$0.96$}}%
      \colorrgb{0.50,0.50,0.50}%
      \put(1163,2712){\makebox(0,0)[r]{\strut{}$0.98$}}%
      \colorrgb{0.50,0.50,0.50}%
      \put(1163,3153){\makebox(0,0)[r]{\strut{}$1$}}%
      \colorrgb{0.50,0.50,0.50}%
      \put(1252,772){\makebox(0,0){\strut{}$0$}}%
      \colorrgb{0.50,0.50,0.50}%
      \put(1693,772){\makebox(0,0){\strut{}$0.02$}}%
      \colorrgb{0.50,0.50,0.50}%
      \put(2133,772){\makebox(0,0){\strut{}$0.04$}}%
      \colorrgb{0.50,0.50,0.50}%
      \put(2574,772){\makebox(0,0){\strut{}$0.06$}}%
      \colorrgb{0.50,0.50,0.50}%
      \put(3014,772){\makebox(0,0){\strut{}$0.08$}}%
      \colorrgb{0.50,0.50,0.50}%
      \put(3455,772){\makebox(0,0){\strut{}$0.1$}}%
    }%
    \gplgaddtomacro\gplfronttext{%
    }%
    \gplbacktext
    \put(0,0){\includegraphics{roc}}%
    \gplfronttext
  \end{picture}%
\endgroup

%% file: fig/confusion-officeI.pdf_tex
\begingroup%
  \makeatletter%
  \providecommand\color[2][]{%
    \errmessage{(Inkscape) Color is used for the text in Inkscape, but the package 'color.sty' is not loaded}%
    \renewcommand\color[2][]{}%
  }%
  \providecommand\transparent[1]{%
    \errmessage{(Inkscape) Transparency is used (non-zero) for the text in Inkscape, but the package 'transparent.sty' is not loaded}%
    \renewcommand\transparent[1]{}%
  }%
  \providecommand\rotatebox[2]{#2}%
  \ifx\svgwidth\undefined%
    \setlength{\unitlength}{115.2bp}%
    \ifx\svgscale\undefined%
      \relax%
    \else%
      \setlength{\unitlength}{\unitlength * \real{\svgscale}}%
    \fi%
  \else%
    \setlength{\unitlength}{\svgwidth}%
  \fi%
  \global\let\svgwidth\undefined%
  \global\let\svgscale\undefined%
  \makeatother%
  \begin{picture}(1,0.84375)%
    \put(0,0){\includegraphics[width=\unitlength]{confusion-officeI.pdf}}%
    \put(0.31760883,0.47775699){\color[rgb]{0,0,0}\makebox(0,0)[b]{\smash{76.3\%}}}%
    \put(0.58521409,0.47775699){\color[rgb]{0,0,0}\makebox(0,0)[b]{\smash{0.3\%}}}%
    \put(0.31785297,0.20692408){\color[rgb]{0,0,0}\makebox(0,0)[b]{\smash{0.8\%}}}%
    \put(0.58572271,0.20692408){\color[rgb]{0,0,0}\makebox(0,0)[b]{\smash{22.6\%}}}%
    \put(0.31738504,0.67089087){\color[rgb]{0,0,0}\makebox(0,0)[b]{\smash{S}}}%
    \put(0.58399338,0.67089087){\color[rgb]{0,0,0}\makebox(0,0)[b]{\smash{C}}}%
    \put(0.16237356,0.47716698){\color[rgb]{0,0,0}\makebox(0,0)[rb]{\smash{S}}}%
    \put(0.16107147,0.20633449){\color[rgb]{0,0,0}\makebox(0,0)[rb]{\smash{C}}}%
    \put(0.74175056,0.20692408){\color[rgb]{0,0,0}\makebox(0,0)[lb]{\smash{23.4\%}}}%
    \put(0.74024508,0.47775699){\color[rgb]{0,0,0}\makebox(0,0)[lb]{\smash{76.6\%}}}%
    \put(0.31760883,0.01414326){\color[rgb]{0,0,0}\makebox(0,0)[b]{\smash{77.1\%}}}%
    \put(0.58572271,0.01414326){\color[rgb]{0,0,0}\makebox(0,0)[b]{\smash{22.9\%}}}%
    \put(0.73511828,0.01414326){\color[rgb]{0,0,0}\makebox(0,0)[lb]{\smash{100\%}}}%
    \put(0.06354729,0.37458765){\color[rgb]{0,0,0}\rotatebox{90}{\makebox(0,0)[b]{\smash{Actual}}}}%
    \put(0.45227273,0.77215243){\color[rgb]{0,0,0}\makebox(0,0)[b]{\smash{Predicted}}}%
  \end{picture}%
\endgroup%

%% file: fig/confusion-officeII.pdf_tex
\begingroup%
  \makeatletter%
  \providecommand\color[2][]{%
    \errmessage{(Inkscape) Color is used for the text in Inkscape, but the package 'color.sty' is not loaded}%
    \renewcommand\color[2][]{}%
  }%
  \providecommand\transparent[1]{%
    \errmessage{(Inkscape) Transparency is used (non-zero) for the text in Inkscape, but the package 'transparent.sty' is not loaded}%
    \renewcommand\transparent[1]{}%
  }%
  \providecommand\rotatebox[2]{#2}%
  \ifx\svgwidth\undefined%
    \setlength{\unitlength}{115.2bp}%
    \ifx\svgscale\undefined%
      \relax%
    \else%
      \setlength{\unitlength}{\unitlength * \real{\svgscale}}%
    \fi%
  \else%
    \setlength{\unitlength}{\svgwidth}%
  \fi%
  \global\let\svgwidth\undefined%
  \global\let\svgscale\undefined%
  \makeatother%
  \begin{picture}(1,0.84375)%
    \put(0,0){\includegraphics[width=\unitlength]{confusion-officeII.pdf}}%
    \put(0.31760883,0.47775699){\color[rgb]{0,0,0}\makebox(0,0)[b]{\smash{77.1\%}}}%
    \put(0.58521409,0.47775699){\color[rgb]{0,0,0}\makebox(0,0)[b]{\smash{0\%}}}%
    \put(0.31785297,0.20692408){\color[rgb]{0,0,0}\makebox(0,0)[b]{\smash{2.8\%}}}%
    \put(0.58572271,0.20692408){\color[rgb]{0,0,0}\makebox(0,0)[b]{\smash{20.1\%}}}%
    \put(0.31738504,0.67089087){\color[rgb]{0,0,0}\makebox(0,0)[b]{\smash{S}}}%
    \put(0.58399338,0.67089087){\color[rgb]{0,0,0}\makebox(0,0)[b]{\smash{C}}}%
    \put(0.16237356,0.47716698){\color[rgb]{0,0,0}\makebox(0,0)[rb]{\smash{S}}}%
    \put(0.16107147,0.20633449){\color[rgb]{0,0,0}\makebox(0,0)[rb]{\smash{C}}}%
    \put(0.74175056,0.20692408){\color[rgb]{0,0,0}\makebox(0,0)[lb]{\smash{22.9\%}}}%
    \put(0.74024508,0.47775699){\color[rgb]{0,0,0}\makebox(0,0)[lb]{\smash{77.1\%}}}%
    \put(0.31760883,0.01414326){\color[rgb]{0,0,0}\makebox(0,0)[b]{\smash{79.9\%}}}%
    \put(0.58572271,0.01414326){\color[rgb]{0,0,0}\makebox(0,0)[b]{\smash{20.1\%}}}%
    \put(0.73511828,0.01414326){\color[rgb]{0,0,0}\makebox(0,0)[lb]{\smash{100\%}}}%
    \put(0.06354729,0.37458765){\color[rgb]{0,0,0}\rotatebox{90}{\makebox(0,0)[b]{\smash{Actual}}}}%
    \put(0.45227273,0.77215243){\color[rgb]{0,0,0}\makebox(0,0)[b]{\smash{Predicted}}}%
  \end{picture}%
\endgroup%

%% file: fig/confusion-cafeteria.pdf_tex
\begingroup%
  \makeatletter%
  \providecommand\color[2][]{%
    \errmessage{(Inkscape) Color is used for the text in Inkscape, but the package 'color.sty' is not loaded}%
    \renewcommand\color[2][]{}%
  }%
  \providecommand\transparent[1]{%
    \errmessage{(Inkscape) Transparency is used (non-zero) for the text in Inkscape, but the package 'transparent.sty' is not loaded}%
    \renewcommand\transparent[1]{}%
  }%
  \providecommand\rotatebox[2]{#2}%
  \ifx\svgwidth\undefined%
    \setlength{\unitlength}{115.2bp}%
    \ifx\svgscale\undefined%
      \relax%
    \else%
      \setlength{\unitlength}{\unitlength * \real{\svgscale}}%
    \fi%
  \else%
    \setlength{\unitlength}{\svgwidth}%
  \fi%
  \global\let\svgwidth\undefined%
  \global\let\svgscale\undefined%
  \makeatother%
  \begin{picture}(1,0.84375)%
    \put(0,0){\includegraphics[width=\unitlength]{confusion-cafeteria.pdf}}%
    \put(0.31760883,0.47775699){\color[rgb]{0,0,0}\makebox(0,0)[b]{\smash{62.2\%}}}%
    \put(0.58521409,0.47775699){\color[rgb]{0,0,0}\makebox(0,0)[b]{\smash{5.8\%}}}%
    \put(0.31785297,0.20692408){\color[rgb]{0,0,0}\makebox(0,0)[b]{\smash{4.3\%}}}%
    \put(0.58572271,0.20692408){\color[rgb]{0,0,0}\makebox(0,0)[b]{\smash{27.7\%}}}%
    \put(0.31738504,0.67089087){\color[rgb]{0,0,0}\makebox(0,0)[b]{\smash{S}}}%
    \put(0.58399338,0.67089087){\color[rgb]{0,0,0}\makebox(0,0)[b]{\smash{C}}}%
    \put(0.16237356,0.47716698){\color[rgb]{0,0,0}\makebox(0,0)[rb]{\smash{S}}}%
    \put(0.16107147,0.20633449){\color[rgb]{0,0,0}\makebox(0,0)[rb]{\smash{C}}}%
    \put(0.74175056,0.20692408){\color[rgb]{0,0,0}\makebox(0,0)[lb]{\smash{32.0\%}}}%
    \put(0.74024508,0.47775699){\color[rgb]{0,0,0}\makebox(0,0)[lb]{\smash{68.0\%}}}%
    \put(0.31760883,0.01414326){\color[rgb]{0,0,0}\makebox(0,0)[b]{\smash{66.5\%}}}%
    \put(0.58572271,0.01414326){\color[rgb]{0,0,0}\makebox(0,0)[b]{\smash{33.5\%}}}%
    \put(0.73511828,0.01414326){\color[rgb]{0,0,0}\makebox(0,0)[lb]{\smash{100\%}}}%
    \put(0.06354729,0.37458765){\color[rgb]{0,0,0}\rotatebox{90}{\makebox(0,0)[b]{\smash{Actual}}}}%
    \put(0.45227273,0.77215243){\color[rgb]{0,0,0}\makebox(0,0)[b]{\smash{Predicted}}}%
  \end{picture}%
\endgroup%

%% file: fig/confusion-outdoor.pdf_tex
\begingroup%
  \makeatletter%
  \providecommand\color[2][]{%
    \errmessage{(Inkscape) Color is used for the text in Inkscape, but the package 'color.sty' is not loaded}%
    \renewcommand\color[2][]{}%
  }%
  \providecommand\transparent[1]{%
    \errmessage{(Inkscape) Transparency is used (non-zero) for the text in Inkscape, but the package 'transparent.sty' is not loaded}%
    \renewcommand\transparent[1]{}%
  }%
  \providecommand\rotatebox[2]{#2}%
  \ifx\svgwidth\undefined%
    \setlength{\unitlength}{115.2bp}%
    \ifx\svgscale\undefined%
      \relax%
    \else%
      \setlength{\unitlength}{\unitlength * \real{\svgscale}}%
    \fi%
  \else%
    \setlength{\unitlength}{\svgwidth}%
  \fi%
  \global\let\svgwidth\undefined%
  \global\let\svgscale\undefined%
  \makeatother%
  \begin{picture}(1,0.84375)%
    \put(0,0){\includegraphics[width=\unitlength]{confusion-outdoor.pdf}}%
    \put(0.31760883,0.47775699){\color[rgb]{0,0,0}\makebox(0,0)[b]{\smash{58.5\%}}}%
    \put(0.58521409,0.47775699){\color[rgb]{0,0,0}\makebox(0,0)[b]{\smash{2.5\%}}}%
    \put(0.31785297,0.20692408){\color[rgb]{0,0,0}\makebox(0,0)[b]{\smash{15.2\%}}}%
    \put(0.58572271,0.20692408){\color[rgb]{0,0,0}\makebox(0,0)[b]{\smash{23.8\%}}}%
    \put(0.31738504,0.67089087){\color[rgb]{0,0,0}\makebox(0,0)[b]{\smash{S}}}%
    \put(0.58399338,0.67089087){\color[rgb]{0,0,0}\makebox(0,0)[b]{\smash{C}}}%
    \put(0.16237356,0.47716698){\color[rgb]{0,0,0}\makebox(0,0)[rb]{\smash{S}}}%
    \put(0.16107147,0.20633449){\color[rgb]{0,0,0}\makebox(0,0)[rb]{\smash{C}}}%
    \put(0.74175056,0.20692408){\color[rgb]{0,0,0}\makebox(0,0)[lb]{\smash{39.0\%}}}%
    \put(0.74024508,0.47775699){\color[rgb]{0,0,0}\makebox(0,0)[lb]{\smash{61.0\%}}}%
    \put(0.31760883,0.01414326){\color[rgb]{0,0,0}\makebox(0,0)[b]{\smash{73.7\%}}}%
    \put(0.58572271,0.01414326){\color[rgb]{0,0,0}\makebox(0,0)[b]{\smash{26.3\%}}}%
    \put(0.73511828,0.01414326){\color[rgb]{0,0,0}\makebox(0,0)[lb]{\smash{100\%}}}%
    \put(0.06354729,0.37458765){\color[rgb]{0,0,0}\rotatebox{90}{\makebox(0,0)[b]{\smash{Actual}}}}%
    \put(0.45227273,0.77215243){\color[rgb]{0,0,0}\makebox(0,0)[b]{\smash{Predicted}}}%
  \end{picture}%
\endgroup%

%% file: fig/conforming-breakdown.pdf_tex
\begingroup
  \makeatletter
  \providecommand\color[2][]{%
    \GenericError{(gnuplot) \space\space\space\@spaces}{%
      Package color not loaded in conjunction with
      terminal option `colourtext'%
    }{See the gnuplot documentation for explanation.%
    }{Either use 'blacktext' in gnuplot or load the package
      color.sty in LaTeX.}%
    \renewcommand\color[2][]{}%
  }%
  \providecommand\includegraphics[2][]{%
    \GenericError{(gnuplot) \space\space\space\@spaces}{%
      Package graphicx or graphics not loaded%
    }{See the gnuplot documentation for explanation.%
    }{The gnuplot epslatex terminal needs graphicx.sty or graphics.sty.}%
    \renewcommand\includegraphics[2][]{}%
  }%
  \providecommand\rotatebox[2]{#2}%
  \@ifundefined{ifGPcolor}{%
    \newif\ifGPcolor
    \GPcolortrue
  }{}%
  \@ifundefined{ifGPblacktext}{%
    \newif\ifGPblacktext
    \GPblacktextfalse
  }{}%
  \let\gplgaddtomacro\g@addto@macro
  \gdef\gplbacktext{}%
  \gdef\gplfronttext{}%
  \makeatother
  \ifGPblacktext
    \def\colorrgb#1{}%
    \def\colorgray#1{}%
  \else
    \ifGPcolor
      \def\colorrgb#1{\color[rgb]{#1}}%
      \def\colorgray#1{\color[gray]{#1}}%
      \expandafter\def\csname LTw\endcsname{\color{white}}%
      \expandafter\def\csname LTb\endcsname{\color{black}}%
      \expandafter\def\csname LTa\endcsname{\color{black}}%
      \expandafter\def\csname LT0\endcsname{\color[rgb]{1,0,0}}%
      \expandafter\def\csname LT1\endcsname{\color[rgb]{0,1,0}}%
      \expandafter\def\csname LT2\endcsname{\color[rgb]{0,0,1}}%
      \expandafter\def\csname LT3\endcsname{\color[rgb]{1,0,1}}%
      \expandafter\def\csname LT4\endcsname{\color[rgb]{0,1,1}}%
      \expandafter\def\csname LT5\endcsname{\color[rgb]{1,1,0}}%
      \expandafter\def\csname LT6\endcsname{\color[rgb]{0,0,0}}%
      \expandafter\def\csname LT7\endcsname{\color[rgb]{1,0.3,0}}%
      \expandafter\def\csname LT8\endcsname{\color[rgb]{0.5,0.5,0.5}}%
    \else
      \def\colorrgb#1{\color{black}}%
      \def\colorgray#1{\color[gray]{#1}}%
      \expandafter\def\csname LTw\endcsname{\color{white}}%
      \expandafter\def\csname LTb\endcsname{\color{black}}%
      \expandafter\def\csname LTa\endcsname{\color{black}}%
      \expandafter\def\csname LT0\endcsname{\color{black}}%
      \expandafter\def\csname LT1\endcsname{\color{black}}%
      \expandafter\def\csname LT2\endcsname{\color{black}}%
      \expandafter\def\csname LT3\endcsname{\color{black}}%
      \expandafter\def\csname LT4\endcsname{\color{black}}%
      \expandafter\def\csname LT5\endcsname{\color{black}}%
      \expandafter\def\csname LT6\endcsname{\color{black}}%
      \expandafter\def\csname LT7\endcsname{\color{black}}%
      \expandafter\def\csname LT8\endcsname{\color{black}}%
    \fi
  \fi
    \setlength{\unitlength}{0.0500bp}%
    \ifx\gptboxheight\undefined%
      \newlength{\gptboxheight}%
      \newlength{\gptboxwidth}%
      \newsavebox{\gptboxtext}%
    \fi%
    \setlength{\fboxrule}{0.5pt}%
    \setlength{\fboxsep}{1pt}%
\begin{picture}(4960.00,2660.00)%
    \gplgaddtomacro\gplbacktext{%
      \colorrgb{0.50,0.50,0.50}%
      \put(667,356){\makebox(0,0)[r]{\strut{}\fns{0\%}}}%
      \colorrgb{0.50,0.50,0.50}%
      \put(667,761){\makebox(0,0)[r]{\strut{}\fns{20\%}}}%
      \colorrgb{0.50,0.50,0.50}%
      \put(667,1166){\makebox(0,0)[r]{\strut{}\fns{40\%}}}%
      \colorrgb{0.50,0.50,0.50}%
      \put(667,1570){\makebox(0,0)[r]{\strut{}\fns{60\%}}}%
      \colorrgb{0.50,0.50,0.50}%
      \put(667,1975){\makebox(0,0)[r]{\strut{}\fns{80\%}}}%
      \colorrgb{0.50,0.50,0.50}%
      \put(667,2380){\makebox(0,0)[r]{\strut{}\fns{100\%}}}%
      \colorrgb{0.50,0.50,0.50}%
      \put(1351,178){\makebox(0,0){\strut{}Office I}}%
      \colorrgb{0.50,0.50,0.50}%
      \put(2266,178){\makebox(0,0){\strut{}Office II}}%
      \colorrgb{0.50,0.50,0.50}%
      \put(3182,178){\makebox(0,0){\strut{}Cafeteria}}%
      \colorrgb{0.50,0.50,0.50}%
      \put(4097,178){\makebox(0,0){\strut{}Outdoor}}%
    }%
    \gplgaddtomacro\gplfronttext{%
      \colorrgb{0.00,0.00,0.00}%
      \put(1351,866){\makebox(0,0){\strut{}\sss{ 50.4\%}}}%
      \colorrgb{0.00,0.00,0.00}%
      \put(2266,735){\makebox(0,0){\strut{}\sss{ 37.5\%}}}%
      \colorrgb{0.00,0.00,0.00}%
      \put(3182,1368){\makebox(0,0){\strut{}\sss{100\%}}}%
      \colorrgb{0.00,0.00,0.00}%
      \put(4097,1368){\makebox(0,0){\strut{}\sss{100\%}}}%
      \colorrgb{0.00,0.00,0.00}%
      \put(1351,1810){\makebox(0,0){\strut{}\sss{ 42.9\%}}}%
      \colorrgb{0.00,0.00,0.00}%
      \put(2266,1747){\makebox(0,0){\strut{}\sss{ 62.5\%}}}%
      \colorrgb{0.00,0.00,0.00}%
      \put(3182,2380){\makebox(0,0){\strut{}}}%
      \colorrgb{0.00,0.00,0.00}%
      \put(4097,2380){\makebox(0,0){\strut{}}}%
      \colorrgb{0.00,0.00,0.00}%
      \put(1351,2312){\makebox(0,0){\strut{}\sss{ 6.7\%}}}%
      \colorrgb{0.00,0.00,0.00}%
      \put(2266,2380){\makebox(0,0){\strut{}}}%
      \colorrgb{0.00,0.00,0.00}%
      \put(3182,2380){\makebox(0,0){\strut{}}}%
      \colorrgb{0.00,0.00,0.00}%
      \put(4097,2380){\makebox(0,0){\strut{}}}%
      \csname LTb\endcsname%
      \put(4106,1816){\makebox(0,0)[r]{\strut{}\sss{Mobile}}}%
      \csname LTb\endcsname%
      \put(4106,2011){\makebox(0,0)[r]{\strut{}\sss{No Measurement}}}%
      \csname LTb\endcsname%
      \put(4106,2206){\makebox(0,0)[r]{\strut{}\sss{Collapsed}}}%
    }%
    \gplbacktext
    \put(0,0){\includegraphics{conforming-breakdown}}%
    \gplfronttext
  \end{picture}%
\endgroup

%% file: fig/collection_time.pdf_tex
\begingroup
  \makeatletter
  \providecommand\color[2][]{%
    \GenericError{(gnuplot) \space\space\space\@spaces}{%
      Package color not loaded in conjunction with
      terminal option `colourtext'%
    }{See the gnuplot documentation for explanation.%
    }{Either use 'blacktext' in gnuplot or load the package
      color.sty in LaTeX.}%
    \renewcommand\color[2][]{}%
  }%
  \providecommand\includegraphics[2][]{%
    \GenericError{(gnuplot) \space\space\space\@spaces}{%
      Package graphicx or graphics not loaded%
    }{See the gnuplot documentation for explanation.%
    }{The gnuplot epslatex terminal needs graphicx.sty or graphics.sty.}%
    \renewcommand\includegraphics[2][]{}%
  }%
  \providecommand\rotatebox[2]{#2}%
  \@ifundefined{ifGPcolor}{%
    \newif\ifGPcolor
    \GPcolortrue
  }{}%
  \@ifundefined{ifGPblacktext}{%
    \newif\ifGPblacktext
    \GPblacktextfalse
  }{}%
  \let\gplgaddtomacro\g@addto@macro
  \gdef\gplbacktext{}%
  \gdef\gplfronttext{}%
  \makeatother
  \ifGPblacktext
    \def\colorrgb#1{}%
    \def\colorgray#1{}%
  \else
    \ifGPcolor
      \def\colorrgb#1{\color[rgb]{#1}}%
      \def\colorgray#1{\color[gray]{#1}}%
      \expandafter\def\csname LTw\endcsname{\color{white}}%
      \expandafter\def\csname LTb\endcsname{\color{black}}%
      \expandafter\def\csname LTa\endcsname{\color{black}}%
      \expandafter\def\csname LT0\endcsname{\color[rgb]{1,0,0}}%
      \expandafter\def\csname LT1\endcsname{\color[rgb]{0,1,0}}%
      \expandafter\def\csname LT2\endcsname{\color[rgb]{0,0,1}}%
      \expandafter\def\csname LT3\endcsname{\color[rgb]{1,0,1}}%
      \expandafter\def\csname LT4\endcsname{\color[rgb]{0,1,1}}%
      \expandafter\def\csname LT5\endcsname{\color[rgb]{1,1,0}}%
      \expandafter\def\csname LT6\endcsname{\color[rgb]{0,0,0}}%
      \expandafter\def\csname LT7\endcsname{\color[rgb]{1,0.3,0}}%
      \expandafter\def\csname LT8\endcsname{\color[rgb]{0.5,0.5,0.5}}%
    \else
      \def\colorrgb#1{\color{black}}%
      \def\colorgray#1{\color[gray]{#1}}%
      \expandafter\def\csname LTw\endcsname{\color{white}}%
      \expandafter\def\csname LTb\endcsname{\color{black}}%
      \expandafter\def\csname LTa\endcsname{\color{black}}%
      \expandafter\def\csname LT0\endcsname{\color{black}}%
      \expandafter\def\csname LT1\endcsname{\color{black}}%
      \expandafter\def\csname LT2\endcsname{\color{black}}%
      \expandafter\def\csname LT3\endcsname{\color{black}}%
      \expandafter\def\csname LT4\endcsname{\color{black}}%
      \expandafter\def\csname LT5\endcsname{\color{black}}%
      \expandafter\def\csname LT6\endcsname{\color{black}}%
      \expandafter\def\csname LT7\endcsname{\color{black}}%
      \expandafter\def\csname LT8\endcsname{\color{black}}%
    \fi
  \fi
    \setlength{\unitlength}{0.0500bp}%
    \ifx\gptboxheight\undefined%
      \newlength{\gptboxheight}%
      \newlength{\gptboxwidth}%
      \newsavebox{\gptboxtext}%
    \fi%
    \setlength{\fboxrule}{0.5pt}%
    \setlength{\fboxsep}{1pt}%
\begin{picture}(3300.00,2040.00)%
    \gplgaddtomacro\gplbacktext{%
      \colorrgb{0.50,0.50,0.50}%
      \put(667,569){\makebox(0,0)[r]{\strut{}\sss{0}}}%
      \colorrgb{0.50,0.50,0.50}%
      \put(667,827){\makebox(0,0)[r]{\strut{}\sss{20}}}%
      \colorrgb{0.50,0.50,0.50}%
      \put(667,1086){\makebox(0,0)[r]{\strut{}\sss{40}}}%
      \colorrgb{0.50,0.50,0.50}%
      \put(667,1344){\makebox(0,0)[r]{\strut{}\sss{60}}}%
      \colorrgb{0.50,0.50,0.50}%
      \put(667,1603){\makebox(0,0)[r]{\strut{}\sss{80}}}%
      \colorrgb{0.50,0.50,0.50}%
      \put(667,1861){\makebox(0,0)[r]{\strut{}\sss{100}}}%
      \colorrgb{0.50,0.50,0.50}%
      \put(756,391){\makebox(0,0){\strut{}\sss{0}}}%
      \colorrgb{0.50,0.50,0.50}%
      \put(1024,391){\makebox(0,0){\strut{}\sss{50}}}%
      \colorrgb{0.50,0.50,0.50}%
      \put(1292,391){\makebox(0,0){\strut{}\sss{100}}}%
      \colorrgb{0.50,0.50,0.50}%
      \put(1559,391){\makebox(0,0){\strut{}\sss{150}}}%
      \colorrgb{0.50,0.50,0.50}%
      \put(1827,391){\makebox(0,0){\strut{}\sss{200}}}%
      \colorrgb{0.50,0.50,0.50}%
      \put(2095,391){\makebox(0,0){\strut{}\sss{250}}}%
      \colorrgb{0.50,0.50,0.50}%
      \put(2363,391){\makebox(0,0){\strut{}\sss{300}}}%
      \colorrgb{0.50,0.50,0.50}%
      \put(2630,391){\makebox(0,0){\strut{}\sss{350}}}%
      \colorrgb{0.50,0.50,0.50}%
      \put(2898,391){\makebox(0,0){\strut{}\sss{400}}}%
    }%
    \gplgaddtomacro\gplfronttext{%
      \csname LTb\endcsname%
      \put(133,1215){\rotatebox{-270}{\makebox(0,0){\strut{}\fns{Runtime (\si\second)}}}}%
      \csname LTb\endcsname%
      \put(1894,124){\makebox(0,0){\strut{}\fns{\# of Participating Identities}}}%
      \csname LTb\endcsname%
      \put(2073,1710){\makebox(0,0)[r]{\strut{}\fns{Total}}}%
      \csname LTb\endcsname%
      \put(2073,1550){\makebox(0,0)[r]{\strut{}\fns{RSST}}}%
      \csname LTb\endcsname%
      \put(2073,1390){\makebox(0,0)[r]{\strut{}\fns{HELLO II}}}%
      \csname LTb\endcsname%
      \put(2073,1230){\makebox(0,0)[r]{\strut{}\fns{HELLO I}}}%
    }%
    \gplbacktext
    \put(0,0){\includegraphics{collection_time}}%
    \gplfronttext
  \end{picture}%
\endgroup

%% file: fig/collection_energy.pdf_tex
\begingroup
  \makeatletter
  \providecommand\color[2][]{%
    \GenericError{(gnuplot) \space\space\space\@spaces}{%
      Package color not loaded in conjunction with
      terminal option `colourtext'%
    }{See the gnuplot documentation for explanation.%
    }{Either use 'blacktext' in gnuplot or load the package
      color.sty in LaTeX.}%
    \renewcommand\color[2][]{}%
  }%
  \providecommand\includegraphics[2][]{%
    \GenericError{(gnuplot) \space\space\space\@spaces}{%
      Package graphicx or graphics not loaded%
    }{See the gnuplot documentation for explanation.%
    }{The gnuplot epslatex terminal needs graphicx.sty or graphics.sty.}%
    \renewcommand\includegraphics[2][]{}%
  }%
  \providecommand\rotatebox[2]{#2}%
  \@ifundefined{ifGPcolor}{%
    \newif\ifGPcolor
    \GPcolortrue
  }{}%
  \@ifundefined{ifGPblacktext}{%
    \newif\ifGPblacktext
    \GPblacktextfalse
  }{}%
  \let\gplgaddtomacro\g@addto@macro
  \gdef\gplbacktext{}%
  \gdef\gplfronttext{}%
  \makeatother
  \ifGPblacktext
    \def\colorrgb#1{}%
    \def\colorgray#1{}%
  \else
    \ifGPcolor
      \def\colorrgb#1{\color[rgb]{#1}}%
      \def\colorgray#1{\color[gray]{#1}}%
      \expandafter\def\csname LTw\endcsname{\color{white}}%
      \expandafter\def\csname LTb\endcsname{\color{black}}%
      \expandafter\def\csname LTa\endcsname{\color{black}}%
      \expandafter\def\csname LT0\endcsname{\color[rgb]{1,0,0}}%
      \expandafter\def\csname LT1\endcsname{\color[rgb]{0,1,0}}%
      \expandafter\def\csname LT2\endcsname{\color[rgb]{0,0,1}}%
      \expandafter\def\csname LT3\endcsname{\color[rgb]{1,0,1}}%
      \expandafter\def\csname LT4\endcsname{\color[rgb]{0,1,1}}%
      \expandafter\def\csname LT5\endcsname{\color[rgb]{1,1,0}}%
      \expandafter\def\csname LT6\endcsname{\color[rgb]{0,0,0}}%
      \expandafter\def\csname LT7\endcsname{\color[rgb]{1,0.3,0}}%
      \expandafter\def\csname LT8\endcsname{\color[rgb]{0.5,0.5,0.5}}%
    \else
      \def\colorrgb#1{\color{black}}%
      \def\colorgray#1{\color[gray]{#1}}%
      \expandafter\def\csname LTw\endcsname{\color{white}}%
      \expandafter\def\csname LTb\endcsname{\color{black}}%
      \expandafter\def\csname LTa\endcsname{\color{black}}%
      \expandafter\def\csname LT0\endcsname{\color{black}}%
      \expandafter\def\csname LT1\endcsname{\color{black}}%
      \expandafter\def\csname LT2\endcsname{\color{black}}%
      \expandafter\def\csname LT3\endcsname{\color{black}}%
      \expandafter\def\csname LT4\endcsname{\color{black}}%
      \expandafter\def\csname LT5\endcsname{\color{black}}%
      \expandafter\def\csname LT6\endcsname{\color{black}}%
      \expandafter\def\csname LT7\endcsname{\color{black}}%
      \expandafter\def\csname LT8\endcsname{\color{black}}%
    \fi
  \fi
    \setlength{\unitlength}{0.0500bp}%
    \ifx\gptboxheight\undefined%
      \newlength{\gptboxheight}%
      \newlength{\gptboxwidth}%
      \newsavebox{\gptboxtext}%
    \fi%
    \setlength{\fboxrule}{0.5pt}%
    \setlength{\fboxsep}{1pt}%
\begin{picture}(3300.00,2040.00)%
    \gplgaddtomacro\gplbacktext{%
      \colorrgb{0.50,0.50,0.50}%
      \put(515,569){\makebox(0,0)[r]{\strut{}\sss{0}}}%
      \colorrgb{0.50,0.50,0.50}%
      \put(515,892){\makebox(0,0)[r]{\strut{}\sss{5}}}%
      \colorrgb{0.50,0.50,0.50}%
      \put(515,1215){\makebox(0,0)[r]{\strut{}\sss{10}}}%
      \colorrgb{0.50,0.50,0.50}%
      \put(515,1538){\makebox(0,0)[r]{\strut{}\sss{15}}}%
      \colorrgb{0.50,0.50,0.50}%
      \put(515,1861){\makebox(0,0)[r]{\strut{}\sss{20}}}%
      \colorrgb{0.50,0.50,0.50}%
      \put(604,391){\makebox(0,0){\strut{}\sss{0}}}%
      \colorrgb{0.50,0.50,0.50}%
      \put(890,391){\makebox(0,0){\strut{}\sss{50}}}%
      \colorrgb{0.50,0.50,0.50}%
      \put(1175,391){\makebox(0,0){\strut{}\sss{100}}}%
      \colorrgb{0.50,0.50,0.50}%
      \put(1461,391){\makebox(0,0){\strut{}\sss{150}}}%
      \colorrgb{0.50,0.50,0.50}%
      \put(1747,391){\makebox(0,0){\strut{}\sss{200}}}%
      \colorrgb{0.50,0.50,0.50}%
      \put(2032,391){\makebox(0,0){\strut{}\sss{250}}}%
      \colorrgb{0.50,0.50,0.50}%
      \put(2318,391){\makebox(0,0){\strut{}\sss{300}}}%
      \colorrgb{0.50,0.50,0.50}%
      \put(2604,391){\makebox(0,0){\strut{}\sss{350}}}%
      \colorrgb{0.50,0.50,0.50}%
      \put(2889,391){\makebox(0,0){\strut{}\sss{400}}}%
    }%
    \gplgaddtomacro\gplfronttext{%
      \csname LTb\endcsname%
      \put(133,1215){\rotatebox{-270}{\makebox(0,0){\strut{}\fns{Consumption (\si\joule)}}}}%
      \csname LTb\endcsname%
      \put(1818,124){\makebox(0,0){\strut{}\fns{\# of Participating Identities}}}%
      \csname LTb\endcsname%
      \put(2455,1701){\makebox(0,0)[r]{\strut{}\fns{Initiator Total}}}%
      \csname LTb\endcsname%
      \put(2455,1523){\makebox(0,0)[r]{\strut{}\fns{Participant Total}}}%
    }%
    \gplbacktext
    \put(0,0){\includegraphics{collection_energy}}%
    \gplfronttext
  \end{picture}%
\endgroup

%% file: fig/classification_time_energy.pdf_tex
\begingroup
  \makeatletter
  \providecommand\color[2][]{%
    \GenericError{(gnuplot) \space\space\space\@spaces}{%
      Package color not loaded in conjunction with
      terminal option `colourtext'%
    }{See the gnuplot documentation for explanation.%
    }{Either use 'blacktext' in gnuplot or load the package
      color.sty in LaTeX.}%
    \renewcommand\color[2][]{}%
  }%
  \providecommand\includegraphics[2][]{%
    \GenericError{(gnuplot) \space\space\space\@spaces}{%
      Package graphicx or graphics not loaded%
    }{See the gnuplot documentation for explanation.%
    }{The gnuplot epslatex terminal needs graphicx.sty or graphics.sty.}%
    \renewcommand\includegraphics[2][]{}%
  }%
  \providecommand\rotatebox[2]{#2}%
  \@ifundefined{ifGPcolor}{%
    \newif\ifGPcolor
    \GPcolortrue
  }{}%
  \@ifundefined{ifGPblacktext}{%
    \newif\ifGPblacktext
    \GPblacktextfalse
  }{}%
  \let\gplgaddtomacro\g@addto@macro
  \gdef\gplbacktext{}%
  \gdef\gplfronttext{}%
  \makeatother
  \ifGPblacktext
    \def\colorrgb#1{}%
    \def\colorgray#1{}%
  \else
    \ifGPcolor
      \def\colorrgb#1{\color[rgb]{#1}}%
      \def\colorgray#1{\color[gray]{#1}}%
      \expandafter\def\csname LTw\endcsname{\color{white}}%
      \expandafter\def\csname LTb\endcsname{\color{black}}%
      \expandafter\def\csname LTa\endcsname{\color{black}}%
      \expandafter\def\csname LT0\endcsname{\color[rgb]{1,0,0}}%
      \expandafter\def\csname LT1\endcsname{\color[rgb]{0,1,0}}%
      \expandafter\def\csname LT2\endcsname{\color[rgb]{0,0,1}}%
      \expandafter\def\csname LT3\endcsname{\color[rgb]{1,0,1}}%
      \expandafter\def\csname LT4\endcsname{\color[rgb]{0,1,1}}%
      \expandafter\def\csname LT5\endcsname{\color[rgb]{1,1,0}}%
      \expandafter\def\csname LT6\endcsname{\color[rgb]{0,0,0}}%
      \expandafter\def\csname LT7\endcsname{\color[rgb]{1,0.3,0}}%
      \expandafter\def\csname LT8\endcsname{\color[rgb]{0.5,0.5,0.5}}%
    \else
      \def\colorrgb#1{\color{black}}%
      \def\colorgray#1{\color[gray]{#1}}%
      \expandafter\def\csname LTw\endcsname{\color{white}}%
      \expandafter\def\csname LTb\endcsname{\color{black}}%
      \expandafter\def\csname LTa\endcsname{\color{black}}%
      \expandafter\def\csname LT0\endcsname{\color{black}}%
      \expandafter\def\csname LT1\endcsname{\color{black}}%
      \expandafter\def\csname LT2\endcsname{\color{black}}%
      \expandafter\def\csname LT3\endcsname{\color{black}}%
      \expandafter\def\csname LT4\endcsname{\color{black}}%
      \expandafter\def\csname LT5\endcsname{\color{black}}%
      \expandafter\def\csname LT6\endcsname{\color{black}}%
      \expandafter\def\csname LT7\endcsname{\color{black}}%
      \expandafter\def\csname LT8\endcsname{\color{black}}%
    \fi
  \fi
    \setlength{\unitlength}{0.0500bp}%
    \ifx\gptboxheight\undefined%
      \newlength{\gptboxheight}%
      \newlength{\gptboxwidth}%
      \newsavebox{\gptboxtext}%
    \fi%
    \setlength{\fboxrule}{0.5pt}%
    \setlength{\fboxsep}{1pt}%
\begin{picture}(3300.00,2040.00)%
    \gplgaddtomacro\gplbacktext{%
      \colorrgb{0.50,0.50,0.50}%
      \put(515,569){\makebox(0,0)[r]{\strut{}\fns{0}}}%
      \colorrgb{0.50,0.50,0.50}%
      \put(515,1000){\makebox(0,0)[r]{\strut{}\fns{5}}}%
      \colorrgb{0.50,0.50,0.50}%
      \put(515,1430){\makebox(0,0)[r]{\strut{}\fns{10}}}%
      \colorrgb{0.50,0.50,0.50}%
      \put(515,1861){\makebox(0,0)[r]{\strut{}\fns{15}}}%
      \colorrgb{0.50,0.50,0.50}%
      \put(604,391){\makebox(0,0){\strut{}\fns{0}}}%
      \colorrgb{0.50,0.50,0.50}%
      \put(1001,391){\makebox(0,0){\strut{}\fns{20}}}%
      \colorrgb{0.50,0.50,0.50}%
      \put(1397,391){\makebox(0,0){\strut{}\fns{40}}}%
      \colorrgb{0.50,0.50,0.50}%
      \put(1794,391){\makebox(0,0){\strut{}\fns{60}}}%
      \colorrgb{0.50,0.50,0.50}%
      \put(2190,391){\makebox(0,0){\strut{}\fns{80}}}%
      \colorrgb{0.50,0.50,0.50}%
      \put(2587,391){\makebox(0,0){\strut{}\fns{100}}}%
      \colorrgb{0.50,0.50,0.50}%
      \put(2676,569){\makebox(0,0)[l]{\strut{}$0$}}%
      \colorrgb{0.50,0.50,0.50}%
      \put(2676,892){\makebox(0,0)[l]{\strut{}$1$}}%
      \colorrgb{0.50,0.50,0.50}%
      \put(2676,1215){\makebox(0,0)[l]{\strut{}$2$}}%
      \colorrgb{0.50,0.50,0.50}%
      \put(2676,1538){\makebox(0,0)[l]{\strut{}$3$}}%
      \colorrgb{0.50,0.50,0.50}%
      \put(2676,1861){\makebox(0,0)[l]{\strut{}$4$}}%
    }%
    \gplgaddtomacro\gplfronttext{%
      \csname LTb\endcsname%
      \put(133,1215){\rotatebox{-270}{\makebox(0,0){\strut{}\fns{Runtime (\si\second)}}}}%
      \csname LTb\endcsname%
      \put(2942,1215){\rotatebox{-270}{\makebox(0,0){\strut{}\fns{Consumption (\si\joule)}}}}%
      \csname LTb\endcsname%
      \put(1595,124){\makebox(0,0){\strut{}\fns{\# of Participating Identities}}}%
      \csname LTb\endcsname%
      \put(1832,1674){\makebox(0,0)[r]{\strut{}\fns{Runtime}}}%
      \csname LTb\endcsname%
      \put(1832,1443){\makebox(0,0)[r]{\strut{}\fns{Energy}}}%
    }%
    \gplbacktext
    \put(0,0){\includegraphics{classification_time_energy}}%
    \gplfronttext
  \end{picture}%
\endgroup

%% file: discussion.tex
\section{Discussion}
\label{sec:discussion}

Sybil classification from untrusted observations is difficult and the
Mason test is not a silver bullet.  Not requiring trusted observations
is a significant improvement, but the test's limitations must be
carefully considered before deployment. As with any system intended
for real-world use, some decisions try to balance system complexity
and potential security weaknesses. In this section, we discuss these
trade-offs, limitations, and related concerns.

\noindent
\textbf{High Computation Time}: The collection phase time is governed
by the 802.11b-induced \SI{12}{\milli\second} per packet latency and
the classification runtime grows quickly, $\bigO(|\setI|^3)$.
Although typically fast (e.g., $<$\SI{5}{\second} for 5--10 nodes),
the Mason test is slower in high density areas (e.g., \SI{40}{\second}
for 100 nodes). However, it should be run infrequently, e.g., once or
twice per hour.  Topologies change slowly (most people change
locations infrequently) and many protocols requiring Sybil resistance
can handle the lag---they need only know a subset of the current
non-Sybil neighbors.

\noindent
\textbf{Easy Denial-of-Service Attack}: An attacker can force the
protocol to abort by creating many identities or jamming
transmissions from the conforming identities.  We cannot on commodity
802.11 devices solve another denial-of-service attack---simply jamming
the channel---so defending against these more-complicated variants is
ultimately useless.  Most locations will at most times be free of such
attackers---the Mason test provides a way to verify this condition,
reject any Sybils, and let other protocols operate knowing they are
Sybil-free.

\noindent
\textbf{Requires Several Conforming Neighbors}: The Mason test
requires true RSSI observations from some neighbors (i.e., 3) and is
easily defeated otherwise.  Although beyond the page limits of this
paper, protocols incorporating the Mason test can mitigate this risk
by (a) a priori estimation of the distribution of the number of
conforming neighbors and (b) careful composition of results from
multiple rounds to bound the failure probability.

\noindent
\textbf{Limit On Total Identities}: This limit (e.g., 400) is
unfortunately necessary to detect when conforming nodes are being
selectively jammed while keeping the test duration short enough that
most conforming nodes remain stationary.  We believe that most
wireless networks have typical node degrees well below 400 anyway.

\noindent
\textbf{Messages Must Be Signed}: Packets sent during the collection
phase are signed, which can be very slow with public key
schemes. However, this is easily mitigated by (a) pre-signing the
packets to keep the delay off the critical path or (b) using faster
secret-key-based schemes. Details are again omitted due to page
constraints.

\noindent
\textbf{Pre-Characterization Reveals RSSIs}: An attacker could
theoretically improve its collapsing probability by pre-characterizing
the wireless environment. We believe such attacks are impractical
because the required spatial granularity is about
\SI{6.25}{\centi\meter}, the device orientation is still unknown, and
environmental changes (e.g., people moving) reduces the usefulness of
prior characterization.

\noindent
\textbf{Prior Rounds Reveal RSSI Information}: The protocol defends
against this. Conforming nodes do not perform classification if their
RSSI observations are correlated with the prior rounds (see
\autoref{sec:classification}).

\noindent
\textbf{High False Positive Rates}: With the motion filter, the false
positive rate can be high, e.g., 20\% of conforming identities
rejected in some environments.  We believe this is acceptable because
most protocols requiring Sybil resistance need only a subset of honest
identities. For example, if for reliability some data is to be spread
among multiple neighbors, it is acceptable to spread it among a subset
chosen from 80\%, rather than all, of the non-Sybils.

%% file: conclusion.tex
\section{Conclusion}
\label{sec:conclusion}

We have described a method to use signalprints to detect Sybil attacks
in open ad hoc and delay-tolerant networks without requiring trust in
any other node or authority.  We use the inherent difficulty of
predicting RSSIs to separate true and false RSSI observations reported
by one-hop neighbors. Attackers using motion to defeat the signalprint
technique are detected by requiring low-latency retransmissions from
the same position.

The Mason test was implemented on HTC Magic smartphones and tested
with human participants in three environments. It eliminates
99.6\%--100\% of Sybil identities in office environments, 91\% in a
crowded high-motion cafeteria, and 96\% in a high-motion open outdoor
environment. It accepts 88\%--97\% of conforming identities in the
office environments, 87\% in the cafeteria, and 61\% in the outdoor
environment. The vast majority of rejected conforming identities were
eliminated due to motion.

%% file: main.bbl
\begin{thebibliography}{10}
\providecommand{\url}[1]{#1}
\csname url@samestyle\endcsname
\providecommand{\newblock}{\relax}
\providecommand{\bibinfo}[2]{#2}
\providecommand{\BIBentrySTDinterwordspacing}{\spaceskip=0pt\relax}
\providecommand{\BIBentryALTinterwordstretchfactor}{4}
\providecommand{\BIBentryALTinterwordspacing}{\spaceskip=\fontdimen2\font plus
\BIBentryALTinterwordstretchfactor\fontdimen3\font minus
  \fontdimen4\font\relax}
\providecommand{\BIBforeignlanguage}[2]{{%
\expandafter\ifx\csname l@#1\endcsname\relax
\typeout{** WARNING: IEEEtran.bst: No hyphenation pattern has been}%
\typeout{** loaded for the language `#1'. Using the pattern for}%
\typeout{** the default language instead.}%
\else
\language=\csname l@#1\endcsname
\fi
#2}}
\providecommand{\BIBdecl}{\relax}
\BIBdecl

\bibitem{hui11nov}
P.~Hui, J.~Crowcroft, and E.~Yoneki, ``{BUBBLE} rap: Social-based forwarding in
  delay tolerant networks,'' \emph{{IEEE} Trans.\ Mobile Computing}, vol.~10,
  no.~11, pp. 1576--1589, Nov. 2011.

\bibitem{xiang12apr}
\student{Y. Xiang}, \fstudent{L. S. Bai}, R.~Piedrahita, R.~P. Dick, Q.~Lv,
  M.~P. Hannigan, and L.~Shang,
  ``\href{\rdpubsurl/xiang12apr.html}{Collaborative calibration and sensor
  placement for mobile sensor networks},'' in \emph{Proc.\ Int.\ Conf.\
  Information Processing in Sensor Networks}, Apr. 2012, pp. 73--84.

\bibitem{gardner-stephen11aug}
P.~Gardner-Stephen, ``The {S}erval project: Practical wireless ad-hoc mobile
  telecommunications,'' Flinders University, Adelaide, South Australia, Tech.
  Rep., Aug. 2011.

\bibitem{douceur02mar}
J.~Douceur, ``The {S}ybil attack,'' in \emph{Proc.\ Int.\ Wkshp.\ Peer-to-Peer
  Systems}, Mar. 2002, pp. 251--260.

\bibitem{newsome04apr}
J.~Newsome, E.~Shi, D.~Song, and A.~Perrig, ``The {S}ybil attack in sensor
  networks: Analysis \& defenses,'' in \emph{Proc.\ Int.\ Conf.\ Information
  Processing in Sensor Networks}, Apr. 2004, pp. 259--268.

\bibitem{levine06oct}
B.~N. Levine, C.~Shields, and N.~B. Margolin, ``A survey of solutions to the
  {S}ybil attack,'' Department of Computer Science, University of Massachusetts
  Amherst, Amherst, MA, Tech. Rep., Oct. 2006.

\bibitem{zhou05nov}
H.~Zhou, M.~Mutka, and L.~Ni, ``Multiple-key cryptography-based distributed
  certificate authority in mobile ad-hoc networks,'' in \emph{Proc.\ Global
  Telecommunications Conf.}, Nov. 2005.

\bibitem{ramkumar05mar}
M.~Ramkumar and N.~Memon, ``An efficient key predistribution scheme for ad hoc
  network security,'' \emph{{IEEE} J.\ Selected Areas in Communications},
  vol.~23, pp. 611--621, Mar. 2005.

\bibitem{borisov06sep}
N.~Borisov, ``Computational puzzles as {S}ybil defenses,'' in \emph{Proc.\
  Int.\ Conf.\ Peer-to-Peer Computing}, Sep. 2006, pp. 171--176.

\bibitem{li12oct}
F.~Li, P.~Mittal, M.~Caesar, and N.~Borisov, ``{S}ybil{C}ontrol: Practical
  {S}ybil defense with computational puzzles,'' in \emph{Proc.\ Wkshp.\
  Scalable Trusted Computing}, Oct. 2012.

\bibitem{yu06sep}
H.~Yu, M.~Kaminsky, P.~B. Gibbons, and A.~Flaxman, ``{SybilGuard}: defending
  against {S}ybil attacks via social networks,'' in \emph{Proc.\ {ACM}
  {SIGCOMM} Computer Communication Review}, Sep. 2006, pp. 267--278.

\bibitem{yu08may}
H.~Yu, P.~Gibbons, M.~Kaminsky, and F.~Xiao, ``{SybilLimit}: A near-optimal
  social network defense against {S}ybil attacks,'' in \emph{Proc.\ Symp.\
  Security and Privacy}, May 2008, pp. 3--17.

\bibitem{rappaport02}
T.~S. Rappaport, \emph{Wireless Communications: Principles \& Practice}.\hskip
  1em plus 0.5em minus 0.4em\relax Prentice-Hall, NJ, 2002.

\bibitem{haeberlen04sep}
A.~Haeberlen, E.~Flannery, A.~M. Ladd, A.~Rudys, D.~S. Wallach, and L.~E.
  Kavraki, ``Practical robust localization over large-scale 802.11 wireless
  networks,'' in \emph{Proc.\ Int.\ Conf.\ Mobile Computing and Networking},
  Sep. 2004, pp. 70--84.

\bibitem{xiao09sep}
L.~Xiao, L.~J. Greenstein, N.~B. Mandayam, and W.~Trappe, ``Channel-based
  detection of {S}ybil attacks in wireless networks,'' \emph{{IEEE} Trans.\
  Information Forensics and Security}, vol.~4, no.~3, pp. 492--503, Sep. 2009.

\bibitem{faria06sep}
D.~B. Faria and D.~R. Cheriton, ``Detecting identity-based attacks in wireless
  networks using signalprints,'' in \emph{Proc.\ Wkshp.\ Wireless Security},
  Sep. 2006, pp. 43--52.

\bibitem{demirbas06jun}
M.~Demirbas and Y.~Song, ``An {RSSI}-based scheme for {S}ybil attack detection
  in wireless sensor networks,'' in \emph{Proc.\ Int.\ Symp.\ on a World of
  Wireless, Mobile, and Multimedia}, Jun. 2006, pp. 564--570.

\bibitem{li06sep-b}
Z.~Li, W.~Xu, R.~Miller, and W.~Trappe, ``Securing wireless systems via lower
  layer enforcements,'' in \emph{Proc.\ Wkshp.\ Wireless Security}, Sep. 2006,
  pp. 33--42.

\bibitem{li07dec}
Q.~Li and W.~Trappe, ``Detecting spoofing and anomalous traffic in wireless
  networks via forge-resistant relationships,'' \emph{{IEEE} Trans.\
  Information Forensics and Security}, vol.~2, no.~4, pp. 793--803, Dec. 2007.

\bibitem{chen10jun-b}
Y.~Chen, J.~Yang, W.~Trappe, and R.~P. Martin, ``Detecting and localizing
  identity-based attacks in wireless and sensor networks,'' \emph{{IEEE}
  Trans.\ Vehicular Technology}, vol.~5, no.~5, pp. 2418--2434, Jun. 2010.

\bibitem{suen05nov}
T.~Suen and A.~Yasinsac, ``Peer identification in wireless and sensor networks
  using signal properties,'' in \emph{Proc.\ Int.\ Conf.\ Mobile Adhoc and
  Sensor Systems}, Nov. 2005, pp. 826--833.

\bibitem{shaohe08dec}
S.~Lv, X.~Wang, X.~Zhao, and X.~Zhou, ``Detecting the {S}ybil attack
  coorperatively in wireless sensor networks,'' in \emph{Proc.\ Int.\ Conf.\
  Computational Intelligence and Security}, Dec. 2008, pp. 442--446.

\bibitem{abbas09dec}
S.~Abbas, M.~Merabti, and D.~Llewellyn-Jones, ``Signal strength based {S}ybil
  attack detection in wireless ad hoc networks,'' in \emph{Proc.\ Int.\ Conf.\
  Developments in {eS}ytems Engineering}, Dec. 2009, pp. 22--33.

\bibitem{bouassida09jul}
M.~S. Bouassida, G.~Guette, M.~Shawky, and B.~Ducourthial, ``{S}ybil nodes
  detection basedon received strength variations within {VANET},'' \emph{Int.\
  J.\ Network Security}, vol.~9, no.~1, pp. 22--33, Jul. 2009.

\bibitem{gesbert03apr}
D.~Gesbert, M.~Shafi, D.~Shiu, P.~J. Smith, and A.~Naguib, ``From theory to
  practice: An overview of {MIMO} space--time coded wireless systems,''
  \emph{{IEEE} J.\ Selected Areas in Communications}, vol.~21, no.~3, pp.
  281--302, Apr. 2003.

\bibitem{liu13dec}
Y.~Liu, D.~R. Bild, and R.~P. Dick, ``Extending channel comparison based
  {S}ybil detection to {MIMO} systems,'' Dept. of Electrical Engineering and
  Computer Science, University of Michigan, Tech. Rep.,
  \href{http://www.davidbild.org/publications/liu13dec.pdf}{http://www.davidbild.org/publications/liu13dec.pdf}.

\bibitem{hashemi92mayII}
H.~Hashemi, D.~Lee, and D.~Ehman, ``Statistical modeling of the indoor radio
  propagation channel -- part {II},'' in \emph{Proc.\ Vehicular Technology
  Conf.}, May 1992, pp. 839--843.

\bibitem{rappaport91may}
T.~S. Rappaport, S.~Y. Seidel, and K.~Takamizawa, ``Statistical channel impulse
  response models for factory and open plan building radio communication system
  design,'' \emph{{IEEE} Trans.\ on Communications}, vol.~39, no.~5, pp.
  794--806, May 1991.

\bibitem{humanbenchmark}
``Reaction time statistics,''
  \url{http://www.humanbenchmark.com/tests/reactiontime/stats.php}.

\end{thebibliography}
